\documentclass[final,3p,times]{elsarticle}
\biboptions{sort&compress}
\usepackage{amsmath,amssymb,amsfonts}
\usepackage{amsthm}
\usepackage{graphicx}
\usepackage{hyperref}
\usepackage{algc}
\usepackage{xcolor}

\definecolor{mycolor}{rgb}{0.7,0.3,0.3}

\usepackage{lineno}

\usepackage{amssymb,amsmath,graphicx}
\usepackage{epigraph}

\def\tr{{\raise0pt\hbox{$\scriptscriptstyle\top$}}}

\begin{document}

\begin{frontmatter}

\title{Dynamic and Thermodynamic Models of Adaptation}

\author[LeicMath,NNU]{A.N. Gorban\corref{cor1}}
\ead{a.n.gorban@le.ac.uk}
 
\author[LeicMath]{T.A. Tyukina}
\ead{tt51@le.ac.uk}

\author[SFU]{L.I. Pokidysheva}
\ead{pok50gm@gmail.com}

\author[SFU]{E.V. Smirnova}
\ead{seleval2008@yandex.ru}

\address[LeicMath]{Department of Mathematics, University of Leicester, Leicester,   UK}
\address[NNU]{Lobachevsky University, Nizhni Novgorod, Russia}

\address[SFU]{Siberian Federal University, Krasnoyarsk, Russia}

\cortext[cor1]{Corresponding author}

\begin{abstract}
The concept of biological adaptation was closely connected to some mathematical, engineering and physical ideas from the very beginning. Cannon in his ``The wisdom of the body'' (1932) systematically used the engineering vision of regulation. In 1938, Selye enriched this approach by the notion of adaptation energy. This term causes much debate when one takes it literally, as a physical quantity, i.e. a sort of energy. Selye did not use the language of mathematics systematically, but the formalization of his phenomenological theory in the spirit of thermodynamics was simple and led to verifiable predictions. In 1980s, the dynamics of correlation and variance in systems under adaptation to a load of environmental factors were studied and the universal effect in ensembles of systems under a load of similar factors was discovered: in a crisis, as a rule, even before the onset of obvious symptoms of stress, the correlation increases together with variance (and volatility). During 30 years, this effect has been supported by many   observations of groups of humans, mice, trees, grassy plants, and on financial time series. In the last ten years, these results were supplemented by many new experiments, from gene networks in cardiology and oncology to dynamics of depression and clinical psychotherapy. Several systems of models were developed: the thermodynamic-like theory of adaptation of ensembles and several families of models of individual adaptation. Historically, the first group of models was based on Selye's concept of adaptation energy and used fitness estimates. Two other groups of models are based on the idea of hidden attractor bifurcation and on the advection--diffusion model for distribution of population in the space of physiological attributes. We explore this world of models and experiments, starting with classic works, with particular attention to the results of the last ten years and open questions.
\end{abstract}


\end{frontmatter}

\section{Introduction \label{Sec:Intro}}

\subsection{Correlation graph in stress and crisis}

Network medicine and biology is undoubtedly a modern vector for the development of biomedical sciences \cite{Loscalzo2017}. Integration and use of  huge collections of individualized data is not possible without network representation. Modern information technologies, such as semantic zooming, are used to organize and analyze information about networks of biological processes \cite{Gomez2020}.
New experimental technologies, such as single cell omics (transcriptomics, epigenomics, and proteomics), open up opportunities to analyze large samples of single cells and provide access to phenomena, which were invisible when studied by standard methods that average the data across the multiple cells \cite{Chappell2018}. They pose new challenges to network data analysis  \cite{Chen2019,Lahnemann2020,Wagner2020} .

Some ideas in network biology have a long history.
Analysis of graphs of correlations between biological traits was proposed in biostatistics in  1931 \cite{Terentjev31}.The vertices of these graphs are attributes, and the edges correspond to sufficiently strong correlations (above a certain threshold). The {\em correlation graphs}  were used to define {\em correlation pleiades} that are clusters of correlated traits. They were used in evolutionary physiology for identification of a modular structure (blocks of connected traits)\cite{Terentjev31,Berg60,Armbruster99,Mitteroecker07}. Terentjev explained his idea as follows: ``Let us take an object as a set of characteristics and consider all conceivable combinations of their correlative relationships. Even superficial observation shows that the uniformity of such correlation of features is a rather rare case. As a rule, the characteristics are grouped into a few `societies' that can be called `correlation pleiades'.'' 

 Technically, it is necessary to distinguish between the core and peripheral elements of the correlation pleiades, where small intersections of different pleiades on the periphery are possible. \cite[Chapter 2]{FehrmanEtAl2019}. (We refer to \cite{XuWunsch2009} for a systematic review of clustering.) Later on, it was discovered that the correlation pleiades of physiological parameters may change under the load. For example, the attributes related to the cardiovascular systems and the attributes of the respiratory systems form different pleiades for healthy people at rest but under significant load (measured in cycle ergometer stress testing) the correlations between attributes of these systems  increase and their pleiades may join in one cluster \cite{Svetlichnaya1997}.
  
Correlation clusters are used in various fields, such as human metrology \cite{Adjeroh2010} (for predicting unknown body measurements and evaluating some soft biometric data), in genomics (for example, for reconstructing RNA-viral quasi-species \cite{Barik2018}), and for many other tasks. 

Network biology of stress and disease based on correlation graph analysis was discovered in 1987 \cite{GorSmiCorAd1st}.  A universal effect in ensembles of similar organisms under the load of similar factors was observed. Represent every organism as a vector of biomarkers (physiological attributes of various types: activities of enzymes,  data of biochemical blood analysis, activities of various genes, etc.). Under stress, as a rule, even before the appearance of obvious symptoms of the crisis, the correlation between the attributes of  organisms in the group increases, and at the same time, the variance also increases. With the development of stress and the approach of a crisis, the connectivity of the correlation network of biomarkers increases. 
This effect has been repeatedly rediscovered. Several theories have been proposed to explain it. Now it is `time to gather stones together'. In this paper we analyze the experimental  and theoretical results on the dynamics of correlation graphs under adaptation and stress. Anticipating a detailed analysis, let us formulate the main conclusions obtained as a result of the review:
\begin{itemize}
\item The dynamics of the correlation graph has been proven to be a useful marker of adaptation and stress.
\item It can be used in development of the `early warning' signals for anticipating of stress and crisis.
\item Some additional experimental data are needed to clarify the dynamics of correlations and variability near deaths. The reports are still ambiguous. The hypothesis is that closer to lethal outcome, the correlations between physiological parameters decay, the variance decreases but the relative variation increases. 
\item Several different theoretical approaches to this dynamics were proposed. They give qualitative explanation of important features but no approach is completely satisfactory and unresolved issues remain. 
\item A general unified theory of the effect is needed, inheriting the achievements of existing models and solving open questions.
\end{itemize}
We discussed both successes and failures of modeling and formulate some open questions.

\subsection{Correlations and variability in the course of adaptation: the main directions of experimental research}

\begin{figure}[t]
\centering
\includegraphics[width=0.6\textwidth]{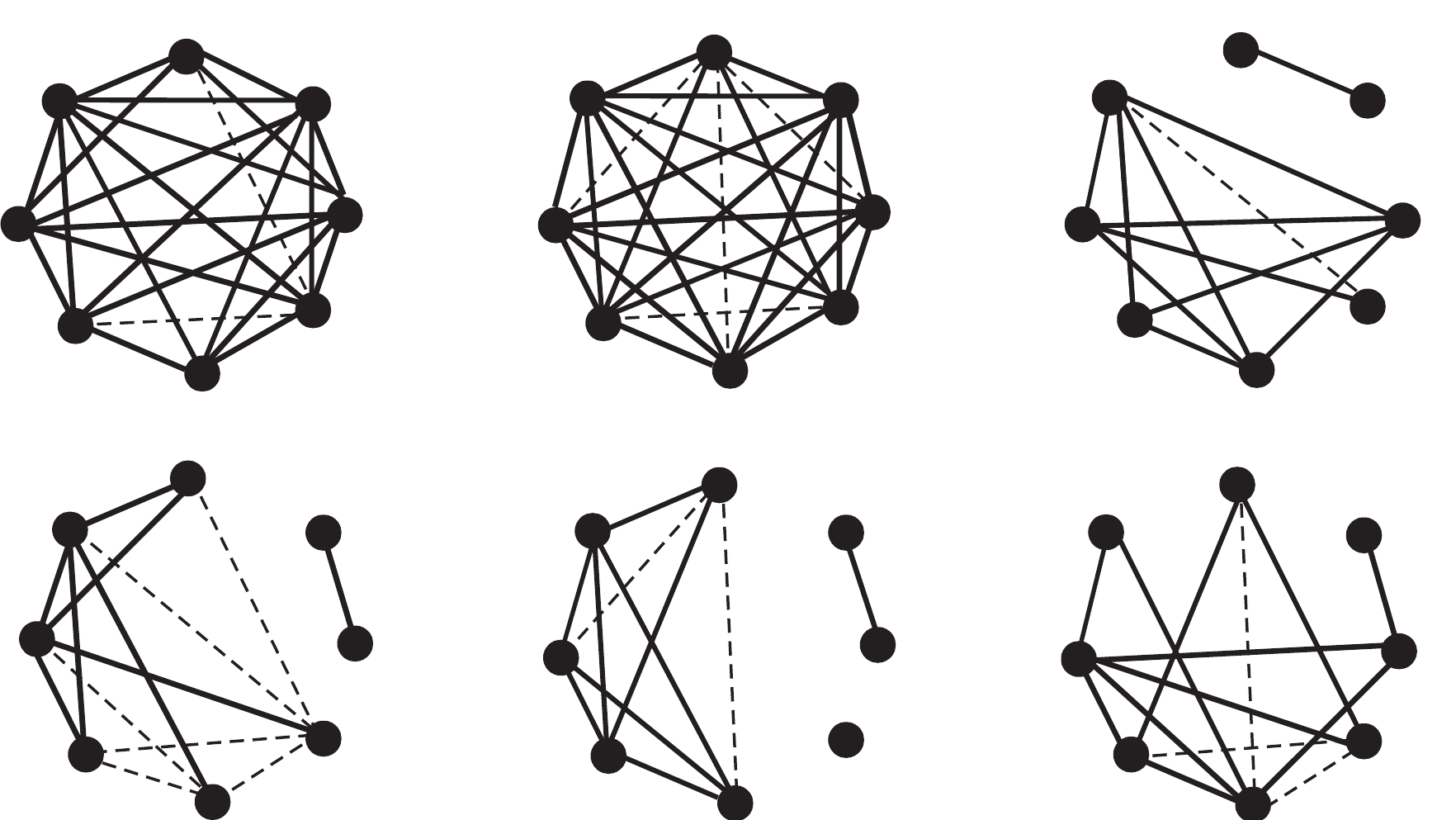}
\caption { Correlation graphs of lipid
metabolism for newborn babies. Eight lipid fractions were
analyzed. Vertices correspond to different
fractions of lipids. Edges drawn by solid lines connect vertices with the correlation
coefficient between fractions $|r_{ij}| \geq 0.5$. Edges drawn by dashed lines
correspond to the correlations $0.5 > |r_{ij}| \geq 0.25$. Upper row -- Far North
(FN), lower row -- the temperate belt of Siberia (TBS). From the
left to the right: 1st-3rd days (TBS -- 123  and FN -- 100
babies), 4th-6th days (TBS -- 98 and FN -- 99 babies), 7th-10th
days (TBS -- 35 and FN -- 29 babies).}
\label{Fig:lipid1987}
\end{figure}

First time, this effect was demonstrated on  the lipid metabolism of healthy newborns \cite{GorSmiCorAd1st}. We studied two groups: (i) babies born in the temperate belt of Siberia (the comfort zone) and (ii) in the migrant families of the same ethnic origin in a Far North city. (The families lived there in the standard city conditions.) A blood test was taken in the morning, on an empty stomach, at the same time every day. All data were collected in the summer. The  correlation graphs are presented in Fig.~\ref{Fig:lipid1987}.  

\begin{figure}[t]
\centering
\includegraphics[width=0.6\textwidth]{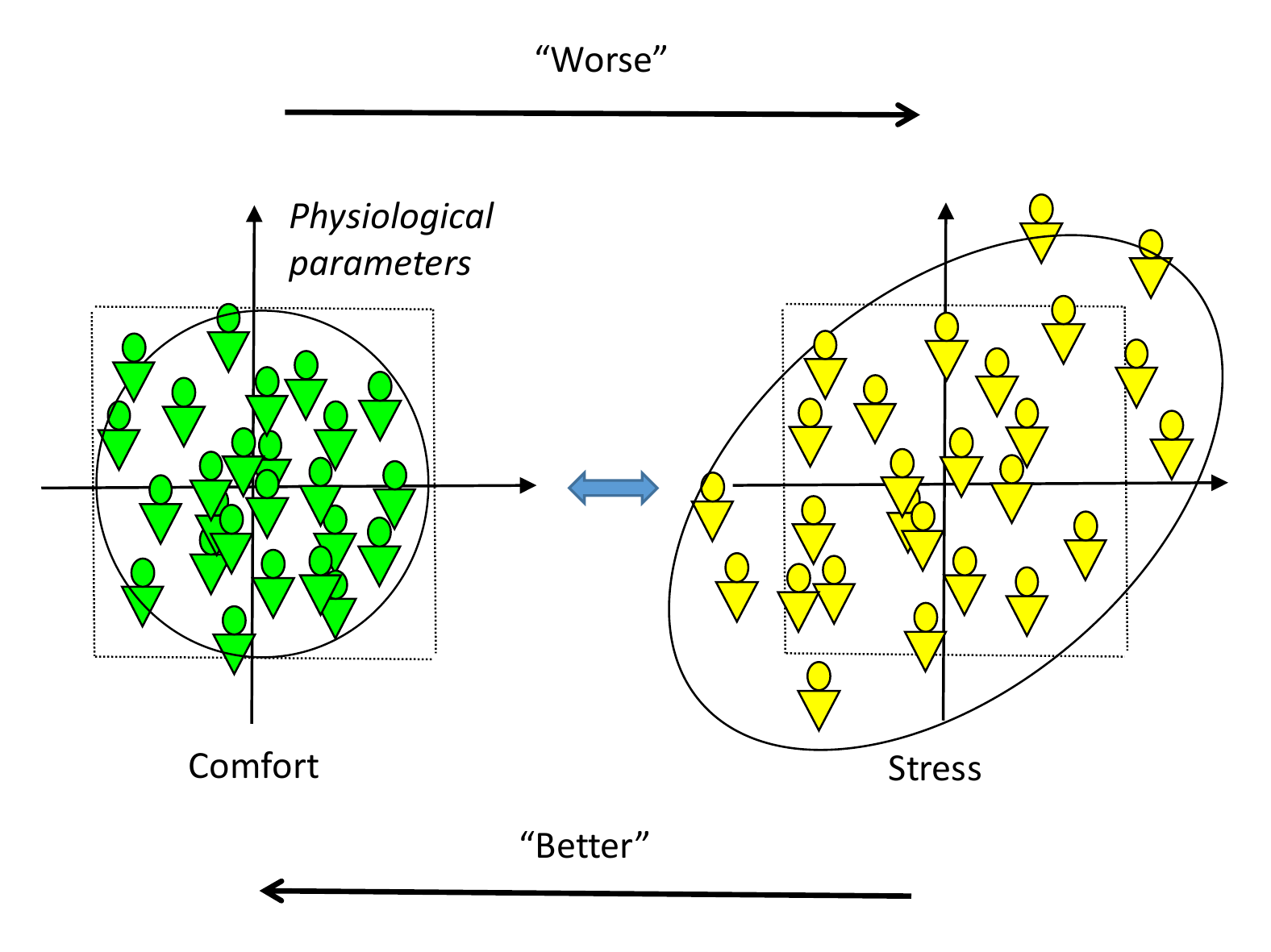}
\caption { Geometric representation of the effect. Each organism is represented as a data point in an $n$-dimensional vector space. The well-adapted
organisms are not highly correlated and after normalization of
scales to unit variance the corresponding cloud of points looks
roughly as a sphere. The organisms under stress  are highly correlated, hence in the same coordinates the data cloud looks like an ellipsoid with remarkable eccentricity.
The largest diameter of this ellipsoid is larger than for
the well--adapted organisms and the variance increases together with
the correlations. The dotted square corresponds to the variance of attributes in the comfort zone normalized to unity.}
\label{Fig:CorrAdapt}
\end{figure}

The observed effect has a clear geometric interpretation (Fig.~\ref{Fig:CorrAdapt}). A typical stress pattern can be expressed symbolically: Correlations$\uparrow$ (increase) and Variance$\uparrow$ (increase). This means that the answer to the following question is ambiguous: Do the organisms  become more or less similar under stress? An increase in variance means more differences in attribute values, while an increase in correlations shows more similarity in the relationships between attributes.

After systematic testing on many groups and various types of stress, this effect was proposed  as a method of screening of the population  in hard living conditions
(Far North city, polar expedition, army recruits, or groups that have changed their living conditions, for example) \cite{Sedov}. Network characteristics appear to be more informative indicators of stress than just attribute values. This was proven by many
experiments and observations of groups of humans, mice, trees and
grassy plants \cite{GorbanSmiTyu2010}. A simple but useful trick is to collect the `population' (Fig.~\ref{Fig:CorrAdapt}) from the same organism at different time moments.   
The effect was successfully applied in psychiatry and  psychotherapy research \cite{Cramer2016,Felice2019,Felice2020}.  In particular, correlations in the network of specific symptoms were used as a marker of major depression: Strong connections correspond to severe depression and deterioration of the patient's condition, while weaker connections characterize stable state or recovery \cite{Cramer2016}.   

The correlation dynamics and Markov models of dyads `patients--psychotherapists' was studied in  \cite{Felice2019,Felice2020,Kleinbub2019,FeliceDodo2019}, where a statistical mechanics -- inspired quantitative approach to evaluation of effectiveness of  psychotherapy was developed using analysis of correlation graph, PCA and cluster analysis. In particular, it was demonstrated that the outcome of psychotherapy  can be qualitatively predicted (good or poor) on the  basis of the correlation pattern between the coarse-grained verbal behavior of patients and psychotherapists \cite{Felice2020}. A physiological approach to measuring of empathy between patients and psychotherapists was developed and tested based on correlations between patient and therapists heart rate and galvanic skin response \cite{Kleinbub2019}. The  data-driven `synchrony index' developed in these works meets the classical idea of  synchrony in psychotherapy (see, for example, the review \cite{Koole2016}).

 Correlation indicators in attribute networks are useful in analyzing large socio-political systems. For example, the dynamics of 19 main public fears that have spread in Ukrainian society during a long stressful period preceding the Ukrainian economic and political crisis of 2014 were analyzed  \cite{Rybnikovs2017}. In the pre-crisis years, there was a simultaneous increase in the total correlation between fears (by about 64\%) and their variance (by 29\%). 

In 1995, this effect (Fig.~\ref{Fig:CorrAdapt}) was also described for the equity markets of seven major countries over the period 1960--1990 
\cite{LonginCorrNonconst1995} and in 1997 for the twelve largest European
equity markets after the 1987 international equity market crash
\cite{MericEuMarket1987}. These observations initiated development of econophysics \cite{Stanley2000}. Analysis or 2008 world financial crisis was performed in   \cite{GorbanSmiTyu2010}.  
  There are attempts to use this effect in seismology to predict earthquakes \cite{Zimatore2017}. It has also been selected as part of a versatile critical transition anticipating toolkit  \cite{SchefferEtal2012}. 
   
In Section~\ref{Sec:CorRiskCri}, ``Correlation, risk and crisis'', we review the relevant data. The collection of examples presented in our previous works \cite{GorbanSmiTyu2010} has been significantly expanded by the findings of the last decade. 

\begin{figure}[t]
\centering
\includegraphics[width=0.6\textwidth]{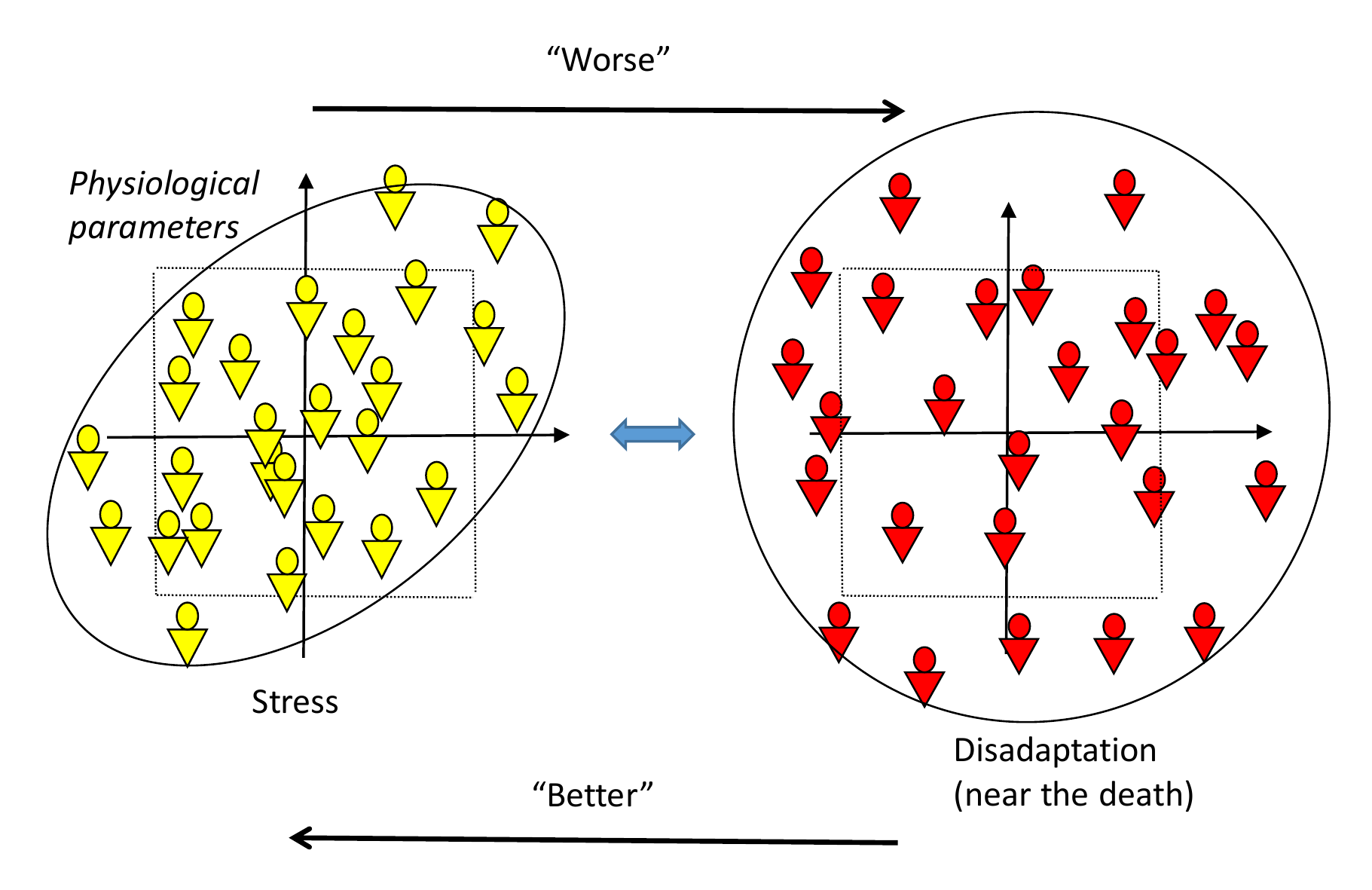}
\caption { Non-monotonicity of correlations as a function of stress intensity: when  organisms are approaching death,  correlations decay, while variance (or relative variation) continues to increase.}
\label{Fig:CorrAdaptNonMon}
\end{figure}

Most of the data, which we collected by ourselves or found in publications, support the hypothesis presented in Fig.~\ref{Fig:CorrAdapt}. Therefore, special attention was needed to search for rear  data that contradict this picture. The effect of non-monotonicity was revealed: when the adaptive abilities are exhausted and the organisms are close to death, the correlations decrease again, and the variance continues to grow (Fig.~\ref{Fig:CorrAdaptNonMon}). Thus, near the fatal outcome, the dynamics of data can be represented as follows:  Correlations$\downarrow$ (decrease) and Variance$\uparrow$ (increase). A simple but significant comment about dimensionless variables is needed. Correlations are dimensionless,  but variances are not. In most physiological data we work with  the variables are positive by their nature (concentrations, activities, etc.). Near the fatal outcome, the means may change drastically, therefore rescaling may be necessary. That is, in such cases we should consider dimensionless {\em relative variation}.

When describing the effect, the idea of stress  and the vague terms ``better'' and ``worse'' were used. What do these ``better" and ``worse" mean? Of course, in classifying specific clinical situations, we relied on the  opinion of medical experts. All the experiments are unbiased in the following sense: the definitions of the ``better--worse" scale were done
before the correlation analysis and did not depend on the results
of that analysis. Hence, one can state, that the expert evaluation
of the stress and crisis can be (typically) reproduced by the
formal analysis of correlations and variance.

However, a deeper analysis of the instantaneous evaluation of the `fitness of the individual' is needed. The theoretical notion `fitness' can be defined in the context of Darwinian selection but requires long time history for evaluation. Even the possibility to consider instantaneous individual fitness  needs clarification. We consider this problem in Section~\ref{SubSec:Fitness}, ``The challenge of defining wellbeing and instantaneous  fitness''.
 
 \subsection{Theoretical approaches `from a bird's eye view}
 
Such a general effect  (Fig.~\ref{Fig:CorrAdapt})  needs also a general theory for its explanation and for prediction of possible exclusions.  The theoretical backgrounds were found in the factors-resource model. Adaptation is modeled as a process of redistributing resources to neutralize harmful factors (the lack of something necessary is also considered a harmful factor). 

The general `factor-resource' models were qualitatively discussed by Selye in his classical works about General Adaptation Syndrome (GAS) \cite{SelyeAEN,SelyeAE1}. In analysis of his experiments, he used a physical (thermodynamic) analogy and proposed a universal adaptation resource -- {\em adaptation energy} (AE). The literal interpretation of this analogy led to criticism of Selye's approach, as the physical energy of adaptation was not revealed.

We considered AE as an internal coordinate on the `dominant path' in the model of adaptation  \cite{GorbanTyukinaDeath2016}.  This approach demystifies the existence of a single resource variable for describing adaptive processes. The dominant path can appear as a manifold of slow motion \cite{GorbanKarlinLNP2005} in dynamics of adaptation. Nevertheless, the additive behavior of this hidden coordinate of adaptation noticed by Selye \cite{SelyeAEN,SelyeAE1} in discussion of his experiments may need additional explanation. 

During last decade, the energetic concept of adaptation cost has been revived \cite{Lan2012}.  A general energy-speed-accuracy relationship was found between the rate of energy dissipation, the rate of adaptation and the maximum accuracy of the adaptation. The theory has been tested for the adaptation of chemoreceptors in the chemosensory system of E. coli.
Analysis of adaptive heat engines that functions under scarce or unknown resources    related the resources of adaptation to the prior information about the environment \cite{Allahverdyan2016}. The  hidden reservoirs of accumulated energy spent on adaptation are recognized as an essential component of adaptation to a not fully known and non-stationary environment \cite{Allahverdyan2016}. This recent resurrection of the thermodynamic foundations of Selye's   adaptation  energy supports the idea of a single adaptive resource and factor-resource models. Simple but universal thermodynamic estimates of the cost of homeostasis are presented in \ref{SubSec:ThermodCost}.

  Section~\ref{Sec:SelyeTherm},  ``Selye's thermodynamics of adaptation'', presents  general models of the adaptation. In the most general factor-resource models, the result of adaptation can be predictable, but not the dynamics of the adaptation process that leads to this result. 

The mechanism of adaptive redistribution of resource is determined by the principle of optimality, that is, by maximizing Darwinian fitness. The optimal mechanism of redistribution strongly depends on the interaction of harmful factors when they affect the organisms. Two opposite main cases are: the Liebig systems and the   synergistic  systems of factors.  The   Liebig systems obey the ``Law of the Minimum" that states that growth and well-being are controlled by the limiting factor (the scarcest resource or the most harmful factor). In the synergistic  systems harmful factors superlinear amplify each other. The optimal redistribution of the adaptation resource leads to a paradoxical result \cite{GorbanPokSmiTyu}:
\begin{itemize}
\item{{\it Law of the Minimum paradox}:
If for a randomly selected pair, (`State of environment -- State
of organism'), the law of the minimum is valid (everything is
limited by the factor with the worst value) then, after
adaptation, many factors (the maximally possible amount of them)
are equally important. In the process of adaptation, such systems evolve towards  breaking the law of the minimum.}
\item{{\it Law of the Minimum inverse paradox}: If for a randomly
selected pair, (`State of environment -- State of organism'), many
factors are equally important  and superlinearly amplify each other
then, after adaptation, a smaller amount of factors is important
(everything is limited by the factors with the worst non-compensated
values). In the process of adaptation, such systems evolve  towards the law of the minimum).}
\end{itemize} 

The classification of interaction in pairs of resources was developed by Tilman  \cite{Tilman1980,Tilman1982}. He studied equilibrium in resource competition.
 Our theory of adaptation to different systems of factors uses very general properties of fitness functions and interactions between factors in combination with the classical idea of evolutionary optimality \cite{GorSmiCorAd1st,GorbanSmiTyu2010,GorbanTyukinaDeath2016,GorbanPokSmiTyu}.  
Since fundamental works of Haldane (see   \cite{Haldane1932}), the principles of evolutionary optimality, their backgrounds and applications are widely discussed in the literature \cite{Smith1982,Dupre1987,Horn1979,GorbanSlect2007,Karev2014}. Despite some limitations and criticism of the `adaptationism' approach, it was decided that the fundamental contribution of optimality models is that they describe what organisms should do in  particular instance \cite{Orzack1994}. Both formulated paradoxes are the consequences of this approach. In short, Law of the Minimum paradox means that the organism responds to situations with limitation by allocating the available regulatory resources to compensate for this limitation. This reaction reduces the influence of the limiting factor and may create a zone of co-limitation, where several factors are important.
 
For example, daphnids may respond to stoichiometrically imbalanced diets after ingestion \cite{Darchambeau2003}, e.g. by down-regulating the assimilation of nutrients in excess and up-regulating the assimilation of limiting nutrients \cite{Hessen2008}. This response reduces limiting  and leads to co-limitation \cite{Sperfeld2012}.

The `inverse paradox' also describes optimal regulation: if two harmful factors superlinearly amplify each other and are both important, then the optimal strategy is to allocate all the available regulatory resources on one of them  and reduce its influence to the minimum. After that, the remaining factor will become limiting. Such interaction was discovered in the theory of  resource competition for pairs of resources \cite{Tilman1980,Tilman1982} and developed later to the general theory  \cite{GorbanSmiTyu2010}. This effect was supported by some consequences like decrease of correlation in stress when the harmful factors are expected to amplify each other. Nevertheless, there is a need for direct experimental testing of this mechanism as it was done for appearance of colimitation  in well-adapted systems. 

An alternative approach to the dynamics of correlation graph under stress, based on the idea {\em  of hidden bifurcations}, was proposed by several authors (see, for example, \cite{Chenetal2012}). It assumes that there exists a dynamical system of physiological regulation. It may not be fully observable, but it does affect  relevant variables. The observed increase of correlation and variance is considered as the result of the bifurcation in this dynamical system. Hypothetically, the disease is interpreted as a new attractor resulting from bifurcation  (Sec.~\ref{subsec:Bifur}).

 The  {\em advection--diffusion model} for distribution of population in the space of physiological attributes assumes that the organisms perform random walk in the space of physiological attributes due to individual differences and various fluctuation of environment \cite{RazzhevaikinZVMF,Razzhevaikin2008}. The area $U$ of this walk is bounded. A harmful factor is represented in this model as a `wind'  that moves the population to the border of $U$. Dimension of the cloud of data reduces under this load because the datapoints under strong pressure are located near the boundary of $U$ (Sec.~\ref{sec:adv}).
A  question remains: how can this minor decrease in the dimension of a data cloud in multidimensional spaces explain a significant increase in correlations?
 
 The adaptation  models introduced and analyzed in this work exploit  common phenomenological properties of  the adaptation process: homeostasis (adaptive regulation), adaptation cost (adaptation resource),  and the idea of optimization. The developed models do  not depend on the particular details of the adaptation mechanisms.  Models that do not depend on many details are very popular in physics, chemistry, ecology, and many other disciplines. They aim to capture the  main phenomena. Thermodynamics is the Queen of these approaches, and it is no accident that language and ideas from thermodynamics are used at important stages of developing basic adaptation models.

Perhaps the most important challenge is assembling a holistic view of the problem of adaptation from the many different successful studies created over the past decades. This synthesis seems to be a difficult task, because the existing models even have   different ontologies. Our vision is that simple and basic adaptation models can provide a useful workbench for this build.

 In Conclusion and outlook we discuss the results and present the open problems and perspectives.

 In  Appendices, we provide formal details of the models that complement the overall picture.
 
 \section{Correlation, risk and crisis \label{Sec:CorRiskCri}}
 
 In this Section, we review the main empirical findings about dynamics of correlation graph in the course of adaptation. Our goal is to demonstrate the universality of the effect for different ensembles of  adaptable systems  with particular attention to two additional questions:
 \begin{enumerate}
 \item Is it possible to convert the observed  correlation graphs into early warning indicators of stress and crisis?
 \item Are there controversies in the interpretations of reported observation? 
 \end{enumerate}
 
 \subsection{Indicators of correlation strength \label{SubSec:Indic}}
 
 For a set of $m$ attributes, $\{x^l | l=1, ... m\}$, we have $m(m-1)/2$ correlation
coefficients $r_{kl}$ between them. The standard choice for numeric variables is the Pearson correlations coefficient:
  \begin{equation}\label{CoCor}
r = \frac{\langle x y \rangle - \langle x\rangle \langle y
\rangle}{\sqrt{\langle(x -\langle x \rangle)^2\rangle}
\sqrt{\langle(y  -\langle y \rangle)^2\rangle}}
\end{equation}
where $x_i$, $y_i$ are the sample values of the attributes $x$, $y$, $\langle ... \rangle $ stands for the sample average value, 
$\langle x \rangle =\frac{1}{n}\sum_i x_i$,   and $n$ is the number of samples.
 
A correlation graph \cite{GorSmiCorAd1st,Whittaker1990,Brillinger1996} is a graph whose vertices correspond to attributes, and edges connect vertices with a correlation coefficient above the threshold  $|r_{kl}|\geq \alpha$. It is standard practice to use multiple   edge types (e.g. very strong correlations and moderately strong correlations - see Fig.~\ref{Fig:lipid1987}.). Such graphs were intensively used in medical research  \cite{Sedov,Butte1999,GorbanSmiTyu2010,GorbanTyukinaDeath2016}. For qualitative attributes, the mutual information can be used  \cite{Butte2000}.  The correlation graph approach was developed in data
mining \cite{Fried2003,Verma2005,Huynh2006,Pedronette2016} with applications in econophysics \cite{GorbanSmiTyu2010,Onella1,Onella2,Soloviev2018,Chakrabarti2017,Leo2018}, genomics \cite{Nereti2007}, social network analysis \cite{Wang2019} and other areas.
  
	To measure the intensity of correlations in general, the weight of the correlation graph is used:
\begin{equation}\label{lpweight}
 G _{p, \alpha} = \left(\sum_{j>k, \ |r_{jk}|> \alpha}
|r_{jk}|^p\right)^{\frac{1}{p}}.
\end{equation}
This a $p$-weight of the $\alpha$-{\it correlation graph}. The usual choice is $p=1$ (i.e., the $l_1$ norm is used). The choice of $\alpha$  is more flexible and depends on the range of correlation coefficients. By default, either $\alpha=0.5$ or $\alpha=0.7$. Normalization or $G _{p, \alpha}$ `per pair of attributes' can be useful for comparing different set of attributes. 

For example, the average absolute value of the correlation coefficient per pair of attributes was systematically used to identify critical transitions in lung injury, liver cancer, and lymphoma cancer by analyzing microarray data for these three diseases \cite{Chenetal2012}. In addition, for each selected feature group, the average absolute value of the internal correlation coefficient and the average absolute value of external correlations were used. The {\em composite index} was proposed
\begin{equation}\label{ChenComposite}
I=\frac{{\rm SD}_d \cdot {\rm PCC}_d}{{\rm PCC}_o},
\end{equation}
where ${\rm PCC}_d$ is   the average absolute value of the Pearson Correlation Coefficient (PCC) of the dominant group, ${\rm PCC}_o$ is the average absolute value of PCC between the dominant group and others, and ${\rm SD}_d$ is the average Standard Deviation (SD) of the dominant group.

The additional idea of \cite{Chenetal2012} is in existence of the dominant group such that near crisis correlations and variance inside the group increase while correlations between this group and other attributes decrease. The composite index combines these three indicators (${\rm PCC}_d \uparrow$, ${\rm SD}_d \uparrow$, and  ${\rm PCC}_o \downarrow$) in one: $I \uparrow$.
 
 There are many correlation indicators based on the eigenvalues of the correlation matrix. If there are no correlations, then all eigenvalues are equal to 1. For strong correlations, the range of the eigenvalues is much wider. Two ideas are used: (i) estimating the number of major principal components \cite{Jolliffe2002} (i.e., the linear dimension of the data) and (ii) estimating the variability of the distribution of eigenvalues. Various definitions of principal components and non-linear generalizations are presented in  \cite{Gorbanatal2008,GorbanZinovyev2009}. Several methods for evaluation of the number of principal components to retain were compared in  \cite{CangelosiBrockStick}. Recently, after testing of many definitions of data dimensionality, we can suggest that the most stable and convenient evaluation of the number of principal components for retaining is based on the condition number of the reduced covariance matrix \cite{gorban2018correction,MirFrac2020}: Intrinsic (linear) dimensionality of dataset is defined as the number of eigenvalues of the covariance matrix exceeding a fixed percent of its largest eigenvalue \cite{Fukunaga1971}. Some other indicators are presented, used and compared in  \cite{GorbanSmiTyu2010}. 
 
 The variability of the distribution of eigenvalues can be estimated by various methods, from standard deviations analysis  to comparison with various `natural' distributions such as the broken stick distribution  \cite{CangelosiBrockStick,GorbanSmiTyu2010} (the length of the pieces of a stick broken at $n-1$ random points distributed uniformly and independently)  or the distributions of eingenvalues of random matrices \cite{Sengupta1999}. This latter approach has become popular in econophysics \cite{Stanley2002,Utsugi2004,Potters2005,Aoyama2020} because in   typical situations where the sample size $n$ and the number of attributes $m$ are both large but comparable spurious correlations appear.  Therefore,  the empirical  correlations should be compared not to the uncorrelated limit but  to the fictitious
correlations, which appear in $m\times n$ data matrices with independent, centralized, normalized and Gaussian matrix elements (the Wishart distribution). If both $m,n \to \infty$ for constant ratio $n/m$ then the limit distribution of eigenvalues   has a simple density, the Marchenko--Pastur distribution \cite{Marchenko1967,Nica2006,GorbanSmiTyu2010} (for generalizations see \cite{Pajor2009}).

\subsection{Humans, physiological data \label{SubSec:Human}}

\subsubsection{Lipid
metabolism of healthy newborn babies}

\begin{figure}[t] \centering{
\includegraphics[width=60mm]{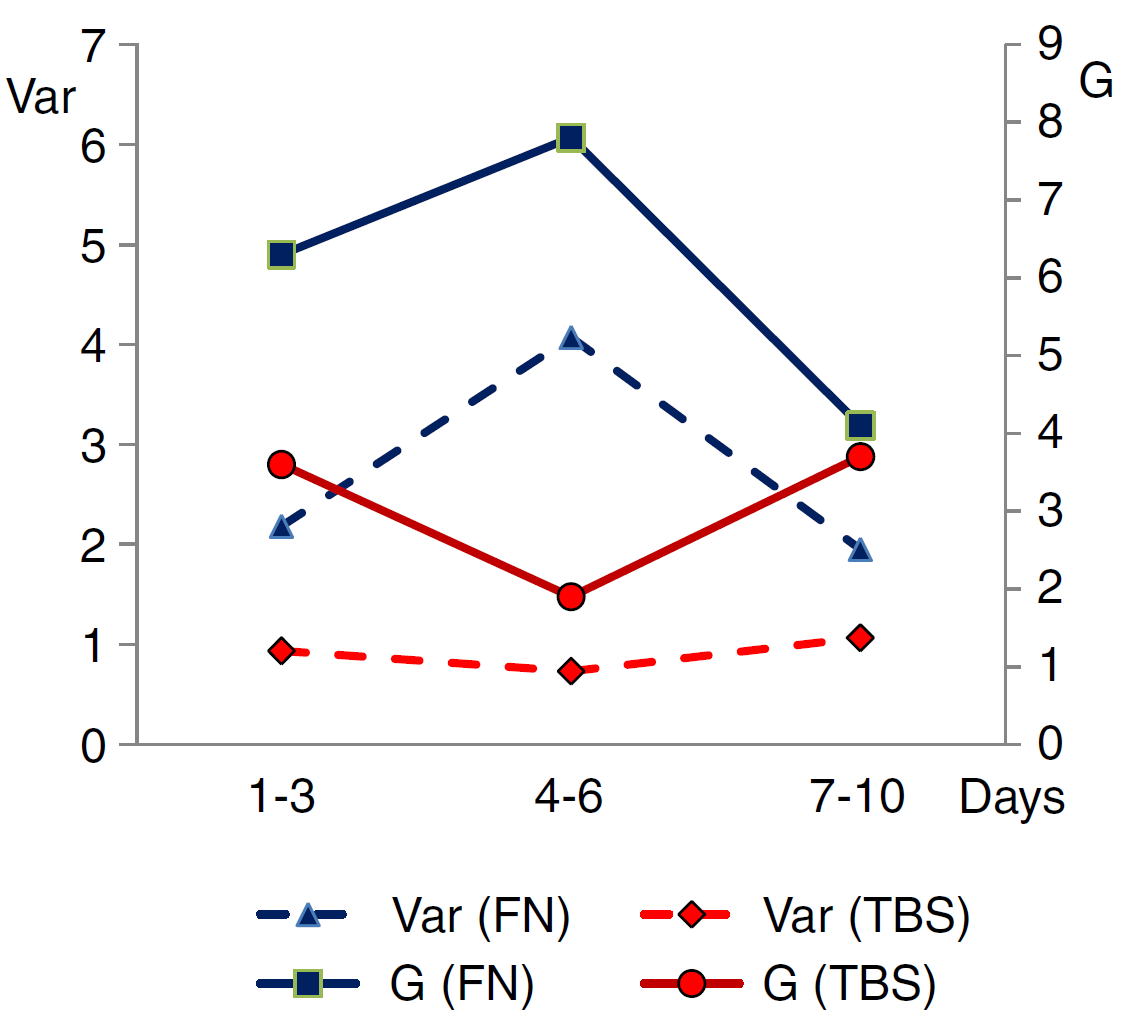}
\caption{\label{Fig:lipid1987b} The weight of the
correlation graphs of eight lipid fractions (solid lines) and the variance (the relative variation) of these fractions (dashed lines)
for healthy newborns born in the  temperate zone of Siberia (TBS), and in families of migrants of the same ethnic origin in the city of the Far North (FN) during the first ten days after birth.}}
\end{figure}

Correlations of eight lipid fractions were studied in healthy newborns born in the  temperate zone of Siberia (TBS), and in families of migrants of the same ethnic origin in the city of the Far North (FN) during the first ten days after birth \cite{GorSmiCorAd1st}. The correlation graphs are presented in Fig.~\ref{Fig:lipid1987}.  On Fig.~\ref{Fig:lipid1987b} we can see that the variance (Var) monotonically increases with the weight of the correlation graph.
The number of babies:  1st-3rd days, TBS -- 123  and FN -- 100
babies; 4th-6th days, TBS -- 98 and FN -- 99 babies; 7th-10th days, TBS -- 35 and FN -- 29 babies. 

 \subsubsection{Adaptation to change of climate zone}
 
We studied people who moved to the Far North six months after moving \cite{Bul1Limf,Bul2Limf}. Two groups were compared: the test group of  54 people  that had any illness during the period of short-term adaptation, and the control group, 98 people without illness during the adaptation period.   We analyzed the activity of enzymes (alkaline phosphatase, acid phosphatase, succinate dehydrogenase, glyceraldehyde3-phosphate dehydrogenase, glycerol-3-phosphate dehydrogenase, and glucose-6-phosphate dehydrogenase) in leucocytes.   The test group demonstrated much higher correlations between activity of enzymes than the control group
 (evaluated after 6 months at Far North), $ G$ = 5.81 in the test group versus $G$ = 1.36 in the control group. For these groups, the dimensionless variance was compared: the enzyme activities were scaled to unit sample mean values, which is necessary, since the normal enzyme activity differs by orders of magnitude.    For the test group  the sum of these relative variations of the normalized enzyme activities was 1.204, and for the control group it was 0.388.

\subsubsection{Gene regulation networks in atrial fibrillation}

\begin{figure}[t] \centering{
\includegraphics[width=0.6\textwidth]{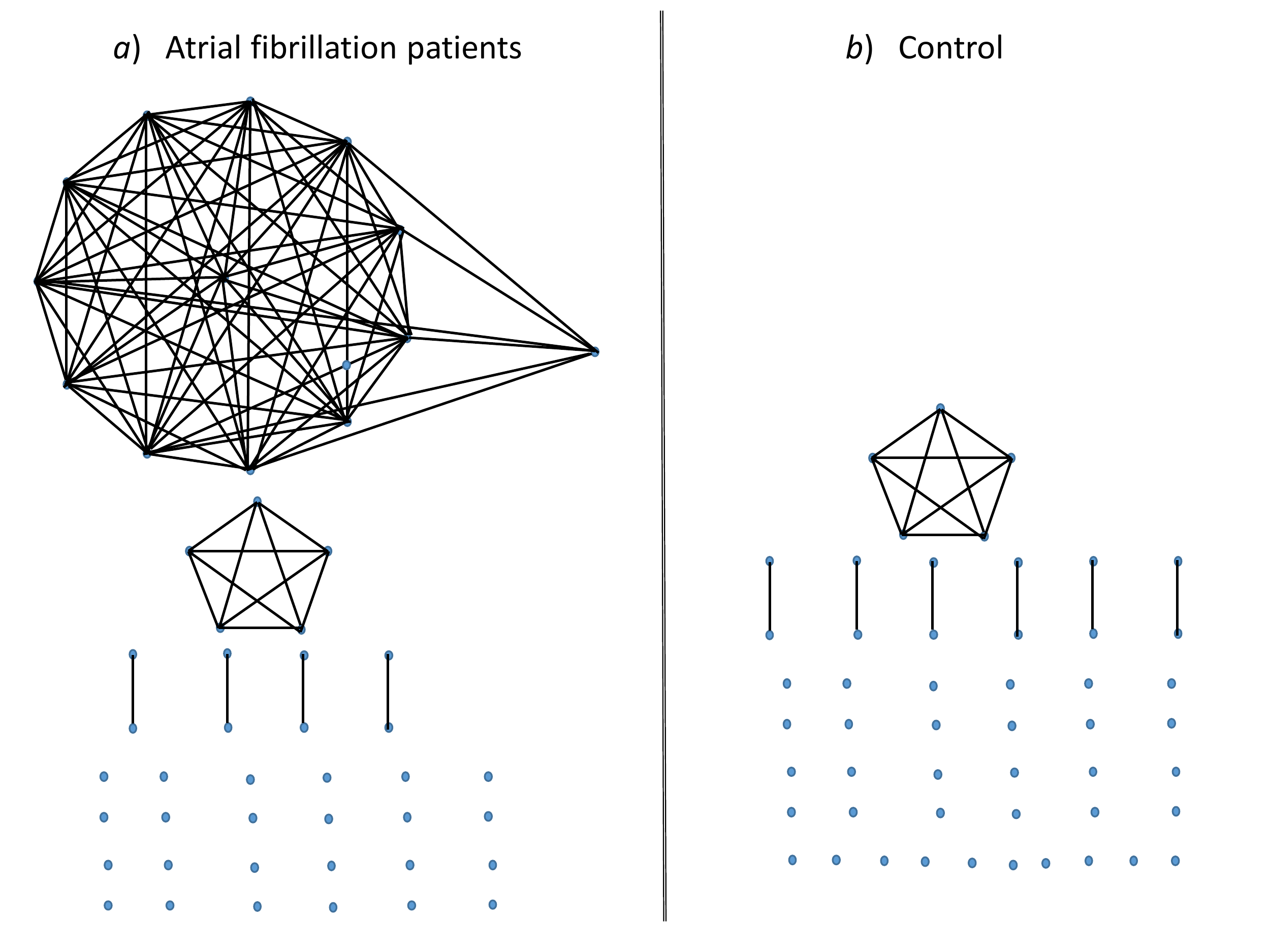}
\caption{\label{Fig:GuliFibrillation}The correlation graph for genes differentially expressed between patients with atrial fibrillation (a) and healthy control (b) (according to data analysis from \cite{Censi2011,Giulianni2014}). The nodes correspond to different genes. Edges correspond to pairwise intra-class between-genes correlations with an absolute value greater than 0.90. }}
\end{figure}

Increasing of connectivity as a response to stressful condition was clearly demonstrated on the correlation network of the gene expression. The differences in gene expression profiles between atrial fibrillation  patients and healthy controls are related to a very small part of the total data variability. Nevertheless,  the factor analysis succeeds in discriminating patients from controls and extracting further genes involved in the pathology \cite{Censi2010}. This approach identified groups of genes
 involved in  structure and organization of cardiac muscle, and in inflammatory processes. Additionally, some genes that allow  completing the transcriptomic deregulation picture of the pathology were detected. 
The correlation graph for the expression of the detected genes is presented for  patients with atrial fibrillation in Fig.~\ref{Fig:GuliFibrillation}a   and for healthy control in Fig.~\ref{Fig:GuliFibrillation}b (according to data analysis from \cite{Censi2011,Giulianni2014}). Edges in (Fig.~\ref{Fig:GuliFibrillation}) correspond to high correlations with an absolute value of the Pearson correlation coefficient greater than 0.90.

 \subsubsection{Gene Network Analysis for Muscular Dystrophy}
 
 In some special cases, the dynamics of the correlation graph contradicts the typical picture shown in Fig.~\ref{Fig:CorrAdapt} (but still supports the idea that the correlation graph can be a better indicator of critical transitions than the attribute values themselves). A seminal case study was performed on  the gene expression data  of skeletal muscle from Duchenne muscular dystrophy (DMD) patients  \cite{Censi2018}. The genome‐wide gene expression profiling of skeletal muscle from 22  DMD children were collected with  14 age-matched controls \cite{Pescatory2007}.   The DMD patients were at the initial or `presymptomatic' phase of the disease. The control group consisted of patients who came to the hospital with a suspect metabolic disorder that was not confirmed by biochemical and histopathological studies.  The data are available \cite{Pescatory2006Data}.
  
 The differences in gene expression profiles between DMD patients and controls are
related to a very small part of the data variability  \cite{Censi2018}. Principal Component Analysis (PCA) helps to overcome this difficulty and to find the relevant signals. The first two principal components show no difference between the two groups. The first principal component explains more than 98\% of the total variance. Nevertheless, in the projection on the third principal component, the  gene expression profiles of the patients with DMD are well separated from the control group. The gene having the highest
score for the third PC was dystrophin (as expected). Correlation network of 100 genes with the highest scores in the third principal components were analyzed.

Correlation was considered significant when above the threshold value, obtained by the surrogate data analysis (0.84). The  activity of important dystrophin gene was not correlated with other genes, both in  DMD patients  and in control. There were 85 significant connections in the DMD patients correlation graph and 133 connections in the control  \cite{Censi2018}. Thus, in general, the connectivity of the correlation graph is lower for the DMD patients.

In more detail, there are two subnetworks in the control group, A and B \cite{Censi2018}. The subnetwork A consists of genes that encode various hemoglobins. The group of DMD patients has practically the same subnetwork. The subnetwork B for the control group  consisted of molecules involved in the extracellular matrix and cytoskeleton organization. For DMD patients, this subnetwork was much less correlated and  split into three disconnected subnets.

 Nevertheless, for the  DMD patients there appear a special new connected component C of correlated genes which are not correlated in the control group. This subnetwork consisted of genes involved in muscle development. If we focus on this small muscle development network, then the dynamics of correlations still follows  Fig.~\ref{Fig:CorrAdapt}.
  
Thus, this case study demonstrated that dynamics of correlations for illness and critical transitions can be multidirectional: some subnetworks may become less correlated, while correlations in  some other subnetworks may increase. For such cases, ``the devil  is in the details'' of choosing subnetworks for correlation analysis.

For example, the same procedure applied to the  mitochondrial network genes in the skeletal muscle of amyotrophic lateral sclerosis patients demonstrates the classical behavior  (Fig.~\ref{Fig:CorrAdapt}).  There are higher levels of correlation  among genes, whose function are aberrantly activated during the progression of muscle atrophy \cite{Bernardini2013}. The genes in this  network    seem  to reflect the perturbation of muscle homeostasis and metabolic balance occurring in affected individuals.

\subsubsection{Myocardial infarction and `no-return' points}
 
The dynamics of correlations between physiological
parameters after myocardial infarction was studied in
\cite{Strygina}. For each patient (more than 100 people),
three groups of parameters were measured: echocardiography-derived
variables (end-systolic and end-diastolic indexes, stroke index,
and ejection fraction), parameters of central hemodynamics
(systolic and diastolic arterial pressure, stroke volume, heart
rate, the minute circulation volume, and specific peripheral
resistance), biochemical parameters (lactate dehydrogenase, the
heart isoenzyme of lactate dehydrogenase LDH1, aspartate
transaminase, and alanine transaminase), and also leucocytes. Two
groups were analyzed after 10 days of monitoring: the patients
with a lethal outcome, and the patients with a survival outcome
(with compatible amounts of group members). These groups do not
differ significantly in the average values of parameters and are
not separable in the space of measured attributes. Nevertheless, the
dynamics of the correlations in the groups are essentially
different. For the fatal outcome correlations were stably low
(with a short pulse at the 7th day), for the survival outcome, the
correlations were higher and monotonically grew. This growth can
be interpreted as return to the ``normal crisis" (the left
position in Fig.~\ref{Fig:CorrAdaptNonMon}). 
Topologically, the correlation graph for the survival outcome
included two persistent triangles with strong correlations: the
central hemodynamics triangle, minute circulation volume -- stroke
volume -- specific peripheral resistance, and the heart
hemodynamics triangle, specific peripheral resistance -- stroke
index -- end-diastolic indexes. The group with a fatal outcome had
no such persistent triangles in the correlation graph.

\begin{figure}[t] \centering{
\includegraphics[width=0.4\textwidth]{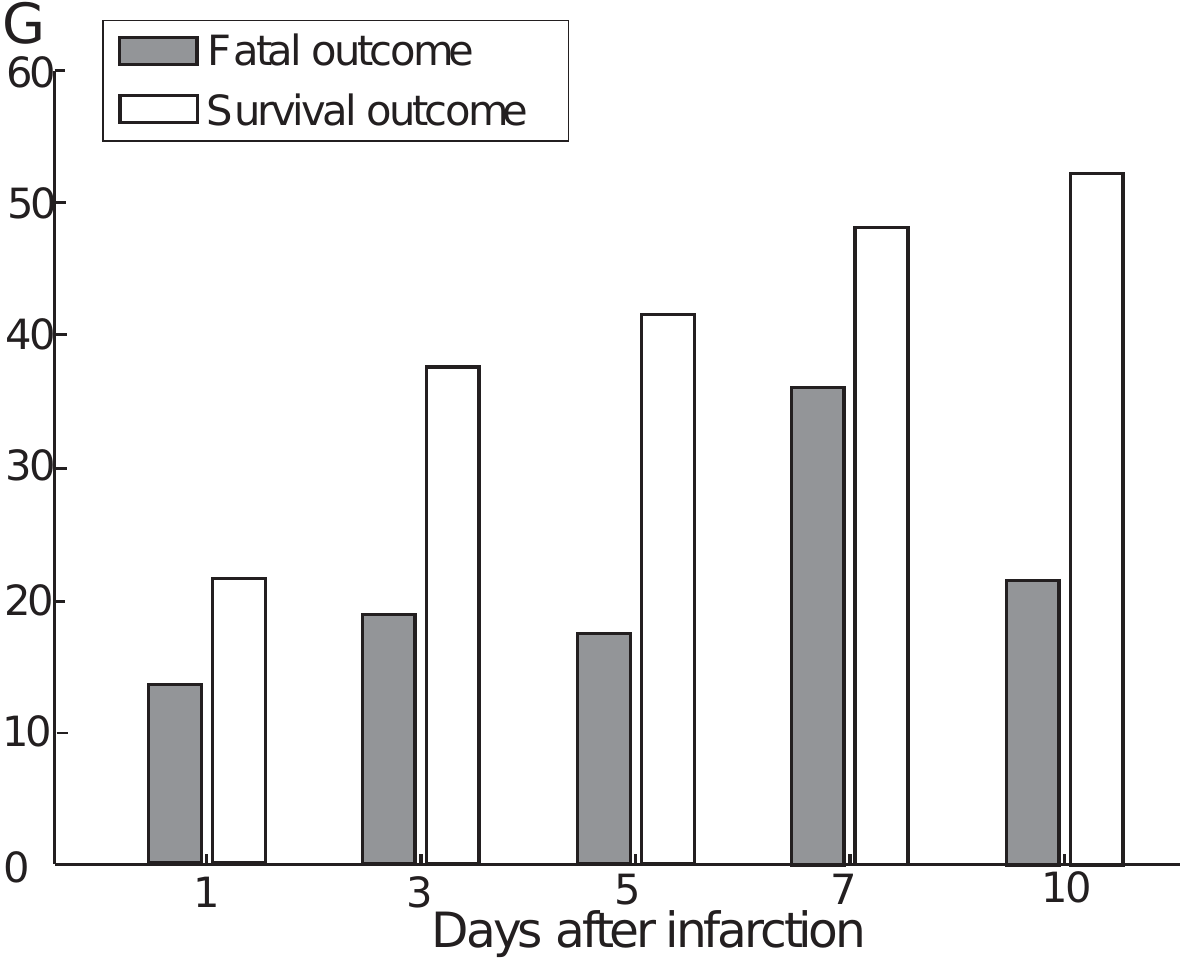}}
\caption{\label{Fig:Infarct}Dynamics of weight of the correlation
graphs of echocardiography-derived variables, parameters of
central hemodynamics, biochemical parameters, and also leucocytes
during 10 days after myocardial infarction for two groups of
patients: for the survival outcome and for the fatal outcome. Here
$G$ is the sum of the strongest Pearson's correlation coefficients $|r_{ij}|>0.4$,
$i\neq j$ \cite{Strygina}.}
\end{figure}

In the analysis of fatal outcomes for oncological patients \cite{MansurOnco} and in
special experiments with acute hemolytic anemia caused by
phenylhydrazine in mice \cite{mice} (see also Sec.~\ref{Sec:Mice1})  one more effect was observed:  the correlations increased for a short
time before death, and then fell down
(see also the pulse in Fig.~\ref{Fig:Infarct}). This pulse
of the correlations (in our observations, usually for one day,  which precedes the fatal outcome) is opposite to the major
trend of the systems in their approach to death. We cannot claim the
universality of this effect and it requires special attention.
 
 \subsubsection{Genomic data about  hepatic lesion by chronic hepatitis B }
 
 Significant numbers of genes were up- or down-regulated in the liver with chronic hepatitis compared with normal liver. The dynamics of correlation  and variance in gene expression at  the  disease and  pre-disease stages were analyzed in \cite{Chenetal2012} using earlier data on differential gene expression  \cite{Honda2001}. There were 12 patients with disease and 6 patients in control. A subnetwork of the dynamical network biomarker (DNB) was identified. According to  \cite{Chenetal2012}, the variance of DNB gene expressions and correlations between them achieve their maxima at the tipping point of the  pre-disease before it is transformed into  the disease. Correlations between DNB gene expressions and the expressions of the genes that do not belong to DNB is minimal at this point.
 
  \begin{figure}[t] \centering{
\includegraphics[width= \textwidth]{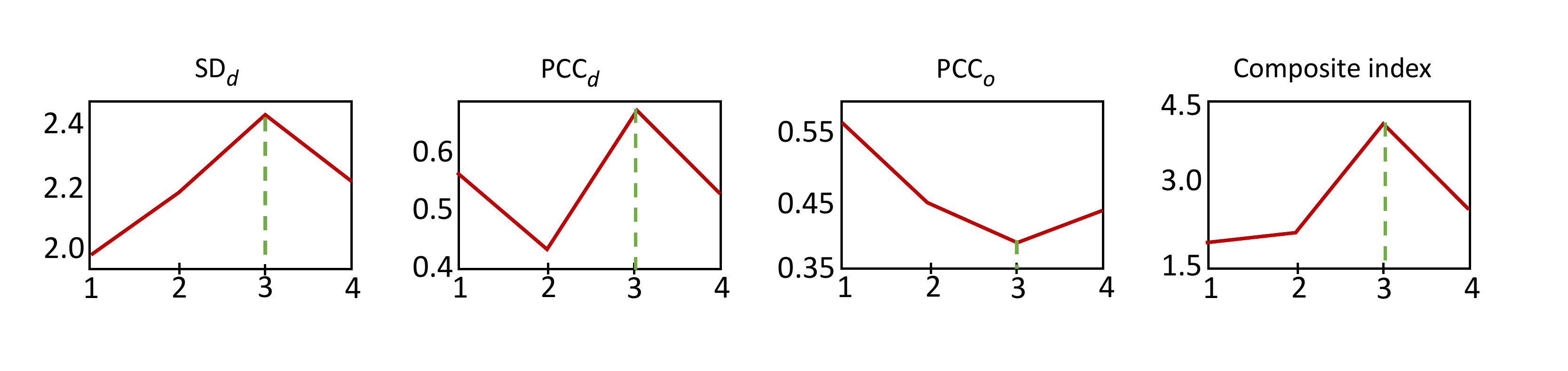}
\caption{\label{Fig:HIV}Early warning signals for the hepatic lesion by
chronic hepatitis B. ${\rm SD}_d $ is the average standard deviation of gene expression in DNB, ${\rm PCC}_d $ is the average Pearson correlation coefficient per pair of gene expressions in DNB, ${\rm PCC}_o$ is the average Pearson correlation coefficient between gene expression in DNB and their external neighborhood gene expression. The composite indicator is given by (\ref{ChenComposite}). The vertical dashed lines indicate the pre-disease state. The figure is drawn using data from  \cite{Chenetal2012}. }}
\end{figure}
 
 The dynamics presented in Fig.~\ref{Fig:HIV} partially supports the hypothesis formulated above (Fig.~\ref{Fig:CorrAdapt}): in the pre-disease stress, both correlation and variance increase. The question appeared about the behavior of correlation and variance after   the tipping point \#3  (Fig.~\ref{Fig:HIV}). According to the observation illustrated in  Fig.~\ref{Fig:CorrAdaptNonMon}, the non-monotonicity of correlations as a function of stress intensity is an expected effect but for a really high stress load (see also \cite{GorbanSmiTyu2010}). 
 
 Another difference is in the dynamics of the SD. According to  \cite{Chenetal2012}, ${\rm SD}_d $ decreases after the tipping point (Fig.~\ref{Fig:HIV}). According to our observations (in different case studies), on the contrary, the average variance continues to increase after the correlations begin  to decrease. Differences (apart from different theoretical models) can be caused by different data preprocessing. We used a dimensionless relative SD (normalized to the unit mean value of the positive attributes). This normalization makes sense because (i) the values of various attributes can differ by orders of magnitude, and without normalization, many SD inputs may practically disappear, and (ii) these values are positive numbers in most cases. The work \cite{Chenetal2012} used the CD ``as it is''. The difference in the emerging dynamics may be caused by a possible decrease in gene expression in DNB after the tipping point. This issue requires more attention in the future.

 \subsubsection{Therapy of obesity}
  The study was conducted on patients   with different levels of obesity
\cite{RazzhevaikinObese2007,Shpitonkov2017}.   The study reported in \cite{Shpitonkov2017} included 235 patients aged 34 to 79 years, suffering from 1-3 degrees of obesity. All patients, depending on the degree of obesity and the nature of concomitant pathology, were divided into 3 groups. The first group of the study included patients with the first degree of obesity. The second group consisted of patients suffering from the second and third degrees of obesity in combination with functional disorders of various organs and systems of the body (dyskinetic disorders of the digestive system, arterial hypertension of the first degree, asthenic syndrome). The third group included patients who had organic lesions (peptic ulcer,  arterial hypertension of the third degree, patients after a heart attack, stroke, etc.) against the background of  the second and third degrees of obesity. 

All patients received a traditional course of treatment for 60 days, aimed at reducing body weight and correcting metabolic and organ disorders. Treatment of patients of the first group was limited  to diet therapy, the second group - with the additional   statin medication (Crestor), the third group — with the additional   drug therapy against comorbidities. The study included the following indicators: body mass index, fat mass, lean mass, total water, total cholesterol, high-density cholesterol, low-density cholesterol, creatinine, and triglycerides. 
The weight of the correlation graph $G$  (the table \ref{Tab:obesity}) was originally high and monotonically dependent on the level of sickness. It decreased during therapy.  

 \begin{table}[!ht]
\centering
\caption{\label{Tab:obesity} The weight $G$ of correlation graph between attributes for patients with different level of obesity before and after treatment.}
\label{tab:2}
\begin{tabular}{|c|c|c|}
\cline{1-3}
	&Before treatment&After treatment \\ \hline
Group 1 &8.24& 7.33\\ \hline
Group 2 &10.74&	8.09  \\ \hline
Group 3 &13.41&	11.72\\\hline
\end{tabular}
\end{table}

\subsubsection{Critical transitions in cardiopulmonary population health  related to air pollution} 

Fluctuations in time series of hospital visits were analyzed to identify a tipping point in the health status of the population  \cite{Wang2018}. Fluctuations were quantified by standard deviations (SD) and autocorrelations at lag-1 (AR-1) and were used as potential early warning indicators. The  hospital visits from a hospital in Nanjing City, China during ten years (2006--2016) were studied for the following cardiopulmonary diseases: cerebrovascular accident disease (CVAD), coronary artery disease (CAD), chronic obstructive pulmonary disease (COPD), lungcancer disease (LCD), and the grouped category of the respiratory system diseases (RESD) with of the cardio-cerebrovascular  system  diseases  (CCD). All these diseases are affected by air pollution.

\begin{figure}[t] \centering{
\includegraphics[width= 0.8 \textwidth]{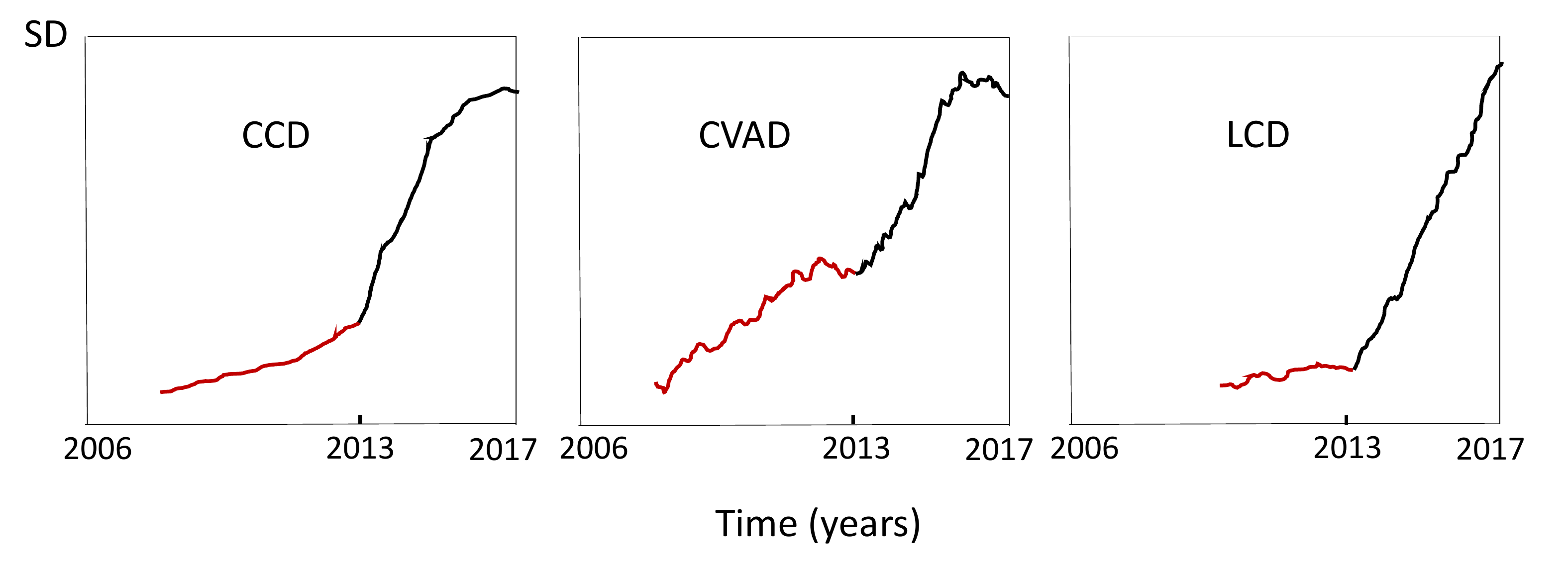}
\caption{\label{Fig:Visits}Dynamics of CD  of daily visits to  a hospital in Nanjing City, China for  CCD, CVAD and LCD during 2006--2016. The figure qualitatively reproduces CD dynamics from   \cite{Wang2018}. All the observed tipping points are close to the beginning of 2013. }}
\end{figure}

The effect of critical transitions with a fast increase in SDs was clearly demonstrated for three diseases, CCD, CVAD and LCD (Fig.~\ref{Fig:Visits}). From the data reported in \cite{Wang2018}, it seems likely that the increase in CDs for these diseases is significantly faster than the increase in visits (smoothed by a moving average filter).

The observed critical transition was interpreted as a health impact of air pollution. Of course, the deeper analysis should combine the data about hospital visits with pathophysiological clinical studies. This is a standard problem of labeling critical transitions found using indirect indicators: what critical transition is observed (for example, in the  population health, or in hospital services, or in the structure and behavior of the population)?

\subsubsection{Network rewiring in cancer: applications to melanoma }

A detailed analysis of rewiring correlation networks of gene expressions was presented in \cite{Ding2018} for melanoma. The authors argue that an approach to analyzing molecular pathophysiological processes based only on gene expression values cannot succeed in predicting the potential effect of a drug because it does not solve the important problem of the relationship between genes.  It is shown that differences in the average expression may be insignificant, while the difference between correlations gives statistically significant $p$-values.

The differences in the topology of rewired networks can be enormous because cancer cell lines may have undergone numerous changes in genetic network correlation and expression patterns.  Gene centrality profiles were used to quantify these differences for future use in drug target determination. These profiles characterize the connections and their strength from a gene to other genes. It is assumed that genes that play a more central role in subsets of genes within a broader relevant network are better targets for drugs in a disease state.  
The authors claim that their results shed light on the understanding of the molecular pathology of melanoma, as well as on the choice of treatment.
  
\subsection{Humans, psychological data \label{SubSec:HumanPsy}}
 
Correlation graph analysis has proven to be useful in psychiatric and psychological research. Here, instead of a population or a group of different organisms shown in Fig.~\ref{Fig:CorrAdapt}, a `population' of states of one person at different points in time is considered. For this section we selected one recent achievement in this area, where individual patients are characterized by their own personal network with unique architecture and resulting dynamics \cite{Cramer2016}. For correlation analysis of dyads `patients--psychotherapists' we refer to  \cite{Felice2019,Felice2020,Kleinbub2019,FeliceDodo2019}.

\subsubsection{Dynamics of depression \label{Depr}}

The network of symptoms of major depression was analyzed in \cite{Cramer2016}. The vertices of this network correspond to 14 symptoms: insomnia, fatigue, concentration problems, depressed mood,  feelings of self-reproach, etc. The main hypothesis is that symptoms can induce each other and individuals differ in the strength of the connections between certain symptoms linked in the network.
The authors of \cite{Cramer2016} went beyond a simple correlation graph analysis and proposed a network model of symptom dynamics  that resembles a probabilistic recurrent neural network. This model appears to be very general and can be applied to dynamical networks of attributes of many diseases.

Each ($i$th) symptom is characterized by a Boolean variable $X_i$ with an obvious interpretation: $X_i=1$ means that the $i$th symptom is `on' (active) and $X_i=0$ means that it is `off' (inactive). We observe the patient in discrete time with a constant time step, $t=1,2,3 \ldots$. The probability that the $i$th symptom is active in the time moment $t+1$ depends on the symptoms and external stress parameters at time $t$. The simple logistic regression was used:
$$\mathbf{P}(X_i(t+1)=1)=\frac{1}{1+\exp(b_i-A_i(t))},$$
where $A_i(t)$ is the signal from the $i$th `input summator':
 $$A_i(t)=\sum_j W_{ij}X_j(t) + S_i(t),$$
$b_i$ is the threshold, $W_{ij}$ is the weight matrix, and $S_i(t)$ is the stress parameter that may be specific for the $i$th symptom.

 There are many methods for evaluation of the weight matrix and   thresholds from  data. An open access R package was used in  \cite{Cramer2016} under assumption that there was no  `authocatalysis' (no self-excitation of symptoms). This assumption, of course, is not necessary. The weight matrix $ W_{ij} $ does not have to be symmetric because it describes effect of  $X_j(t)$ on the $i$th symptom at the next time step, $X_i(t+1)$.
 For strongly connected networks (large values of $W_{ij}$), the long self-sustained major depression is possible even without  exposure to external stress. On the contrary, in weakly connected networks, the depression decays without external stress if the thresholds are not too small. 

This model sheds light on the mechanisms of development of major depression, explaining and predicting the dynamics of mutual induction of symptoms, for example, insomnia increases fatigue, which leads to problems with concentration, etc. The critical transitions to major depression are described using the connectivity of the network of symptoms.  But the main result of \cite{Cramer2016} is, from our point of view, the introduction of a new class of simple and identifiable disease models with dynamic symptoms that can appear and disappear. We should also mention that the weight matrix  $W_{ij}$ is, in its essence, the correlation matrix, and the proposed model demonstrates how the correlation graph can be used for dynamic modeling of interactions between varying attributes. 
 
 \subsection{Mice \label{SubSec:Mice}}
 
 Most data demonstrate that under stress and at the onset of illness the correlation and the variance increase (with an important remark that the choice  of   vertices for constructing  a relevant  graph can be a non-trivial task). This is just a beginning of the `medical history'. The question of a system near the fatal outcome is also important: what does the dynamics of Y correlations look like near the point of no return? 
 
  This problem about the dynamics of correlations near the point of no return was studied by  analysis of fatal outcomes in oncological \cite{MansurOnco} and cardiological
\cite{Strygina} clinics, and also in special experiments with
acute hemolytic anemia caused by phenylhydrazine in mice
\cite{mice} and phosgene inhalation lung injury \cite{Chenetal2012,Scuito2005}. The main result here is: when approaching the no-return point, correlations destroy ($G$ decreases). 

There exists no  formal conventional criterion to recognize the situation ``on
the other side of crisis". Nevertheless, the labeling of such situation is
needed. The common sense 
``general practitioner point of view" \cite{GP_AE1952,GP_AE1952C} can help in the labeling. From this   point of view, the situations described below are close to fatal outcome and should be considered as `the other side of crisis': the acute 
hemolytic anemia caused by phenylhydrazine in mice with lethal
outcome and phosgene inhalation lung injury with high mortality rate.

\subsubsection{Destroying of Correlations ``on the Other Side of Crisis": Acute Hemolytic Anemia in Mice \label{Sec:Mice1}}
 
\begin{figure}[t] \centering{
\includegraphics[width=0.3\textwidth]{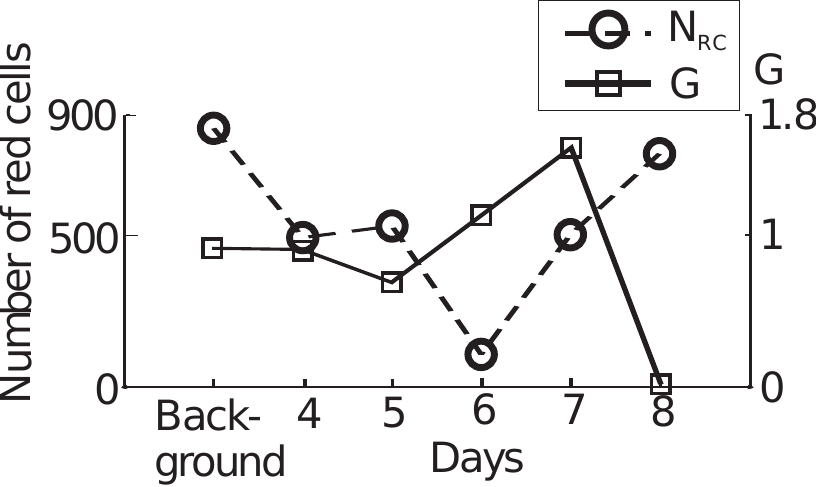}}
\caption{\label{Fig4:mice}Adaptation and disadaptation dynamics
for mice after phenylhydrazine injection. Weight $G$ of the correlation graph and number 
 of erythrocytes $N_{\rm RC}$}
\end{figure}

This effect was demonstrated in special experiments \cite{mice}.
Acute hemolytic anemia caused by phenylhydrazine was studied in
CBAxlac mice. After phenylhydrazine injections (60 mg/kg, twice a
day, with interval 12 hours) during first 5-6 days the number of
red cells decreased (Fig.~\ref{Fig4:mice}), but at the 7th and 8th
days this number increased because of spleen activity. After 8
days most of the mice died. Dynamics of correlation between
hematocrit, reticulocytes, erythrocytes, and leukocytes in blood
is presented in Fig.~\ref{Fig4:mice}. 
An increase in $ G $ preceded an active adaptation response, but
$G$ decreased to zero before death. 
The number of red cells increased also at the last day.

\subsubsection{Phosgene inhalation lung injury}

Carbonyl chloride (phosgene) poisoning leads to life-threatening pulmonary edema and irreversible acute lung damage. A genomic approach was used to investigate the molecular mechanism of phosgene-induced lung damage in mice \cite{Scuito2005}. CD-1 male mice were exposed whole body to either air (control) or phosgene for 20 min. Lung tissue was collected from air- or phosgene-exposed mice at nine time moments: 0, 0.5, 1, 4, 8, 12, 24, 48, and 72 h postexposure.  The strongest physiological effects occur  within the first 8 hours after exposure. They lead to increased levels of the BALF protein, severe pulmonary edema, and increased mortality \cite{Scuito2005}. In particular, after 12 hours, the mortality rate reached 50\% -- 60\%, and after 24 hours it was 60\% -- 70\%.
Oligonucleotide microarrays were used to determine global changes in gene expression in the lungs of mice after exposure to phosgene. These data reveal biological processes involved in phosgene toxicity including GSH biosynthesis (previously known), angiogenesis, cell death, cholesterol biosynthesis, cell adhesion and regulation of the cell cycle. PCA was applied to determine the greatest sources of data variability.  The genes most significantly changed as a result of phosgene exposure weer identified and categorized based on molecular function and biological process  \cite{Scuito2005}. 

\begin{figure}[t] \centering{
\includegraphics[width= \textwidth]{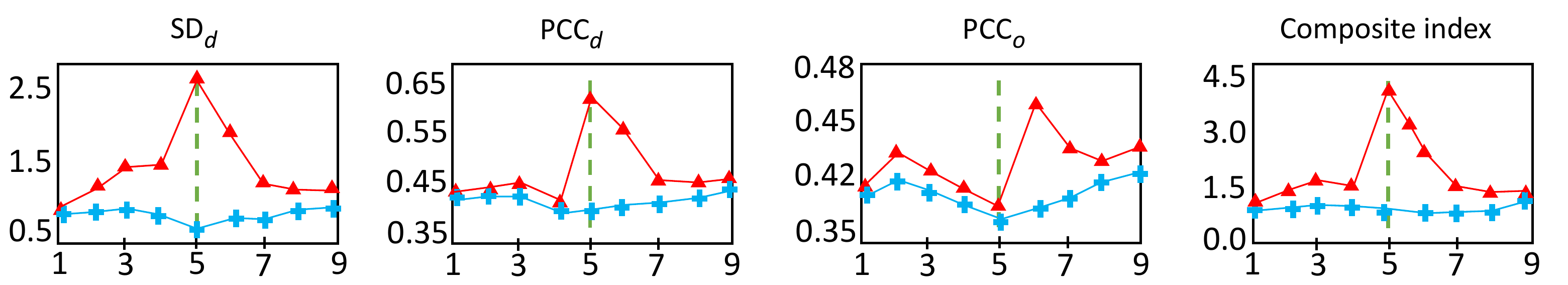}
\caption{\label{Fig:MiceChen}
Early-warning signals for phosgene-induced lung damage in mice. The horizontal axis represents the time period $t$. Red triangles   correspond to the case group, while blue crosses represent the control group. The time scale is not uniform, and the 9 time moments are:  0, 0.5, 1, 4, 8, 12, 24, 48, and 72 h postexposure. Here, ${\rm SD}_d $ is the average standard deviation of gene expression in DNB, ${\rm PCC}_d $ is the average Pearson correlation coefficient per pair of gene expressions in DNB, ${\rm PCC}_o$ is the average Pearson correlation coefficient between gene expression in DNB and their external neighborhood gene expression. The composite indicator is given by (\ref{ChenComposite}). The vertical dashed lines indicate  the pre-disease state. The figure is drawn using data from  \cite{Chenetal2012,Scuito2005}.
 Clearly, the composite indicator starts increasing significantly from the 4th time period
(4h), and reaches the peak in the 5th time period (8h), while the OPCC decreases. This fact
was interpreted by  \cite{Chenetal2012} as a demonstration   that the pre-disease state is located near the 5th time period (between the fourth and the fifths sampling points), and the phase transition into the disease state occurs after the 5th time moment.  }}
 \end{figure}

These data were analyzed further in \cite{Chenetal2012}.  A relevant subnetwork of the dynamical network biomarker (DNB) was identified  (220 genes and 1167 links). Dynamics of the standard deviation and correlations in the DNB, correlations  between the DNB and other molecules, and  the composite index were presented in Fig.~\ref{Fig:MiceChen}.

 \subsection{Plants \label{SubSec:Plants}}
 
\subsubsection{Grassy Plants Under Trampling Load}

\begin{table}[t]\caption{Weight $G$ of the correlation graph for different grassy plants
under various trampling load  \label{GrassTrampl}} \centering{
\begin{tabular}{|l|c|c|c|} \hline
~Grassy Plant~ & ~Group 1~ & ~Group 2~ & ~Group 3~ \\
\hline
Lamiastrum   &     1.4   & 5.2  &  6.2  \\
 Paris (quadrifolia) & 4.1            & 7.6 & 14.8 \\
Convallaria   & 5.4    & 7.9               &  10.1  \\
Anemone   & 8.1                    & 12.5               &  15.8 \\
Pulmonaria   &  8.8        & 11.9     &  15.1   \\
Asarum & 10.3  & 15.4   & 19.5   \\
 \hline
\end{tabular}
}\end{table}

The effect exists for plants too. The grassy plants in oak
tree-plants are studied \cite{RazzhevaikinTrava1996}. For analysis
the fragments of forests are selected, where the densities of trees
and bushes were the same. The difference between those fragments was
in damaging of the soil surface by trampling. Tree groups of
fragments are studied:
\begin{itemize}
\item{Group 1 -- no fully destroyed soil surface;}
\item{Group 2 -- 25\% of soil surface are destroyed by trampling;}
\item{Group 3 -- 70\% of soil surface are destroyed by trampling.}
\end{itemize}

The studied physiological attributes were: the height of sprouts,
the length of roots, the diameter of roots, the amount of roots, the
area of leafs, the area of roots. Results are presented in
Table~\ref{GrassTrampl}.

\subsubsection{Scots Pines Near a Coal Power Station}

The impact of emissions from a heat power station on Scots pine was
studied \cite{KofmantREES}. For diagnostic purposes the secondary
metabolites of phenolic nature were used. They are much more stable
than the primary products and hold the information about past impact
of environment on the plant organism for longer time.

The test group consisted of Scots pines (Pinus sylvestric L)  in a
40 year old stand of the II class in the emission tongue 10 km
from the power station. The station had been operating on brown
coal for 45 years. The control group of Scots pines was from a
stand of the same age and forest type, growing outside the
industrial emission area. The needles for analysis were one year
old from the shoots in the middle part of the crown. The samples
were taken in spring in bud swelling period. Individual
composition of the alcohol extract of needles was studied by high
efficiency liquid chromatography.  26 individual phenolic
compounds were identified for all samples and used in analysis.

No reliable difference was found in the test group and control
group average compositions. For example, the results for
Proantocyanidin content (mg/g dry weight) were as follows:
\begin{itemize}
\item{Total 37.4$\pm$3.2 (test) versus 36.8$\pm$2.0 (control);}
\end{itemize}
Nevertheless, the variance of compositions of individual compounds
in the test group was significantly higher, and the difference in
correlations was huge: $G=17.29$ for the test group versus
$G=3.79$ in the control group.

\subsubsection{Drought stress response  in sorghum}

The  drought-specific subnetwork for sorghum was extracted and described in detail in \cite{Woldesemayat2018}. First, authors found 14 major drought stress related hub genes (DSRhub genes). These genes were identified by combination of various approaches and analysis of data from different sources like analysis of gene expression, regulatory pathways, sorghumCyc, sorghum protein-protein interaction, and gene ontology. Then, investigation of  the DSRhub genes led to revealing distinct regulatory genes such as ZEP, NCED, AAO,   MCSU and CYP707A1. Several other protein families were found to be involved in the response to   drought stress:  aldehyde and alcohol dehydrogenases, mitogene activated protein kinases , and Ribulose-1,5-biphosphate carboxylase.

 This analysis resulted in construction a drought-specific subnetwork, characterized by unique candidate genes that were associated with DSRhub genes \cite{Woldesemayat2018}. Authors analyzed connections in this network and constructed pathway cross-talk network for 69 significantly enriched pathways.

In particular, they found that  the stress increases the overall association and connectivity of the gene regulation network, even if different stress conditions, such as desiccation, salt and oxidative stresses in addition to cold or heat,  occur in combination. For more detail and specific presentation of the drought-specific subnetwork of genes and pathway cross-talk network we refer to the original work \cite{Woldesemayat2018}.

 \subsection{Social systems \label{SubSec:Soc}}
 
Anticipating critical changes in social systems is a big problem. Measuring `social stress' and tracking its dynamics is a top priority in public administration. Unfortunately, very often social stress is measured post-factum after devastating social upheavals. The problem of reliably measuring social stress before critical events is still open.  This problem was attacked recently using the analysis of correlation graph and variance under basic hypothesis presented in Fig.~\ref{Fig:CorrAdapt} \cite{Rybnikovs2017}. In particular, they clearly demonstrate that the dramatic events in Ukraine in December 2013 and February 2014 were preceded by a monotonous increase in social tension during several years. This was not just a sudden `social explosion'.

For analysis, the data collected by the Sociological Pollster Group ‘Rating’ were used.  In 2009--2012, this group conducted several surveys, in which respondents  named three most important threats faced by Ukraine. Data were aggregated for four main regions of Ukraine:  for six geographical regions of the country:
(a) {\it The west} (Chernivtsi, Ivano-Frankivsk, Lviv, Rivne, Ternopil, Volyn
and Zakarpattia provinces); (b)  {\it The centre} (Cherkasy, Khmelnytsky,
Kirovohrad, Poltava and Vinnytsia provinces); (c)  {\it The north} (Chernihiv,
Kyiv, Sumy and Zhytomyr provinces); (d)  {\it The south} (Crimea, Kherson,
Mykolaiv and Odessa provinces); (e)  {\it The east} (Dnipropetrovsk, Kharkiv
and Zaporizhia provinces); and (f)  {\it Donbas} (Donetsk and Luhansk provinces).

Nineteen main fears were identified for analysis: (1) Economic regress, (2) Rise in unemployment, (3) Depreciation of national currency, 
(4) Arbitrary rule, (5) Degeneracy of population, (6) Health services' worsening,
 (7) Environmental accidents, (8) Rise in crime, (9) Mass exodus, 
 (10) Demographic crisis, (11) Schism of the state, (12) Losing sovereignty,
 (13) Civil war, (14) Education services' worsening, 
(15) Losing control over the gastransport system,  (16) Coup d'etat,
 (17) Military aggression from Russia,  (18) Terrorism,
 (19) Military aggression from West.

\begin{figure}[t] \centering{
\includegraphics[width= 0.7\textwidth]{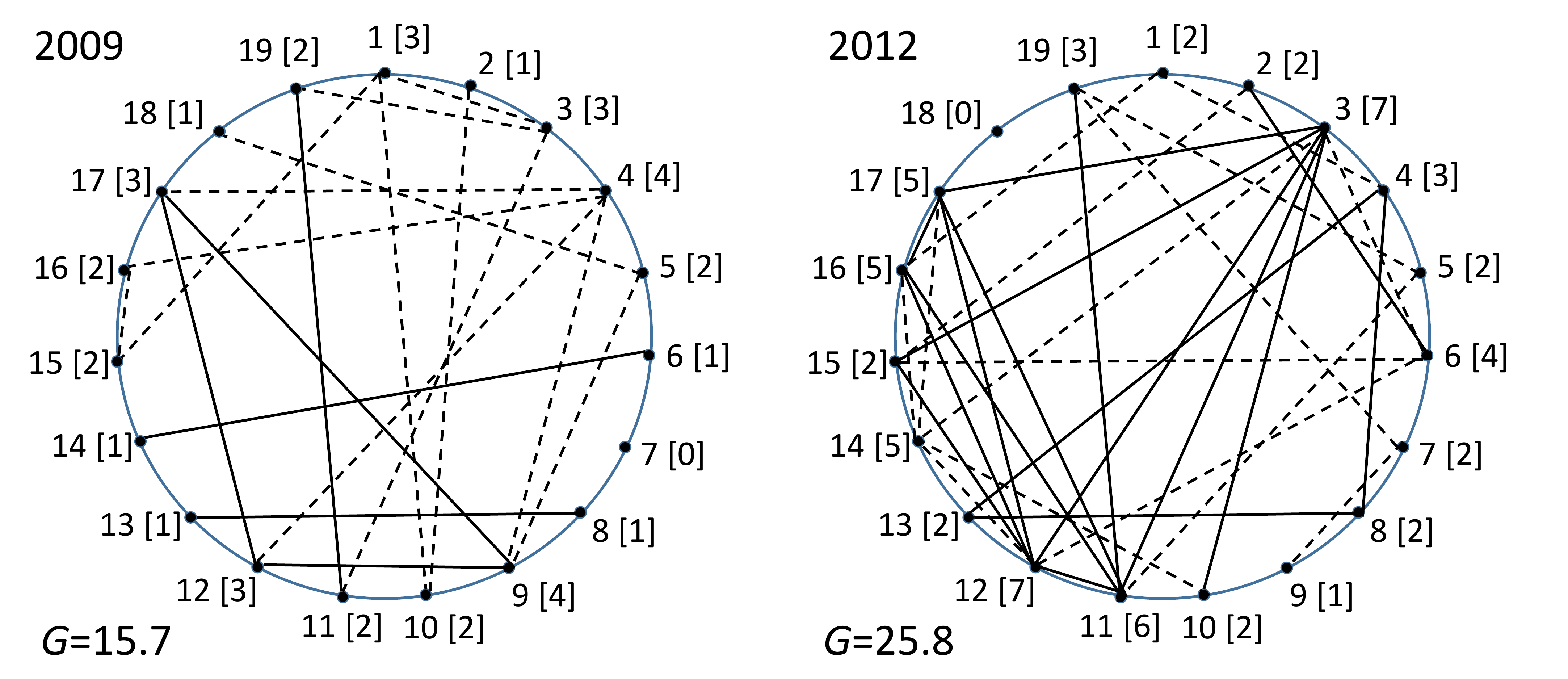}
\caption{\label{Fig:UkraineFears}
 The correlation graph of the prevalence of  19 main fears in regions of Ukraine for 2009 ($G=15.7$) and 2012 ($G=25.8$). The vertices with the PCC $|r_{ij}|>0.7$ are connected. The solid line means that   $r_{ij}>0.7$ (strong positive correlation), the dashed line means that $r_{ij}<-0.7$ (strong negative correlation). The numbers in brackets are the orders of the vertices (the number of links). The data from \cite{Rybnikovs2017} are used.}}
 \end{figure}

For each region, a 19-dimensional vector of the prevalence rates of the public fears was identified. The correlations between fears were analyzed. For this small sample (6 regions) the probability to  find the correlation coefficient between two given attributes $r>0.7$ by chance is $p\approx 0.06$. The $l_1$ weight of the correlation graph $G$ (\ref{lpweight}) was defined with the threshold $\alpha=0.7$. 
The connectivity of the correlation graph definitely increased between 2009 and 2012 (Fig. ~\ref{Fig:UkraineFears}). The variance of the prevalence of various fears also increased \cite{Rybnikovs2017}.

This simple analysis can serve as a prototype measurer of social stress. It seems promising to combine the correlation analysis of social fears with a dynamic  model of network of symptoms developed for individual depression  \cite{Cramer2016} (see Subsection~\ref{Depr}). 
 
 \subsection{Markets and finance \label{SubSec:Market}}
 
 Intensive study of transformation of correlation graph under stress in physiology was started at the 1980s \cite{GorSmiCorAd1st,Sedov}.  Some years later, in 1995, the seminal work of Longin and Solnik \cite{LonginCorrNonconst1995} demonstrated that similar effects exist  in market dynamics: correlations in  equity  market   are  also non-constant and near crisis correlations increase.  
 
During 25 years after this work, the effect of increase of correlations in crisis was demonstrated for many financial time series. It belongs now to the basis of econophysics \cite{Stanley2000,Stanley2002,Utsugi2004,Aoyama2020}, and analysis of correlation graphs is an important instrument for investigation of financial market.

Special attention was paid to the analysis of crises \cite{GorbanSmiTyu2010}. Here are some important and well-represented examples. Two phases of the Asian crisis were identified:  (i) an increase in correlation (contagion) and (2) a continued high correlation (herding). A  shift in variance during the crisis period was also detected. These effects have raised doubts about the benefits of international portfolio diversification \cite{Chiang2007}. Sudden and gradual changes in correlation between stocks, bonds and commodity futures returns were estimated. Most correlations start the 1990s at a low level but increased around the early 2000s and reached peaks during the crisis. Diversification benefits to investors were significantly reduced  \cite{Silvennoinen2013}. The crisis has put pressure on emerging markets in the Middle East and North Africa (MENA) region in particular. Correlations between MENA stock markets and the more developed financial markets, and the intra-regional financial linkages between MENA countries' financial markets were analyzed in \cite{Neaime2012}. A multiscale correlation analysis of the stock market  during the global financial crisis  was carried out for the    G7 and BRIC countries  and  a new multiscale correlation contagion statistic test was developed for to support  decision-making on the global portfolio diversification in crisis  \cite{Wang2017}. Partial correlation financial networks were introduced and applied to analysis of market craches \cite{Millington2020}. Structural breaks of eight national stock markets  associated with the Asian and the  Global financial crises were analyzed. Significant cross effects and long range volatility dependence were revealed \cite{Karanasos2016}.   Analysis of Russian banking and monetary system in global crisis was performed through the analysis of correlation graph and an estimation of conjugacy of monetary and banking policy. Hidden internal patterns were detected in this conjugacy and hidden systemic crisis of Russian banking and monetary system was revealed in 2007 data \cite{PokidyshevaE2010}. This crisis can be indicated by a sudden increase (`explosion') in the values of variance.

The correlation graph approach works also for analysis of a single company. The method of the structure and indicators analysis of a company business processes based on the analysis of  correlations in historic series of expenses is developed \cite{Masaev2010}. This method was used to identify   periods of stress in the company, to optimize and reengineer the management process and to allocate and reallocate resources between company functions. It can be also used by audit companies to identify  cheating in  a company's financial statements. Correlation graph analysis also helps to uncover fraudulent behavior hidden in the business distribution channels, where colluding partners enter into fake big deals in order to obtain lower product prices -- behavior that is considered extremely detrimental to the sales ecosystem. \cite{Luo2020Cheating}.  

\subsubsection{Correlation anatomy of Global financial crisis} 

 In this Subsection, we give just one illustration and refer to special literature for more detail. Let us  illustrate the general rule ( Fig.~\ref{Fig:CorrAdapt}) with the dynamics of the stock market, close to the famous financial crisis of 2008, following our work \cite{GorbanSmiTyu2010}.  We used the daily closing values over the time period 03.01.2006 -- 20.11.2008 for companies that are registered in the FTSE 100 index (Financial Times Stock Exchange Index). The FTSE 100 is a
 weighted index representing the performance
of the 100 largest UK-domiciled blue chip companies.  Thirty companies that had the highest value of the capital on Jan 1, 2007, were selected for  the analysis of correlations in financial time series. The companies, the corresponding abbreviations, and the business type are listed in Table~\ref{CompList}.  Data for these companies were available on the Yahoo!Finance.
  
 \begin{table}[t]\caption{Thirty  companies from the FTSE 100 index chosen for analysis
 \label{CompList}} \centering{\small
\begin{tabular} {|l|l|l|l|}
  \hline
   Number &Business type & Company &  Abbreviation \\
  \hline \hline
  1& Mining & Anglo American plc& AAL\\
  2& & BHP Billiton& BHP\\
   \hline
  3& Energy (oil/gas) & BG Group & BG \\
  4& & BP & BP\\5&  & Royal Dutch Shell & RDSB \\
  \hline
  6 & Energy (distribution) & Centrica& CNA\\7& & National Grid & NG\\
  \hline
8&  Finance (bank) & Barclays plc & BARC\\9&   & HBOS & HBOS\\10& & HSBC HLDG & HSBC \\
 11& & Lloyds & LLOY\\
   \hline
  12& Finance (insurance)  & Admiral& ADM \\13& & Aviva & AV\\14& & LandSecurities&LAND\\
15&   & Prudential& PRU\\16& & Standard Chartered& STAN\\
    \hline
    17& Food production & Unilever &ULVR\\
    \hline
18& Consumer  & Diageo & DGE\\
19& goods/food/drinks & SABMiller& SAB\\
20& & TESCO &TSCO\\
\hline
21& Tobacco & British American Tobacco &BATS\\
22& & Imperial Tobacco & IMT\\
\hline
23& Pharmaceuticals& AstraZeneca &AZN\\
24& (inc. research)& GlaxoSmithKline & GSK\\
\hline
 25& Telecommunications & BT Group &BTA\\
 26& & Vodafone &VOD\\
\hline
27&Travel/leasure& Compass Group & CPG\\
\hline
28&Media (broadcasting) & British Sky Broadcasting & BSY\\
\hline
29& Aerospace/ & BAE System & BA\\
30& defence & Rolls-Royce& RR\\
\hline \hline
\end{tabular}
}\end{table}

\begin{figure}[t] \centering{
a)\includegraphics[width=0.5\textwidth]{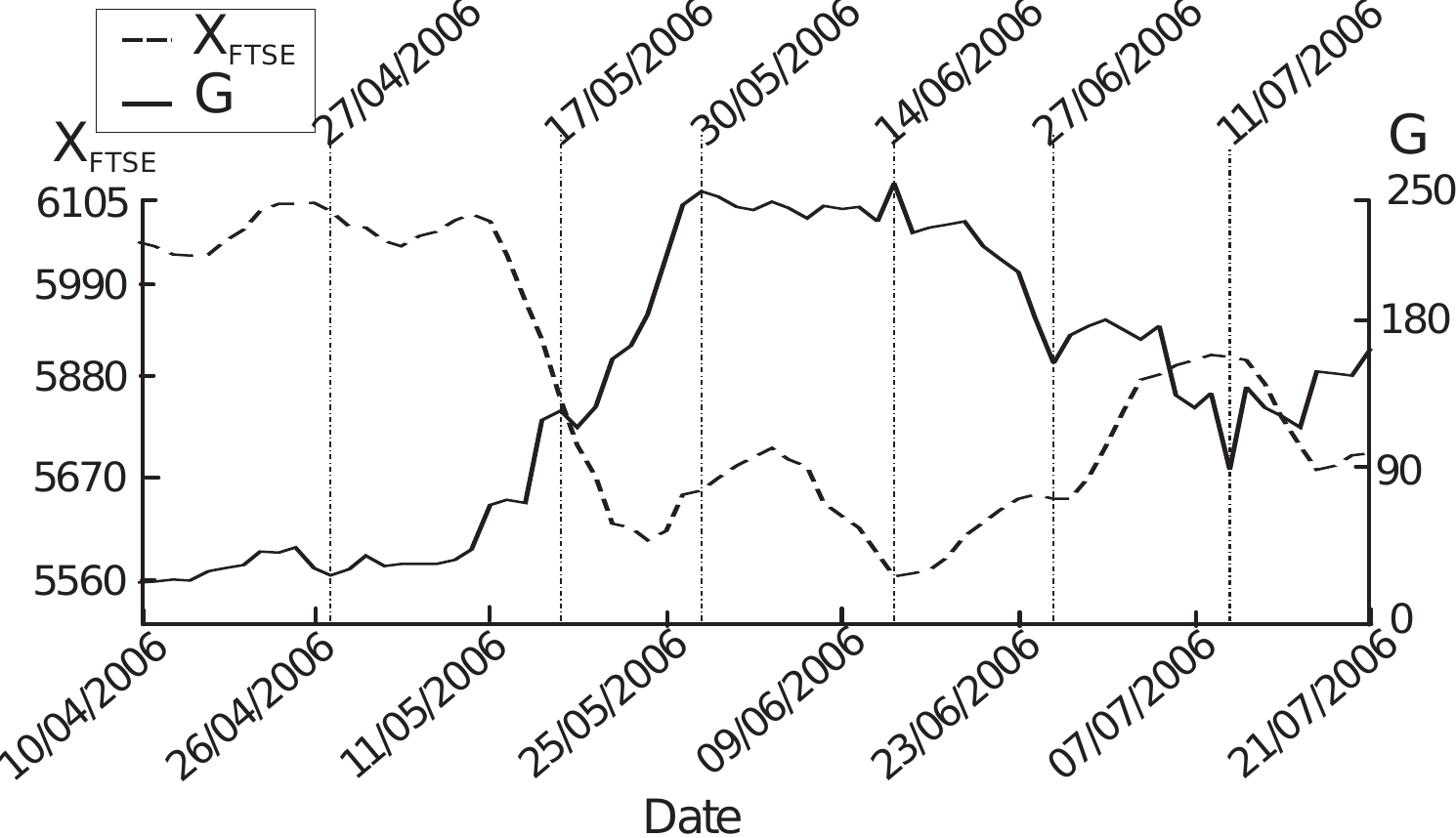} 
b)\includegraphics[width=0.7\textwidth]{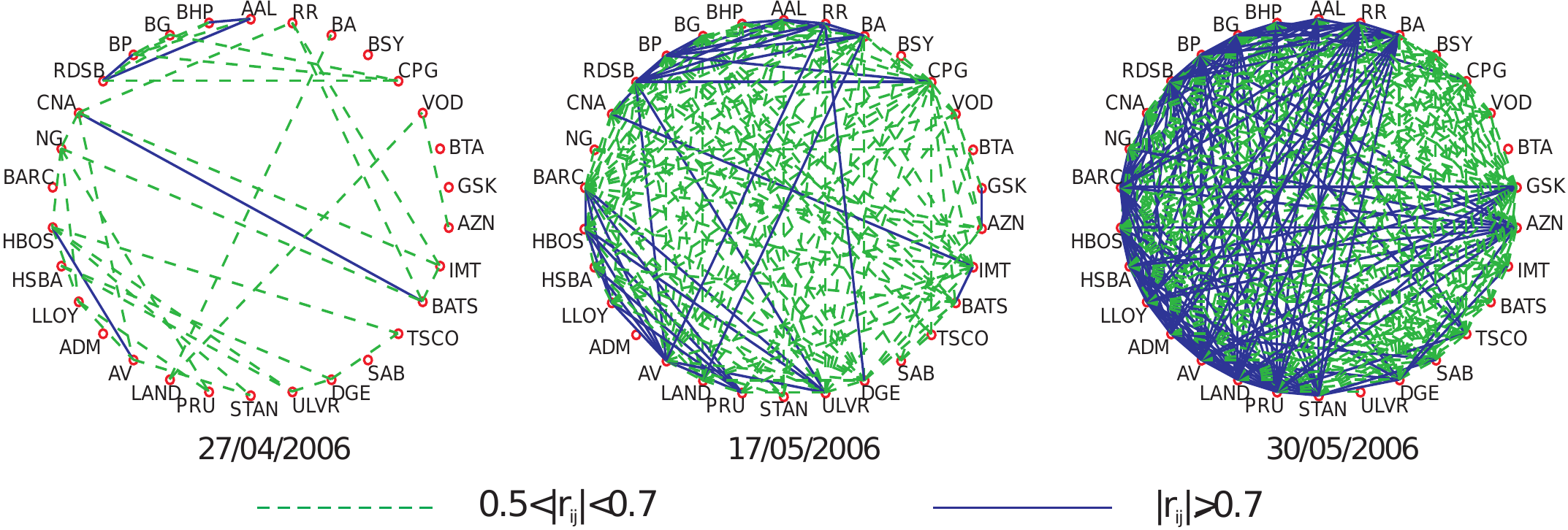} \\
c)\includegraphics[width=0.7\textwidth]{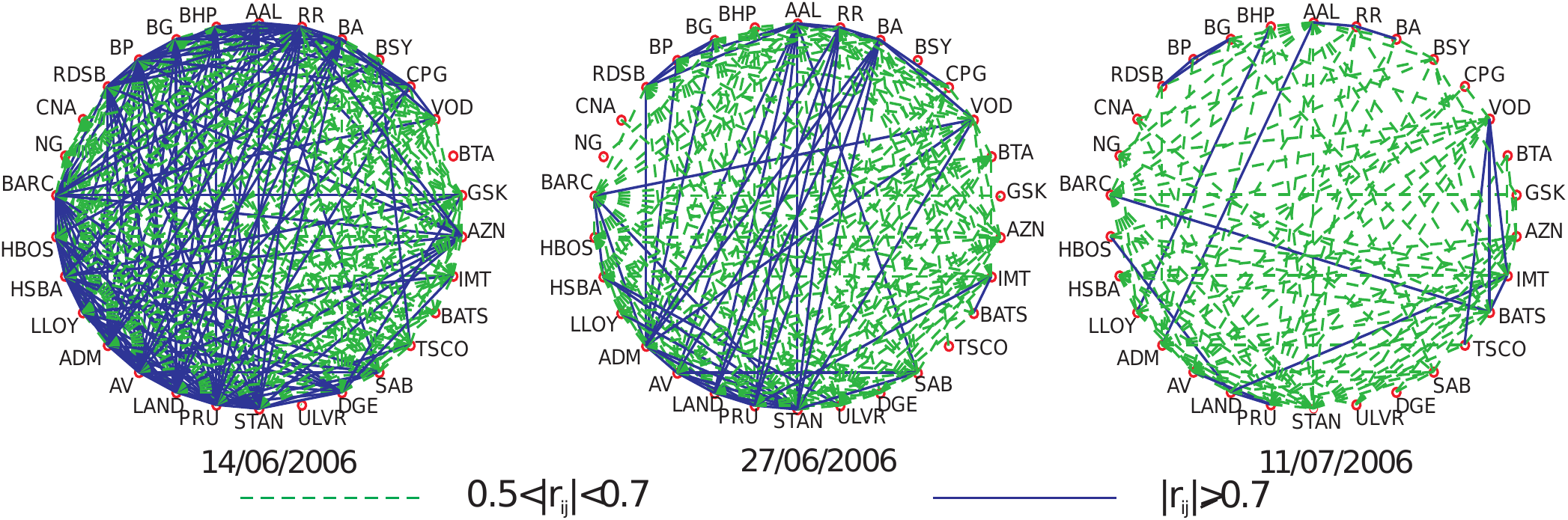}
\caption{Correlation graphs for six positions of sliding time window
 on interval 10/04/2006 - 21/07/2006.
a)~Dynamics of FTSE100 ($X_{\rm FTSE}$, dashed line) and of  $G=\sum_{j>i,|r_{ij}|>0.5}|r_{ij}|$ (solid line), vertical lines correspond to the points that were used
for the correlation graphs.
b)~The correlation graphs for the
first three points, FTSE100 decreases, the correlation graph becomes
more connective. 
c)~The correlation graphs for the last three
points, FTSE100 increases, the correlation graph becomes less
connective.\label{graph_int1}}}
\end{figure}

\begin{figure}[t] \centering{
a)\includegraphics[width=0.5\textwidth]{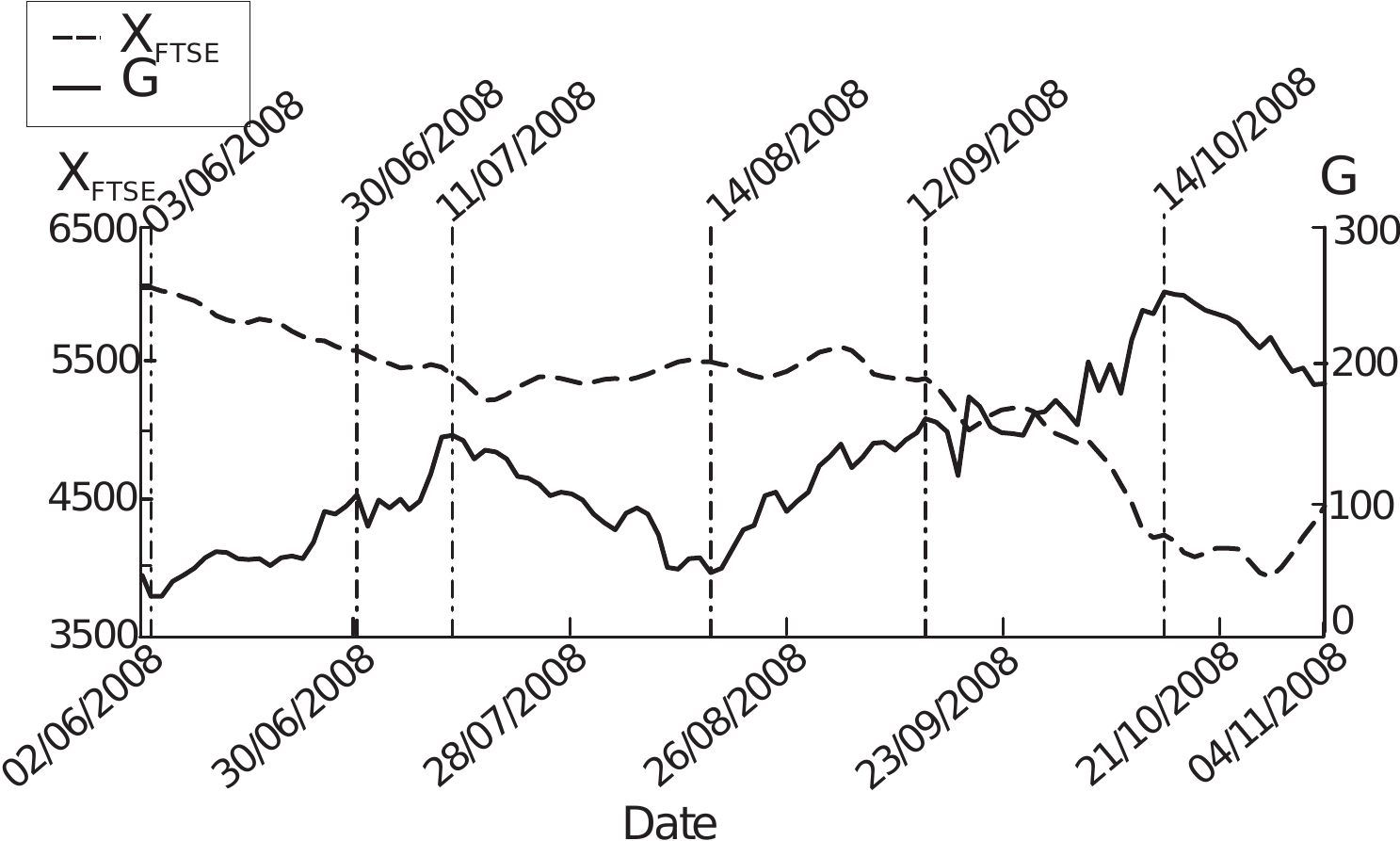} 
b)\includegraphics[width=0.7\textwidth]{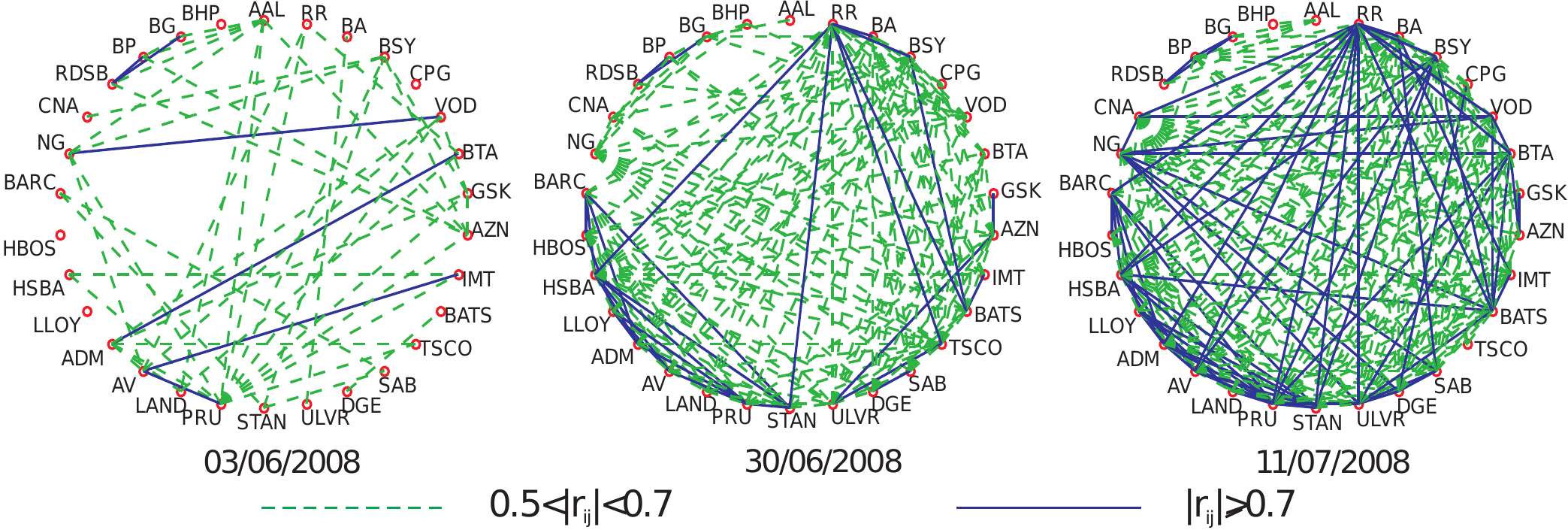}\\
c)\includegraphics[width=0.7\textwidth]{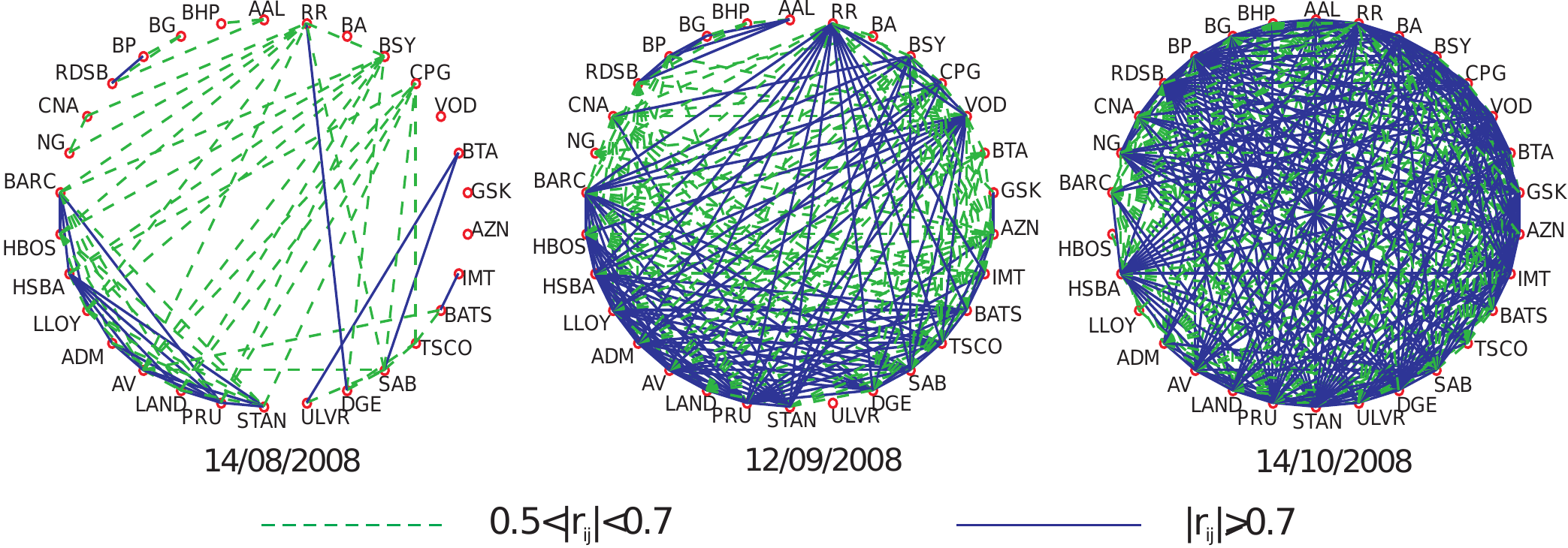}
\caption{\label{graph_int3} Correlation graphs for six positions
of the sliding time window  on interval 02/06/2008 - 04/11/2008.
 a)~Dynamics of FTSE100 ($X_{\rm FTSE}$, dashed line) and of  $G=\sum_{j>i,|r_{ij}|>0.5}|r_{ij}|$  (solid line), vertical lines correspond to the points that were used  for the correlation graphs.
 b)~The correlation graphs for 03/06, 30/06 and 11/07, 2008: FTSE100 monotonically
decreased, and the correlation graph monotonically became more connective. 
c)~The correlation graphs for the last three points. There was a plateau in the FTSE100 dynamics  between 11/07 and 14/08  and the correlation graph on 14/08 became more rarefied. Then the crisis escalated and the connectivity grew stronger.}}
\end{figure}

Data presented in Figs.~\ref{graph_int1}, \ref{graph_int3} demonstrate how  correlations of the daily closing values in the sliding window (20 days) increase in crisis. Dynamics of correlations can be used as an indicator of critical transitions. Moreover, sometimes the changes in correlation occur earlier than in index and there is a chance to use the correlation graph as an early warning signal. The correlation graphs can be used for detailed analysis of the crisis spreading. Thus, we can see how the  strong correlation appeared in the financial sector and then propagated to other types of business. The trajectory of the reverse process is very different as we can  from the comparisons  of cliques and strongly correlated clusters in Fig.~\ref{graph_int1}b and c (especially in the correlation graphs for 17/05/2006 and 27/06/2006). The asymmetry between the ups and downs of the financial market  was also noticed when analyzing the empirical financial correlation matrix of 30 companies included in Deutsche Aktienindex  (DAX) \cite{DrozdComCollNoise2000}.

 \subsection{What can we learn from examples? \label{SubSec:WhatLearn}}

\begin{itemize}
\item The correlation graph and variance are sensitive indicators of stress and illness.
\item In the initial stages of stress and illness, correlations and variances tend to increase.
\item This effect could be sensitive to selection of attributes for analysis and special procedures are needed for selection of relevant attributes. Correlations between relevant and irrelevant attributes may even decrease, while the dynamics of correlations between irrelevant attributes may be  not as clearly defined. 
\item At the disadaptation stage and near the fatal outcome the correlation may decrease (the effect of  `points of no return'). The dynamics of variance near the `points of no return' is reported controversially: in some experiments the variance increases, while in others it decreases, which may be caused by different normalization and dynamics of the means. The hypothesis is that the relative variation should decrease.
\item  The effect is observed in the ensembles of similar systems that adapt to the load of the same factors (organisms under load of harmful factors, enterprises  in adverse conditions, etc.).
\item These ensembles of similar systems can be collected from the history of one system at different time periods.
\item Construction of early warning indicators of stress, illness and various crises on the basis of correlation graph and variance of relevant attributes is possible. 
\item For each class of systems, family of  harmful factors  and type of crisis, relevant data collection and a special data mining procedure are required to develop a stress and crisis metrology.
\end{itemize}
 
 \section{Selye's thermodynamics of adaptation \label{Sec:SelyeTherm}}
 
 Hans Selye did not study the correlation networks. Nevertheless, his experiments and ideas shed light on dynamics of adaptation and on the network effects in adaptation. He discovered the General Adaptation Syndrome (GAS). GAS was described as the universal answer of an organism to every harmful factor or `noxious agent'. Discovery of GAS  focused  several research programs on unspecific universal reactions. Analysis of Selye's experiments led him to introduce a general model of adaptation as a redistribution of the available adaptive resource to neutralize various harmful factors \cite{SelyeAEN,SelyeAE1}. These ideas were formalized and applied to analysis of adaptation and stress in a series of works \cite{GorSmiCorAd1st,GorbanSmiTyu2010,GorbanTyukinaDeath2016,GorbanPokSmiTyu}. In particular, increase of correlations under stress was predicted, the dynamic models of individual adaptations were created. Here we review these works with the addition of several new results. We start from the classical work of Selye's predecessor, Walter Cannon.
 
 \subsection{Cannon's `Wisdom of the body' and industrial controllers \label{SubSec:CnnonsControl}}
 
\begin{figure}[t] \centering{
a)\includegraphics[height=0.25\textwidth]{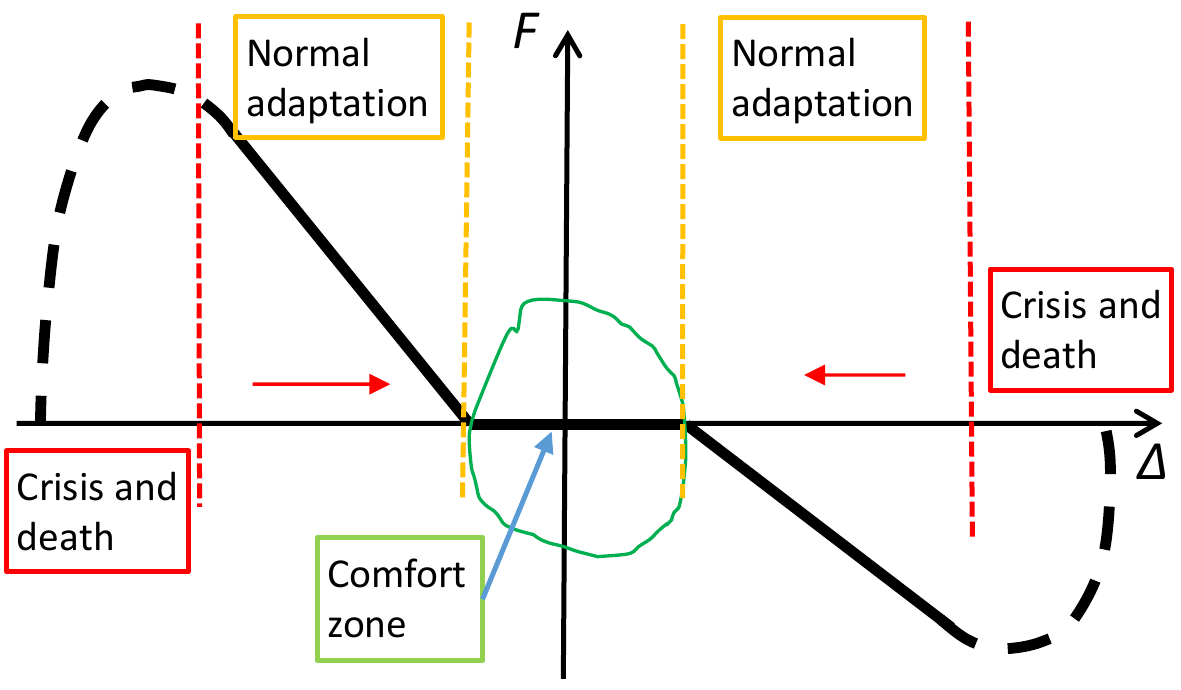} \; \; \; 
b)\includegraphics[height=0.25\textwidth]{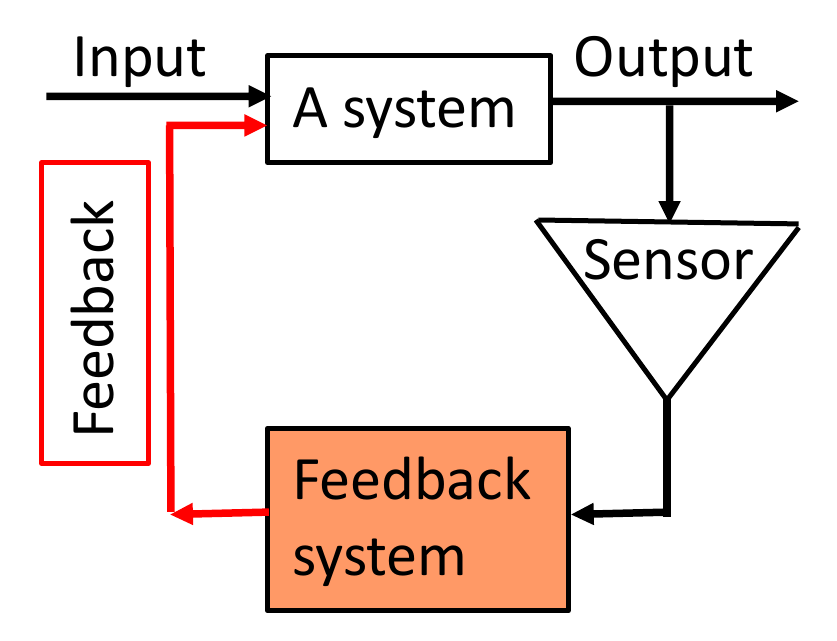} 
\caption{\label{Fig:controllers}The simple schemes of governors and controllers: a)  Restoring force $F$ as a function of  deviation $\Delta$ of the controlled parameter from the target value. The vertical lines separate different regions of deviations: the comfort zone with the tolerance to small deviation from the target, the zone of the normal adaptation where the restoring force increases with the deviation, the crisis and death area where the possibility of adaptation are exhausted; b) The elementary scheme of the closed loop feedback  control.}}
\end{figure}

Idea of control and feedback loop was systematically used by Cannon  \cite{Cannon1932}. There are no explicit formulas in his classical book but the idea of regulation and closed loop control can be easily extracted from the homeostasis description. Two schemes in Fig.~\ref{Fig:controllers} illustrate these ideas. The restoring force returns the system to the vicinity of the comfort zone (Fig.~\ref{Fig:controllers} a). The deviation from the comfort zone can be caused by the external force. If this deviation is too large then the system cannot  be returned to the normal functioning.  This scheme follows Cannon's definition:
``Whenever conditions are such as to affect the organism harmfully, factors appear within the organism itself that protect or restore its disturbed balance'' \cite{Cannon1932}.

The feedback closed loop controller 
 (Fig.~\ref{Fig:controllers} b) is an elementary brick of most controller schemes. It is well understood that the homeostasis theory, this central unifying concept of physiology,   is an application of universal ideas developed in control engineering to living organisms \cite{Schneck1987,Carpenter2004,Billman2020}.  The evolution of our understanding of homeostasis and the role of physiological regulation and dysregulation in health and disease are discussed in modern review \cite{Billman2020}.  Homeostasis his defined there  as a ``self-regulating process by which an organism can maintain internal stability while adjusting to changing external conditions.''
 
We inherited from Cannon the following picture of regulation and control of the body
\begin{itemize}
\item Organism is represented as a structure of relatively independent subsystems (groups of parameters);
\item This decomposition is dynamic in nature and may change under the influence of harmful factors;
\item Homeostasis is provided by a rich structure of feedback loops.
\end{itemize}

More and more detailed regulatory mechanisms are being disclosed and this work is supported by a data-driven approach. Small models provide qualitative understanding, while large models provide us by quantitative insights. Both mechanisms driven and data driven approaches to homeostasis  modeling are developed. Among popular tools there are combinations of system analysis, dynamics, control theory, and modern data mining.  

 \subsection{Factor -- resource model of adaptation \label{SubSec:FactorResource}}
 
 \subsubsection{Adaptation energy  --- generalized adaptability \label{SubSecAEdef}}

Selye  introduced a special currency to pay for the cost of adaptation, the Adaptation Energy (AE). He demonstrated that ``during adaptation to a certain stimulus the resistance to other stimuli decreases.'' His main experimental observations supported this idea showing ``that rats pretreated with a certain agent will resist such doses of this agent which would be fatal for not pretreated controls. At the same time, their resistance to toxic doses of agents other than one to which they have adapted decreases below the initial value'' \cite{SelyeAE1}.

He came to the conclusion:
\begin{itemize}
\item ``These findings are tentatively interpreted by the assumption that the resistance of the organism to various damaging stimuli is dependent on its adaptability. 
\item ``This adaptability is conceived to depend upon adaptation energy of which the organism possesses only a limited amount, so that if it is used for adaptation to a certain stimuli will necessarily decrease. 
\item ``We conclude that adaptation to any stimulus, is always acquired at a cost, namely, at the cost of adaptation energy.
\end{itemize}
 
\begin{figure}[t]
\centering{
\includegraphics[width=0.6\textwidth]{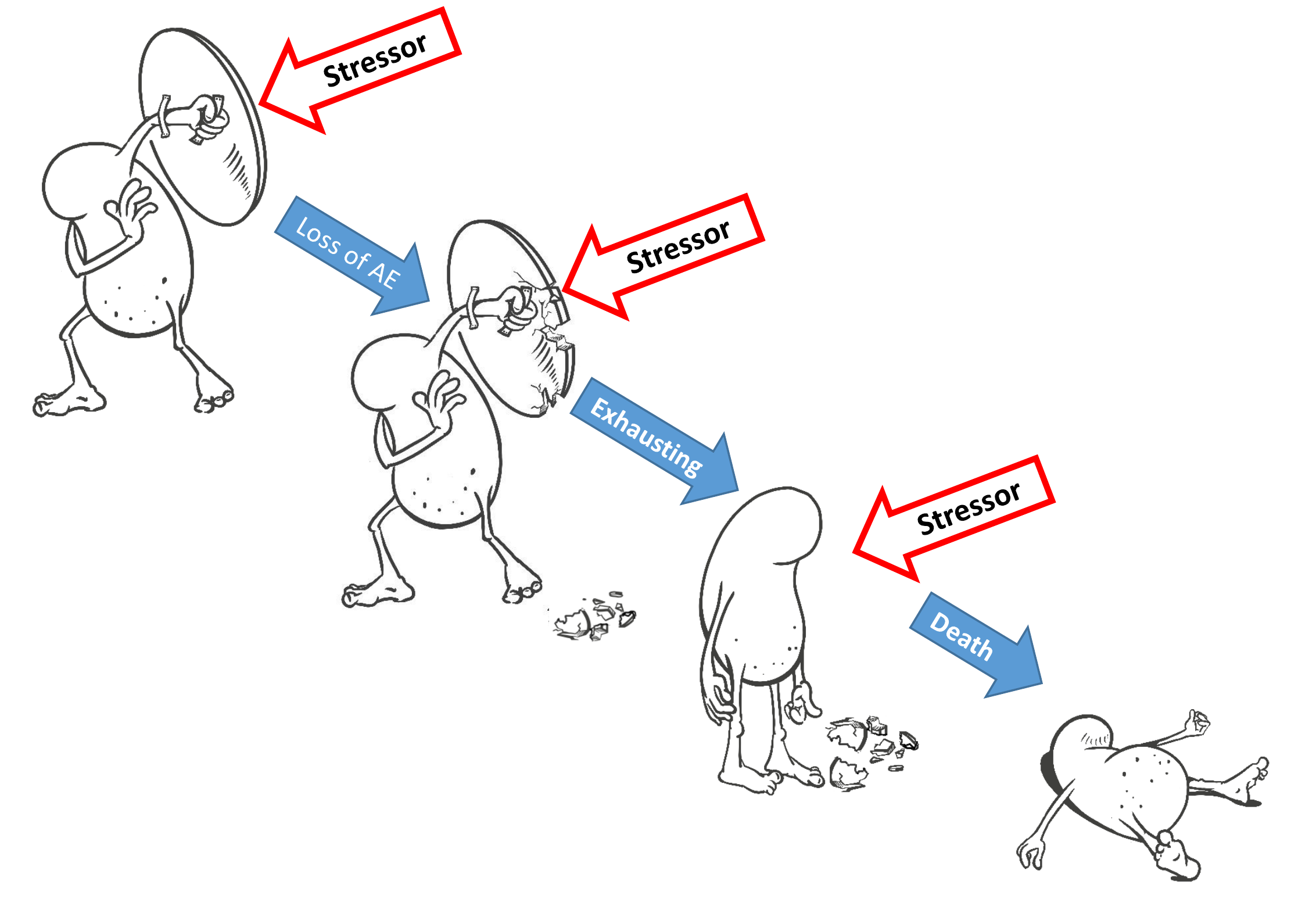}
\caption{Schematic representation of AE spending according to Selye's axioms. The shield of adaptation spends AE for protection from each stress. Finally, AE becomes exhausted, the animal cannot resist stress and dies.  (The character `organism' was drawn by Anna Gorban.)  \label{schemeSel} }}
\end{figure}

The idea of AE that shields the organism against stressors and is spending in this shielding is illustrated in Fig.~\ref{schemeSel}.
	
Selye formulated `axioms' of  AE. This system of axioms was slightly edited and analyzed by his successors  \cite{SchkadeOccAdAE2003}. For modern detailed analysis we refer to \cite{GorbanTyukinaDeath2016}. The first axiom met with the most objections: {\em AE is a finite supply, presented at birth.}
 
In 1952, Goldstone published detailed analysis of Selye's AE axiomatics from the General Practitioner (GP) point of view \cite{GP_AE1952,GP_AE1952C}. From his point of view, the main problem to solve is: How one stimulus will affect an individual's power to respond to a different stimulus? Goldstone  proposed the concept of a production of AE. This AE  may be stored (up to a limit), as a capital reserve of adaptation. He demonstrated that
this concept best explains the clinical observations and Selye's own laboratory findings.  It seems to be possible  that,  had Selye's experimental animals been asked to
spend adaptation at a rate below their AE income, they might have been able to cope successfully with their stressor indefinitely or until their AE production drops below the critical level.

\begin{figure}[t]
\centering{
\includegraphics[width=0.6\textwidth]{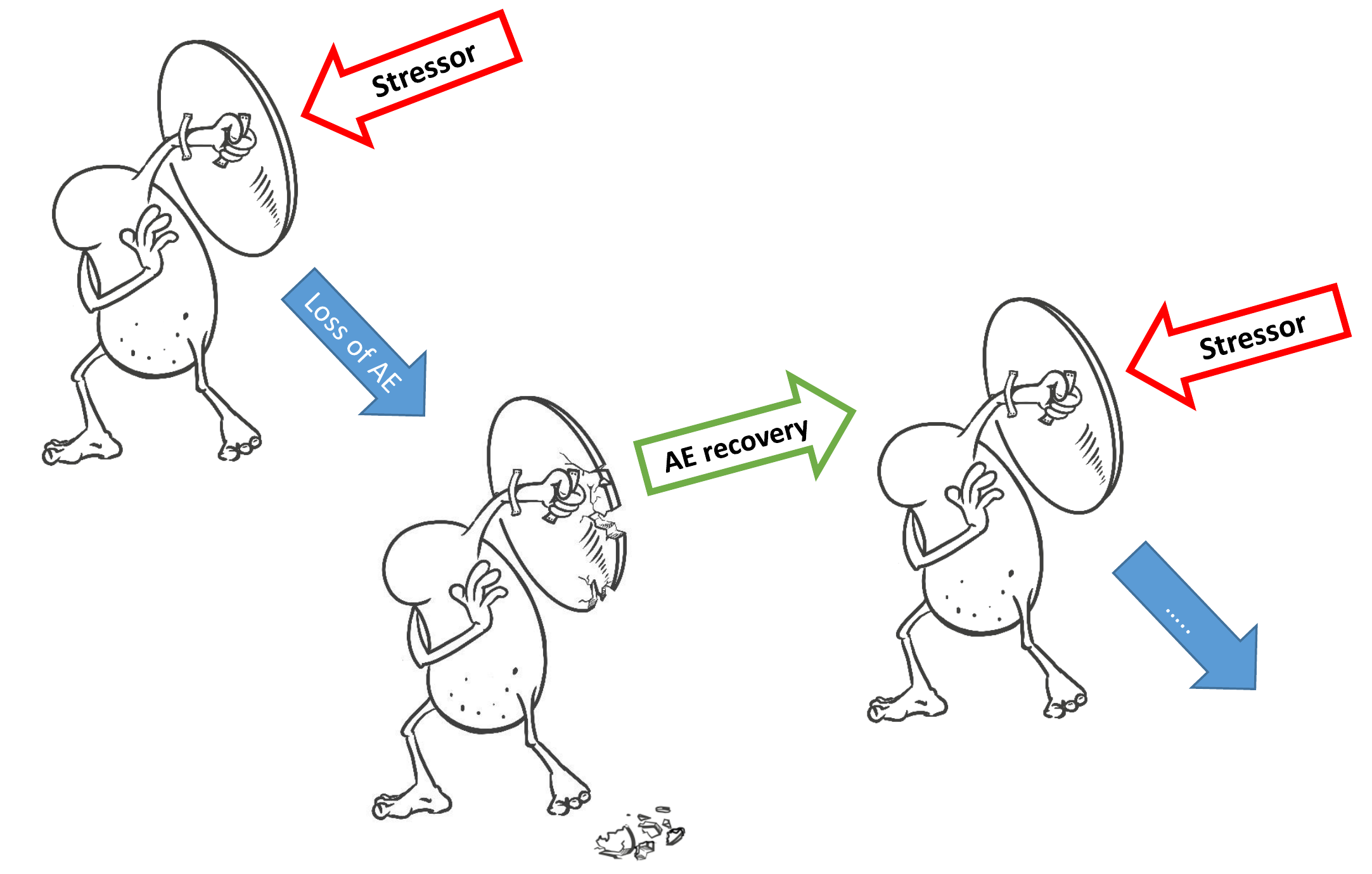}
}\caption{Schematic representation of Goldstone's modification of Selye's axioms: AE can
be restored and adaptation shield may persist if there is enough time and reserve for
recovery. Surviving depends on the balance between AE loss and recovery. \label{schemeGold}}
\end{figure}

These findings may be formulated as a modification of Selye's axiom~1. We call this modification {\em Goldstone's axiom~1' } \cite{GorbanTyukinaDeath2016}:
\begin{itemize}
\item AE can be created, though the income of this energy is slower in old
    age;
\item It can also be stored as Adaptation Capital, though the storage capacity has a
    fixed limit.
\item If an individual spends his AE faster than he creates it,
    he will have to draw on his capital reserve;
\item When this is exhausted he dies.
\end{itemize}
Their difference from Selye's axiom~1 is illustrated by Fig.~\ref{schemeGold} (compare to
Fig.~\ref{schemeSel}). 

The further development of Selye's axiomatics was performed by Garkavi with coauthors
 \cite{Garkavi1979}. They  developed the activation therapy, which was applied in clinic,
aerospace and sport medicine.
 
The AE for Selye was a generalized measure of adaptability and not a physical quantity, not a type of energy. Nevertheless, the request to  demonstrate the physical nature of this `energy' was very popular. Even in the  `Encyclopedia of Stress' we can read: `As for adaptation energy, Selye was never able to measure it...' \cite{AEencicl}.  Despite of common use of this notion  as a general  `adaptation resource' (see, for example, \cite{BreznitzAEappl,SchkadeOccAdAE2003}) the metaphor of energy stimulated criticism of the concept: people took it literally and demanded a direct measurement of this `energy'.

It is worth noting that the physical meaning of the free energy of adaptation has recently been revived in physical models of adaptable systems \cite{Allahverdyan2016}. In any case, the AE can be considered as an internal coordinate on the `dominant path' in the adaptation model, regardless of whether it has a physical interpretation or not  \cite{GorbanTyukinaDeath2016}.
 
The basic model of an adaptation resource and its use should include few details and be as simple as possible, as we expect from a model of such generality. Let us represent the life history of an organism  as a sequence of adaptation events. Each event is a distribution of the available adaptation resource for neutralization of current values of harmful factors.
This view reproduces the structure of Selye's experiments: a comfortable life -- the action of harmful agents -- a comfortable life... (See Figs.~\ref{schemeSel}, \ref{schemeGold}.) 

\subsubsection{Factors--resource quasistatic models of adaptation}

 We represent the systems, which are adapting to stress, as the systems which optimize
distribution of available amount of resource for the neutralization of different harmful  factors (we also consider deficit of anything needful as a harmful factor). 

 Formally, consider systems that are under the influence of
several factors $F_1,... F_q$.  Each factor $F_i$ is characterized its intensity, $f_i$
($i=1,...q$). We accept the convention of the scale orientation:  all these factors are
harmful.  If the {\em  fitness function}  $W$ is known then fitness decreases with increasing factor intensity.  

The adaptation system  is a `shield' that protects the organism and decreases the influence of these factors. Let the organism have  an available adaptation resource, $R$. If some amount of  the resource is distributed for neutralization of factors then the effective pressure of the $i$th factor is $\psi_i=f_i-a_i r_i \geq 0$ , where   $a_i>0$ is the coefficient of efficiency of neutralization of factor   $F_i$ and  $r_i\geq 0$ is the amount of resource  assigned for the neutralization of factor $F_i$. The condition  $\sum_i r_i \leq R$ should hold. The zero value of $\psi_i$ is optimal, and further compensation is impossible. We can take all $a_i=1$ after rescaling, and  $\psi_i=f_i- r_i$

Thus, two quantities describe interaction of the system with a factor $F_i$: the factor uncompensated pressure $\psi_i$ and the amount of resource $r_i$ assigned for the factor neutralization.

Adaptation should optimize the allocation of resources to neutralize factors. To model such optimization, the objective function is needed.   Even an assumption about the existence of an objective function and about its general properties helps in the analysis of the adaptation process. Assume that adaptation should
maximize an objective function (`fitness') $W(\psi_1, \ldots , \psi_q)$ which depends on the
uncompensated values of factors, $\psi_i=f_i- r_i\geq 0$ for the given
amount of available resource:
\begin{equation}\label{Optimality0}
\left\{ \begin{array}{l}
 W(f_1- r_1, f_2- r_2, ... f_q-  r_q) \ \to \ \max \ ; \\
 r_i \geq 0$, $f_i-  r_i \geq 0$, $\sum_{i=1}^q r_i \leq R   .
\end{array} \right.
\end{equation}
It is worth to mention that the optimal distribution of resources given by the optimization problem (\ref{Optimality0}) is invariant with respect to monotonically increasing transformations of the objective functions $W \to \varphi(W)$.

The total amount of AE, $ R $, changes in the chain of adaptation events. Selye believed that this was only consumed, but could not be produced. Goldstone proposed a more flexible concept with the body's ability to generate AE \cite{GP_AE1952,GP_AE1952C}. The first kinetic models of AE production and consumption were developed in \cite{GorbanTyukinaDeath2016}.

The function $W$ should monotonically decrease in the following sense:
 if $\phi_i\geq \psi_i \geq 0$ for  all $i=1, \ldots , q$ then $W(\phi_1, \ldots , \phi_q) \leq W(\psi_1, \ldots , \psi_q)$. That is, the factors are harmful, indeed. Using this monotonicity, we can reformulate the optimality problem. Let $\sum_i{f_i} -R> 0$. Find $\psi_1, \ldots , \psi_q  $ such that

\begin{equation}\label{Optimality}\boxed{
\left\{ \begin{array}{l}
W(\psi_1, \ldots , \psi_q) \ \to \ \max \ ; \\
f_i \geq \psi_i \geq 0$, $\sum_i  {\psi_i} =\sum_i {f_i} -R   .
\end{array} \right.}
\end{equation}
 
 If $\sum_i{f_i} -R\leq 0$ then all factors can be fully compensated by the allocation of resource  $r_i={f_i} $.

The question remains: why  adaptation follows any optimality principle? Optimality is proven for microevolution processes and ecological succession.
The  backgrounds   of `natural
selection' in these situations are well established after works
of Haldane (1932) \cite{Haldane1932} and Gause (1934) \cite{Gause}.
Various concepts of {\it fitness} (or `generalized fitness') are elaborated in many
details (see, for example, review papers \cite{Bom02,Oechssler02,GorbanSlect2007,Lion2018}).
It seems productive to accept the idea of optimality and use it, while it does not contradict the data. We discuss optimality principle in the next  Sec.~\ref{SubSec:Fitness}.

\subsubsection{The challenge of defining wellbeing and instantaneous  fitness \label{SubSec:Fitness}}
 
In the optimality principle (\ref{Optimality0}), (\ref{Optimality}), an individual's fitness $W$ is used. It measures the wellbeing (or performance) of an organism. This is an instantaneous  value, defined for at an  instant in time. Defining of the instantaneous fitness for measuring an individual's performance is a highly non-trivial task.  In mathematical biology the term `fitness' is widely used in essentially another sense based on the averaging  of reproduction rate over a long time \citep{Haldane1932,Maynard-Smith1982,MetzNisbetGeritz1992,GorbanSlect2007,Karev2014}. This is Darwinian fitness. Darwinian fitness is non-local in time. It is the average reproduction coefficient in a series of generations and does not characterize an instantaneous  state of an individual organism. 

Merely highlighting the difference between  individual instantaneous fitness used in the optimality principle  (\ref{Optimality0}), (\ref{Optimality}) and Darwinian fitness is not very useful. Optimality principles in biology have one source, the evolutionary optimality based on Darwinian fitness and on the space of possibilities where  selection works (this space often deserves the epithet ``mysterious space of possibilities'' because of the difficulty of characterizing it).  Thus, in order to use the optimality principle with individual instantaneous fitness as an objective function, we must build a bridge between this quantity and Darwinian fitness. We cannot expect this bridge to be completely constructive, but the logic of its construction should be outlined.

The synthetic evolutionary approach to optimality starts with the analysis of genetic variation and studies the phenotypic effects  of that  variation  on physiology. Then it goes to the performance of organisms in the series of generations (with supplementary analysis of various forms of environment) and, finally, it must return to Darwinian  fitness \cite{Lewontin1974}. The  physiological ecologists  focus their attention,  on  the  observation  of  variation  in individual performance \cite{Pough1989}. In this approach we have to measure the individual performance and then link it to  Darwinian fitness. Empirically, Darwinian fitness is built from individual instantaneous  performances, not the other way around.

The connection between individual performance and Darwinian fitness is not trivial. In particular, the dependence between them is not necessarily  monotone because the individual success and success in effective reproduction do not coincide \cite{MacArthurWilson1967,Pianka1970}. Fisher  \cite{Fisher1930}  formulated the problem of balancing individual performance and reproductive success as follows:  ``It  would  be instructive  to know  not  only  by what physiological  mechanism  a  just apportionment  is  made  between  the  nutriment  devoted  to  the  gonads  and that  devoted  to the  rest  of  the  parental
organism,  but  also  what  circumstances  in  the  life-history  and environment
would  render  profitable  the  diversion  of  a  greater  or lesser  share  of  the
available  resources  towards  reproduction.'' The optimal balance between individual performance and fecundity depends on environment. This is well-known since the works of Dobzhansky \cite{Dobzhansky1950} who stated that in  the  tropical  zones  selection typically  favors   lower  fecundity  and  slower  development,  whereas  in  the temperate  zones  high  fecundity  and  rapid development could  increase  Darwinian fitness.

Despite some controversy, the idea that the states of an organism can be linearly ordered from bad to good (well-being) is popular and useful in applied physiology. The coordinate on this scale is also called `fitness'. Several indicators are proposed for fitness assessment  and then the fitness is evaluated as a composite of many attributes and competencies. For example,  in sport physiology these competencies include physical, physiological and psychomotor factors \cite{ReillyDoran2003}. The combination of various components for evaluation of sport-related instantaneous  individual fitness depends upon the specific sport, age, gender, individual history and even on the role of the player  in the team.

In ecological physiology the evaluation of  `performance'  is also `task--dependent' \cite{Wainwright1994}. It refers to an organism's ability to carry out specific  tasks (e.g., capture prey, escape predation, obtain mates). Direct instantaneous  measurement of Darwinian fitness is impossible. Nevertheless, it is possible to measure various instantaneous  performances several times and consider them as the components of fitness in the chain of generations. Several criteria for selection of the good  measure  of performance in the evolutionary study were proposed \cite{Arnold1983}.

\begin{figure}[t]
\centering{
\includegraphics[width=0.45\textwidth]{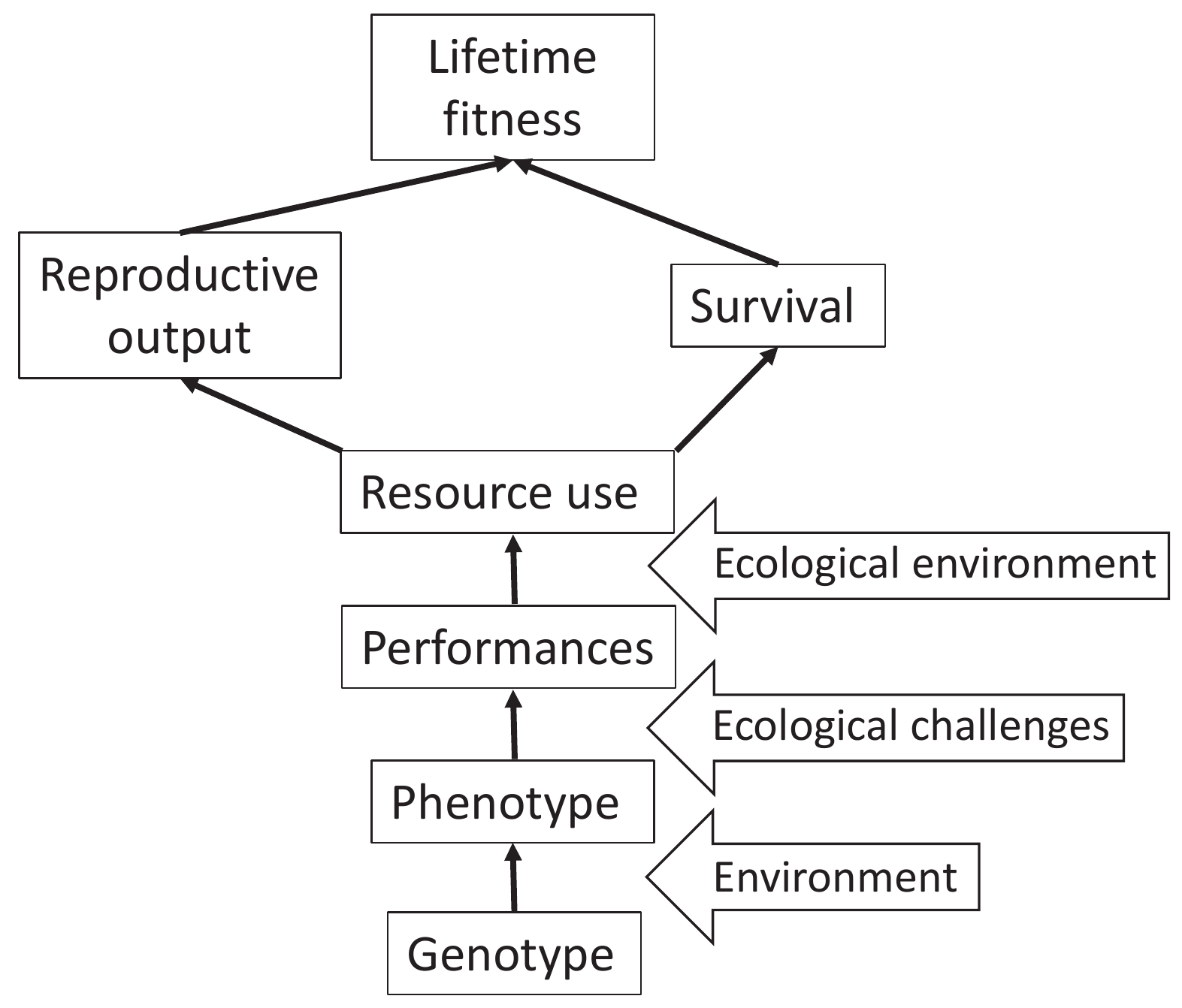}
}\caption{The paths through from genotype to Darwinian fitness. 
 \label{PerfScheme}}
\end{figure}

The scheme of relations between performance and lifetime fitness is sketched on flow-chart (Fig.~\ref{PerfScheme}) following \cite{Wainwright1994} with minor changes:
Genotype in combination with environment determines the organismal design (the phenotype) up to some individual variations.  Phenotype determines  the limits of an individual's ability to perform day-to-day behavioral answer to main ecological challenges (performances). Performance capacity interacts with the given ecological environment and determines the resource use, which is  the key internal factor determining reproductive output and survival. Darwinian fitness may be defined as the lifetime fitness averaged in a sequence of generations.

We use the instantaneous  individual fitness (wellbeing) $W$ as a characteristic of the current state of the organism, reflecting the non-optimality of its performance. The following normalization is convenient: $W=1$ means the maximal possible performance, while $W=0$ means inviability (death). If the organism lives at some level of $W$ then we can consider $W$ as a factor in the lifetime fitness.  The more detailed review of the problem of defining instantaneous individual fitness is presented in \cite{GorbanTyukinaDeath2016}. 

For the detailed philosophical discussion of various notions of fitness we refer to \cite{Trivino2016}. It is argued that fitness is something different from both the physical traits of an organism and the number of offspring it leaves and a definition of fitness as a causal dispositional property is developed.
 
Finally,  we have to stress that the quantitative definition of   $W$ is related to its place in the equations \cite{GorbanTyukinaDeath2016}. The change of the basic equation will cause the change of the quantitative definition.  Moreover, it is plausible that for different purposes we may need different definitions of $W$.
  
  \subsection{Liebig's law of the minimum and optimal resource allocation \label{SubSec:Liebig}}
  
  The optimality principle (\ref{Optimality}) in its general form without specifying the fitness function $ W (\psi_1, \ldots, \psi_q) $ does not add much light to understanding the adaptation process. The situation changes  if we add plausible hypotheses about the interaction of factors.  
  
  The fruitful idea is limitation: the worst factor determines the fitness value. That is,  
\begin{equation}\label{LawMin}\boxed{
  W (\psi_1, \ldots, \psi_q) = w\left(\max_{i=1, \ldots , q}\{\psi_1, \ldots, \psi_q\}\right), }
\end{equation}
where $w$ is a monotonically decreasing function of one variable.

\begin{figure}[t] \centering{
\includegraphics[width=0.3 \textwidth]{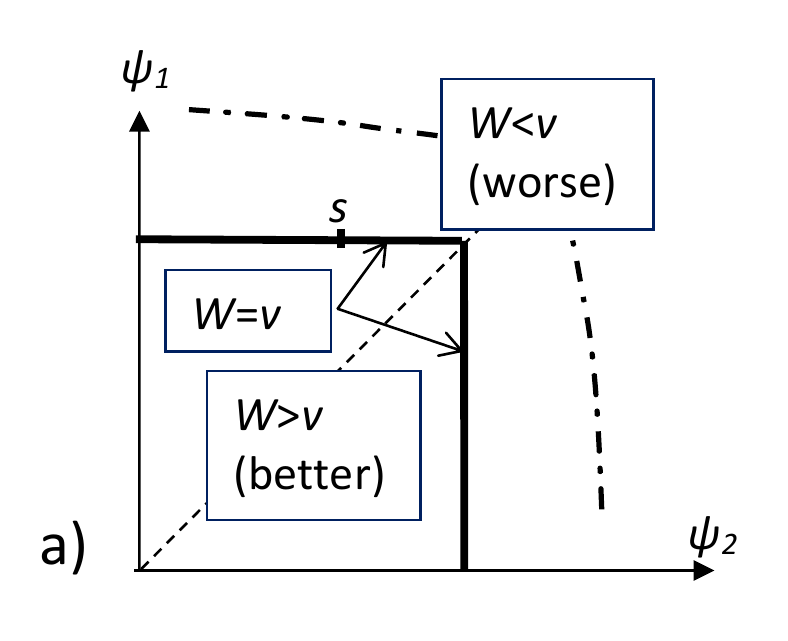}  \;\;\;
\includegraphics[width=0.3 \textwidth]{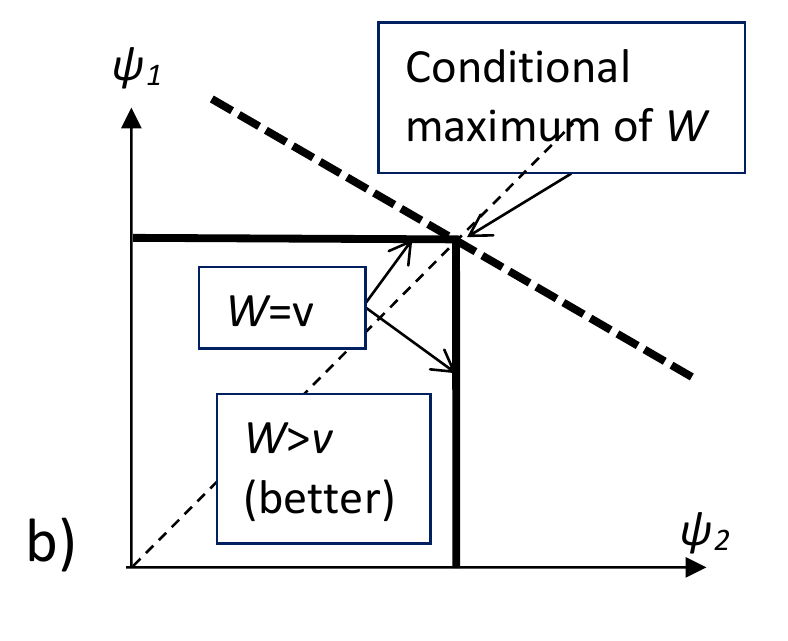}}
\caption{\label{Liebig} (a) The law of the minimum. Coordinates $\psi_1$,
$\psi_2$ are normalized values of non-compensated factors. For a given state,
$s=(\psi_1(s),\psi_2(s))$, the bold solid line
$\min\{\psi_1,\psi_2\}=\min\{\psi_1(s),\psi_2(s)\}$ separates the states with
better conditions (higher fitness) from the states with worse
conditions (a). On this line the limiting factor has the same value. The dot
dash line shows the border of survival. On the dashed line (the diagonal) the
factors are equally important ($\psi_1=\psi_2$). (b)  Conditional maximization of $W$  with linear constrains. The dashed line represents the plane of linear constrains with positive normal. The conditional maximum of $W$ on this plane is achieved on the diagonal, where the pressure of all non-compensated factors is the same.}  
\end{figure}  
  
  This form (\ref{LawMin}) of  the {\em law of the minimum} assumes correct normalization of factors and evaluation of the pressure of factors in the same units of measurement.
  
\begin{figure}[t] \centering{
\includegraphics[width=0.65\textwidth]{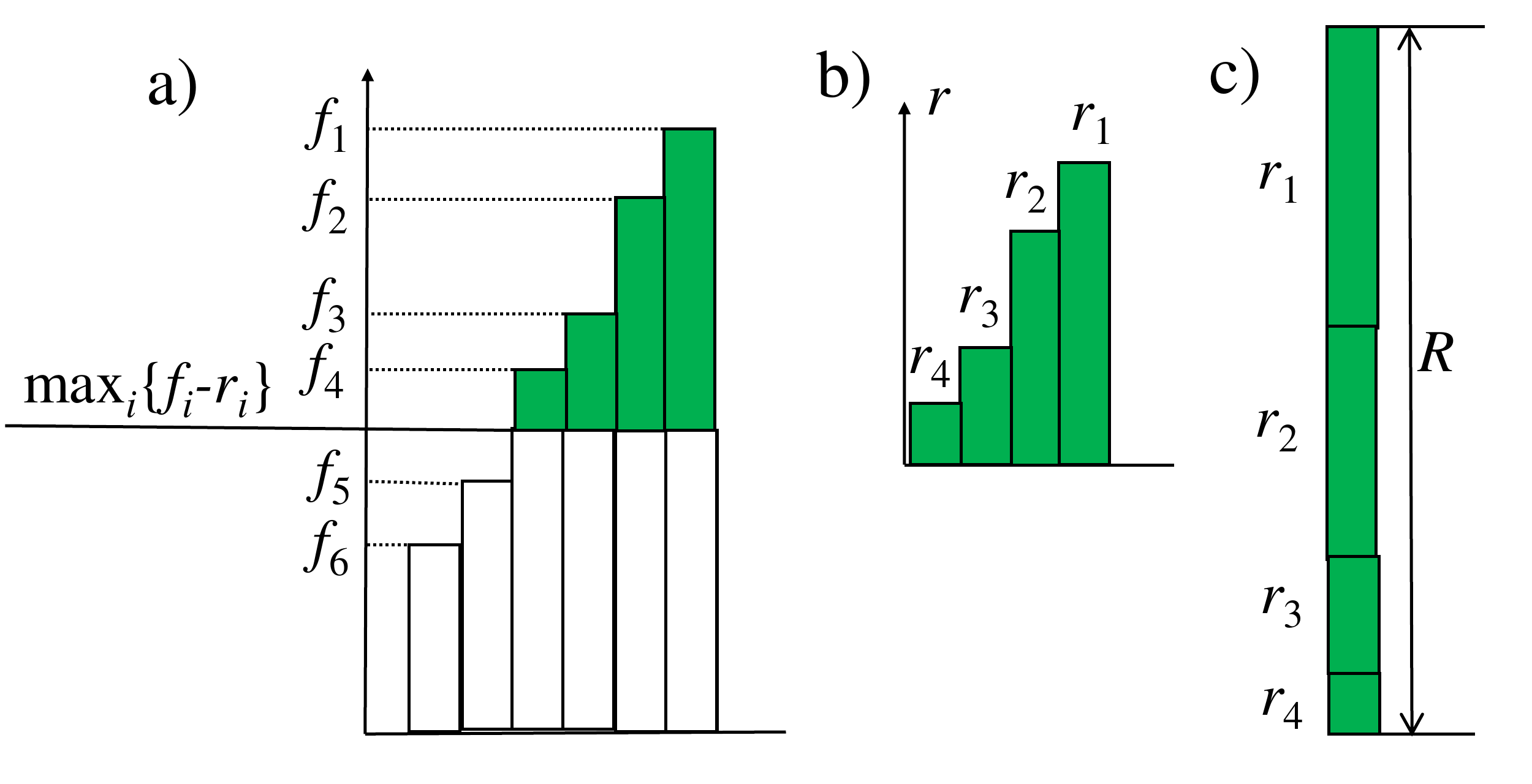}}
\caption{\label{Fig:FactorDistrib} Optimal distribution of resource
for neutralization of factors under the law of the Minimum. (a)
Histogram of factors intensity (the compensated parts of factors are
highlighted, $k=3$), the distribution of compensated factors became more uniform than the initial distribution of factor loads, (b) a histogram of resource allocation, (c) the sum of allocated resources. }
\end{figure}

 The function  $w\left(\max_{i=1, \ldots , q}\{\psi_1, \ldots, \psi_q\}\right)$ is  not fully defined because it includes unknown monotonically decreasing function $w$ of one variable. Nevertheless, the level sets do not depend on the specific  $w$.  The superlevel set given by the inequality $W>v$ for a value $v$ is a cube (see Fig.~\ref{Liebig}). Maximum of $W$ in the positive orthant on a hyperplane with positive normal belongs to the diagonal, where all the factor pressures  are equal (Fig.~\ref{Liebig} b). The optimality problem  (\ref{Optimality}) for functions (\ref{LawMin}) is a bit more complex and requires to maximize $W$ in the intersection of the parallelepiped  $f_i \geq \psi_i \geq 0$ with the hyperplane $L$ given by the equation 
 $\sum_i  {\psi_i} =\sum_i {f_i} -R$. Nevertheless, it can be solved in explicit form. 
 For any objective function $W$ that satisfies the law of the minimum in the form (\ref{LawMin}) the optimizers $r_i$ in the problem (\ref{Optimality0}), (\ref{Optimality}) for functions (\ref{LawMin}) are defined by the following algorithm.
\begin{enumerate}
\item{Re-enumerate factors in the order of their intensities starting from the worst: $f_{1} \geq f_{2} \geq ...
f_{q}$.}
 \item{Calculate differences $\Delta_j =f_{j} -f_{j+1}$
(take formally $\Delta_0=\Delta_{q+1}=0$).}
 \item{Find such $k$ ($0 \leq
k \leq q$) that
  $$\sum_{j=1}^k j \Delta_j
  \leq R \leq \sum_{j=1}^{k+1}j  \Delta_j \ . $$
 For $R< \Delta_1$ we put $k=0$, for $R > \sum_{j=1}^{k+1} j
 \Delta_j$ we take $k=q$.}
\item{If $k < q$ then the optimal amount of resource $r_{j_l}$ is
\begin{equation}\label{OptimDistr}
r_{l}=\left\{\begin{array}{ll}
 & {\Delta_l}  + \frac{1}{k}
 \left(R- \sum_{j=1}^k j \Delta_j\right)\
 , \  \ {\rm if} \ \ l \leq k+1\ ; \\
 &0\ , \ \ \ \ \ {\rm if } \  \ l> k+1\ .
\end{array}
 \right.
\end{equation}
If $k=q$ then $r_i = f_i$ for all $i$.  }
\end{enumerate}
  After all we have to restore the initial enumeration $f_{i_1}\geq
f_{i_2} \geq \ldots f_{i_q}$. This optimization is illustrated in
Fig.~\ref{Fig:FactorDistrib}.

   We demonstrated here  {\it the law of the minimum paradox }\cite{GorbanPokSmiTyu}: if for a randomly chosen pair ``organism--environment" the law of the minimum typically holds (\ref{LawMin}), then, in a well-adapted system, 
we have to expect violations of this law and the pressure of factors tend to be less different (see Fig.~\ref{Fig:FactorDistrib}). In particular, the factors with the strongest pressure can become equally important (factors $f_{1-4}$ in Fig.~\ref{Fig:FactorDistrib}).

\begin{figure}[t] \centering{
\includegraphics[width=0.95\textwidth]{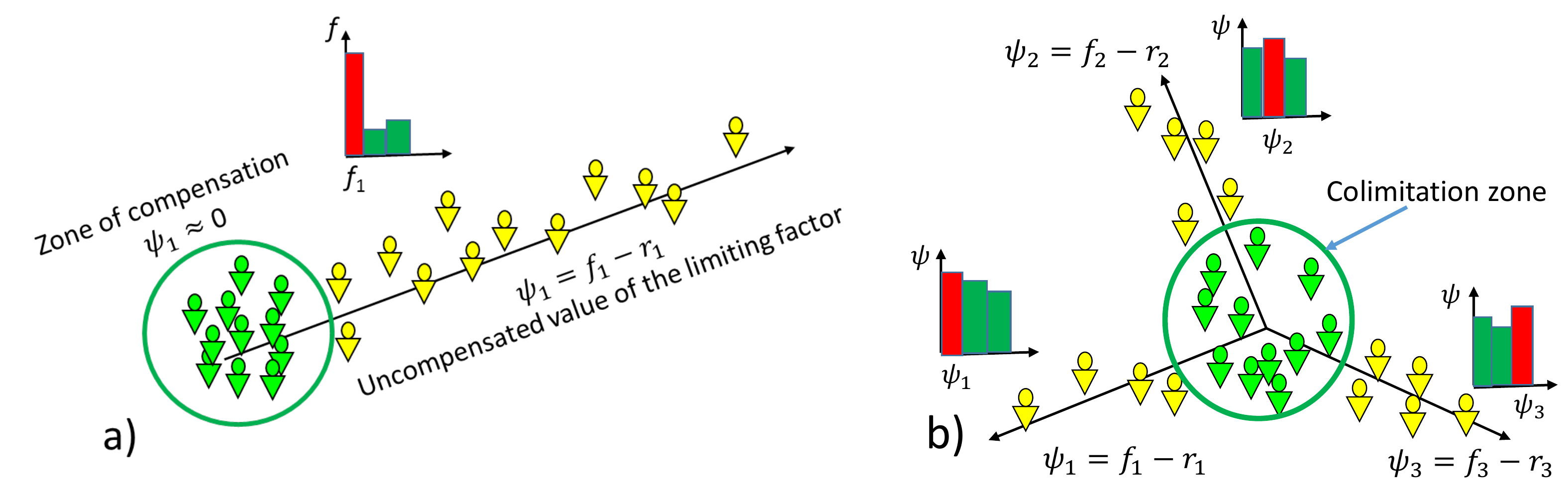}}
\caption{\label{Fig:LimFluct}High correlations in the situation with strong limitation and uncompensated  limiting factor (a) and low correlations near colimitation, where factor pressures are almost equal (b).}
\end{figure}

This effect also provides a mechanism of increase of correlations and variance under strong load of an external harmful factor.  For strong limitation and strong load of the limiting factor the main difference between organisms is the uncompensated pressure of the limiting factor (Fig.~\ref{Fig:LimFluct}a) and near colimitation different factors can become limiting `by chance' and the organisms deviate from the comfort zone in different directions (Fig.~\ref{Fig:LimFluct}b). According to the general results about quasiorthogonality of random vectors in high-dimensional spaces \cite{GorbTyuProSof2016}, these directions are almost orthogonal (with high probability) and the {\em dimension of the data cloud can be effectively estimated as the number of independent co-limiting factors.}

The observation that adaptation makes the
limiting factors equally important was supported by many data of
human adaptation to the Far North conditions \cite{GorSmiCorAd1st}. At the same time, disadaptation causes inequality of factors and leads to appearance
of single limiting factor. The results are used for monitoring of human
populations in Far North \cite{Sedov}.

 The ``law of the minimum" in its original form stated that growth is controlled by the
scarcest resource (limiting factor) \cite{Liebig1}. This law was published by Justus von Liebig's  in  1840  and is very often called ``Liebig's law''. Twelve years  earlier, Carl Sprengel published an article that contained in essence the law of the minimum. Thus, it can be called the Sprengel--Liebig law of the minimum \cite{van der Ploeg1999}. The history, of this law, its applications  and related controversies were reviewed in \cite{GorbanPokSmiTyu}.
 
The law of the minimum is one of the most important tools for mathematical modeling of ecological systems. This provides the clue to creating the first  model of multi-component and multi-factor systems. This clue sounds pretty simple: first of all, we must take into account the most important factors, which are probably limiting factors. Everything else should be eliminated and returned back only if `sufficient  reason' is found.
This law was considered as an ``archetype" for ecological modeling \cite{Nijlandatal2008}.
   
 The economical metaphor of ecological concurrency was elaborated for combination of the idea of limitation with optimization approach \cite{BloomChapinMooney1985,ChapinSchulzeMooney1990}. In this approach,  a
plant should adjust allocation so that, for a given expenditure in
acquiring each resource, it achieves the same growth response:
Growth is equally limited by all resources. This is a result of
adjustment: adaptation makes the limiting factors equally important.
 
Three inter-related issues which form the core of
evolutionary ecology were considered in \cite{SihGleeson1995}: 
 (i) key environmental factors,
(ii) organismal traits that are responses to the key factors, and  
(iii) the evolution of these key traits. 
 Adaptation leads to optimality and
equality of traits as well as of factors but under variations some
traits should be more limiting than others,  therefore the concept of 'limiting traits'
was introduced  \cite{SihGleeson1995}.  

Analysis of species-specific growth and mortality of juvenile trees at
several contrasting sites suggests that light and other resources
can be simultaneously limiting, and challenges the application of
the Law of the Minimum to tree sapling growth \cite{Kobe1996}. This is the manifestation of the  ``colimitation" phenomenon: limitation of growth and surviving by a
group of equally important factors and traits.
 
There are  regions in the world ocean
where chlorophyll concentrations are lower than expected
concentrations given the phosphate and nitrate levels.
Limitations of phytoplankton  growth by other
nutrients like silicate or iron was studied and supported
by experiments \cite{Aumontatal2003}.  

The double--nutrient--limited growth appears also as a transition
regime between two regimes with single limiting factor. This phenomenon was studied for bacteria and yeasts at a constant dilution rate in the chemostat, and three
distinct growth regimes were recognized:  a  carbon-limited regime,  a nitrogen-limited regime, and the double--nutrient--limited growth regime where both the carbon and
the nitrogen source were below the detection limit. The double--nutrient--limited zone is very narrow at high growth rates and becomes broader during slow growth
\cite{EgliZinn2003,ZinnWitholtEgli2004}.
 
There is colimitation on community level
even if organisms and populations remain limited by single factors.
Communities adjust their stoichiometry by competitive
exclusion and coexistence mechanisms.  This conclusion was supported by a  
  model and  experiments with  microcosms  \cite{Dangeratal2008}.
Various types of colimitation were identified and studied  and  the potential nutrient colimitation pairs in the marine environment were described in \cite{SaitoGoepfert2008}.
  
  The  multiple-resource limitation responses in plant communities was summarized using  datasets from 641 studies of nitrogen (N) and phosphorus (P) in freshwater, marine and terrestrial systems. Synergistic response to N and P addition was found in more than half of the studies \cite{Harpole2011}. 
  
The `trade-off' between law of the minimum and co-limitation includes optimization arguments and becomes paradoxical \cite{GorbanPokSmiTyu,Sperfeld2012}: if organisms in  a randomly selected  environment 
satisfy the law of the minimum (everything is
limited by the factor with the worst value) then, after
adaptation, these factors tend to be  equally important.  Conversely, if for an organism  in  a randomly selected  environment importance of factors is comparable and  factors superlinearly amplify each other then,  in  the process of adaptation, such systems evolve  towards the law of the minimum and less factors become important.
 
The main consequences of the law of the minimum for the topic of our paper are: 
\begin{itemize}
\item In well-adapted systems the most harmful  external factors are approximately equally important. The behavior of the system depends on fluctuations of many factors, and  therefore, the data cloud   is expected to be multidimensional. This conclusion applies  both to  data on populations (groups) of similar well-adapted systems and to  data on different snapshots from the same system.
\item  Conversely, when the pressure from a factor suddenly increases, then the behavior of systems mainly depends on the uncompensated pressure of this factor. The expected data cloud has a lower dimension (with a large dominance of this  factor). It can even become almost one-dimensional with accumulation of almost all variance of the relevant attributes in the first principal component.
\item The number of independent co-limiting factors gives the effective estimate of the dimension of the data cloud.
  \end{itemize}

\subsection{Generalized law of the minimum \label{SubSec:GenLieb}}

\begin{figure}[t] \centering{
\includegraphics[width=0.3 \textwidth]{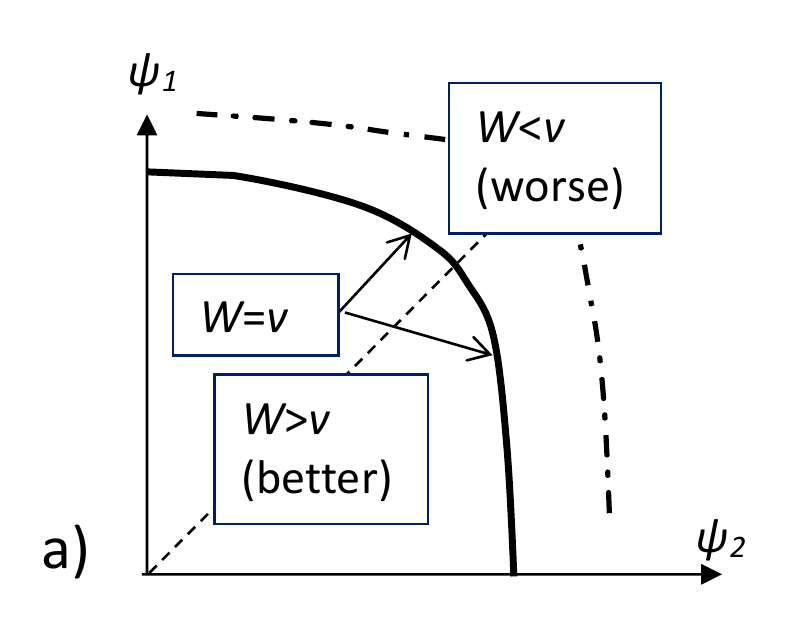}  \;\;\;
\includegraphics[width=0.3 \textwidth]{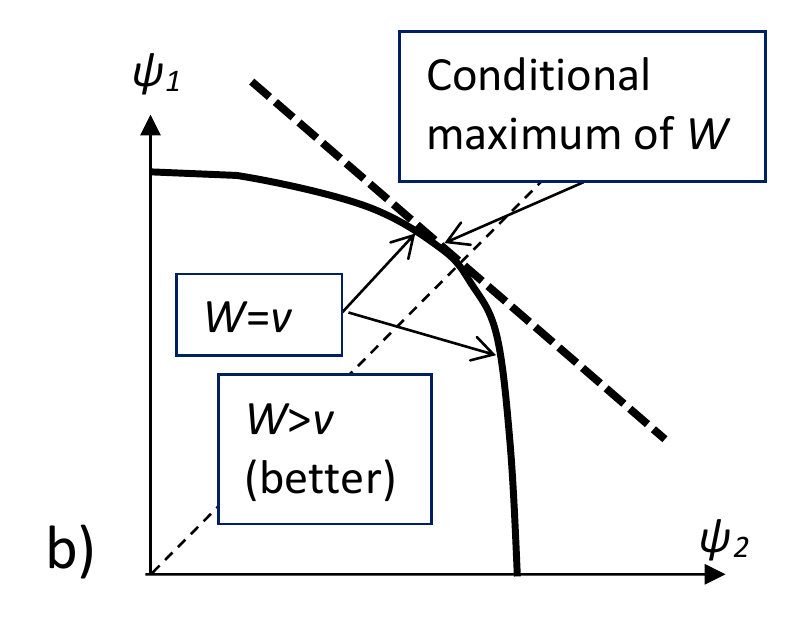}}
\caption{\label{LiebigGen} (a) The generalized law of the minimum. Coordinates $\psi_1$,
$\psi_2$ are normalized values of non-compensated factors. The fitness function $W$ is quasiconcave that means convexity of the superlevel sets ($W\geq v$).  The dot
dash line shows the border of survival.  The bold solid line
is the level set of fitness, $W=v$. It separates the states with
better conditions (higher fitness) from the states with worse
conditions. On the dashed line (the diagonal) the factors are equally important ($\psi_1=\psi_2$). (b)  Conditional maximization of $W$ with linear constrains. The dashed line  represents the plane of linear constrains $L$ with positive normal. The conditional maximum of $W$ on this plane is achieved at an extreme point of the convex superlevel set of $W$   where $L$ is a support hyperplane.}  
\end{figure}

 In fact, when the pressure of several factors is comparable, then we can expect that all of them affect the situation, and the law of the minimum has its original form asymptotically, when the pressure of one factor is much greater than the others. 
The  levels of the fitness functions (Fig.~\ref{Liebig}) should be smoothed  near the bisector of the positive octant, where the pressures of factors are close to each other (Fig.~\ref{LiebigGen}). In this area we expect colimitation. 

 Fig.~\ref{LiebigGen} illustrates  the {\em generalized law of the minimum:  when the pressure of the factors differs significantly, then fitness is determined by the worst factor, but for closer values of pressure several factors matter} because fitness is a smooth function of the pressure of the factors. This qualitative idea needs proper formalization. Notice that  in both Figs.~\ref{Liebig} and \ref{LiebigGen} the set where $W>v$ is convex. This property that the superlevel set is convex may be crucial. Compare: in thermodynamic convexity/concavity of thermodynamic potentials is a very seminal idea.
 
 We know that the superlevel sets of concave functions are concave. But concavity of functions requires more: a function defined on a convex set $U$ is convex if and only if the region below its graph is a convex set. This definition involves both the arguments of the function and its values: $W(\boldsymbol{x})$ is concave if the set of pairs $(\boldsymbol{x},v)$ is convex, where $v\leq W(\boldsymbol{x})$ and  $\boldsymbol{x} \in U$.
 
 In physics, the value of energy, entropy, free energy and other thermodynamic potentials play important role. In the principle of fitness maximization, the values of fitness are not much important, but the order of these values matters. It measures which combination of traits is `better'. All the solutions of the fitness maximization problem will remain the same after a transformation of the scale $W \to \vartheta(W)$ with monotonically increasing function $\vartheta$. Concavity of the function is not invariant with respect to this transformation, whereas its quasiconcavity persists. A function  $W(\boldsymbol{x})$ defined on a convex set $U$ is quasiconcave if and only if its superlevel sets given by inequalities   $W(\boldsymbol{x})>v$ are convex. The equivalent definition is: $W(\boldsymbol{x})$ defined on a convex set $U$ is   quasiconcave if for all $\boldsymbol{x}, \boldsymbol{y} \in U$ and $\lambda \in [0,1]$ inequality holds
$$f(\lambda x + (1 - \lambda)y)\geq\min\big\{f(x),f(y)\big\}.$$

The generalized  law of the minimum for the system of harmful factors is defined as follows: the fitness  $  W (\psi_1, \ldots, \psi_q) $ defined in the non-negative orthant ($\psi_i \geq 0$)  has three following properties:
 \begin{enumerate}
 \item Monotonicity.  If $\phi_i\geq \psi_i \geq 0$ for  all $i=1, \ldots , q$ then $W(\phi_1, \ldots , \phi_q) \leq W(\psi_1, \ldots , \psi_q)$.
 \item Quasiconcavity. The superlevel sets defined by the inequalities $ W(\psi_1, \ldots , \psi_q) \geq v$ are convex.
\item For any $i$, $\frac{\partial   W(\psi_1, \ldots , \psi_q)}{\partial \psi_i} \to 0$ when  $\psi_i  \to 0$, other values of $\psi_j $ ($j\neq i$) are constant,  $\psi_j \geq 0$ and at least one $\psi_j > 0$.
 \end{enumerate}
 An additional requirement \#3 is needed to ensure the asymptotic validity of the  law of the minimum for a situation where the pressure of factors differs significantly: the effect of the factors with small pressure vanishes.
 
Such a  function is presented in Fig.~(\ref{LiebigGen} a). The conditional maximum of $W$ on a hyperplane $L$ with positive normal is achieved at an extreme point of the convex superlevel set of $W$   where $L$ is a support hyperplane ( Fig.~(\ref{LiebigGen} b).  
We can call the system of factors that satisfy the generalized law of the minimum the {\em generalized Liebig system}. 

\begin{figure}[t]
 \centering{
\includegraphics[width=0.3\textwidth]{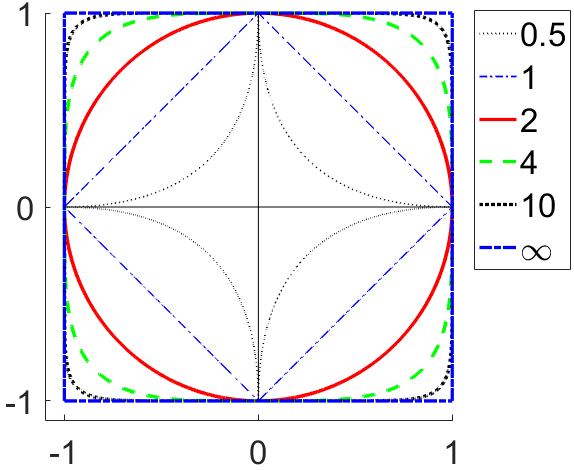}
\caption{\label{Fig:UnitCircles}Unit circles for $l_p$ norms with $p=1, 2, 4, 10,\infty$ and for  a quasinorm with $p=1/2$.}}
\end{figure}

Typical examples of generalized Liebig systems give the fitness functions defined through the distance of the vector of uncompensated  factor loads $\boldsymbol{\psi}=(\psi_1, \ldots , \psi_q)$ from the origin that is the state with fully compensated loads:
\begin{equation}\label{l_pFitness}
W(\boldsymbol{\psi})=\left\{
\begin{array}{ll}
 1-\frac{\|\boldsymbol{\psi}\|_{p}}{\psi_0}, &\mbox{ if }\|\boldsymbol{\psi}\|_{p}\leq \psi_0;\\
  0, & \mbox{ if } \|\boldsymbol{\psi}\|_{p}>\psi_0.
\end{array} \right.
\end{equation}
Here, $ \|\boldsymbol{\psi}\|_{p}$ for $p\geq 1$ is the $l_p$ norm:
\begin{equation}\label{l_p}
 \|\boldsymbol{\psi}\|_{p}=\left(\sum_i |\psi_i|^p\right)^{1/p}.
 \end{equation}
 Unit  circles of $l_p$ norms are presented in Fig.~\ref{Fig:UnitCircles}. For $0<p<1$, formula (\ref{l_p}) does not define a norm  because the `ball' becomes not convex. Such uniform functionals without the triangle inequality are called  `quasinorms'.

For $p>1$, the fitness (\ref{l_pFitness}) has the properties 1-3. For $p=1$, the property 3 is not  satisfied because in this case the partial derivatives of $W$ do not depend on (positive values of $\psi_i$. The classical Liebig law corresponds to the case $p=\infty$ because  $\|\boldsymbol{\psi}\|_{\infty}=\max_i \{|\psi_i|\}$.

 This model fitness (\ref{l_pFitness}) $W=1$  when the factor pressures are compensated, $\boldsymbol{\psi}=0$, and goes to zero when  $ \|\boldsymbol{\psi}\|_{p}$ becomes larger than a threshold $\psi_0>0$. Zero fitness means death.

\subsection{Synergistic systems of  factors \label{SubSec:Synerg}}

The  law of the minimum  means that everything is defined by the worse factor. The generalized law of the minimum gives qualitatively similar but `smoothed' picture (see Fig.~\ref{LiebigGen}). Importance of the Liebig principle and its consequences immediately leads us for the question: how the strong deviations of the Liebig principles ook like and what are their consequences for adaptation in the frame of the Selye's thermodynamic approach given by the optimality principle (\ref{Optimality}).

According to the generalized law of the minimum, fitness is a quasiconcave function and its sensitivity to minor factors vanishes. The opposite situation is also possible, when factors amplify the harm from other factors. 

The system of harmful  factors is called  {\em synergistic} if the fitness  $  W (\psi_1, \ldots, \psi_q) $ defined in the non-negative orthant ($\psi_i \geq 0$)  has two following properties:
 \begin{enumerate}
 \item Monotonicity.  If $\phi_i\geq \psi_i \geq 0$ for  all $i=1, \ldots , q$ then $W(\phi_1, \ldots , \phi_q) \leq W(\psi_1, \ldots , \psi_q)$.
 \item Quasiconvexity. The sublevel sets defined by the inequalities $ W(\psi_1, \ldots , \psi_q) \leq v$ are convex.
 \end{enumerate}
   
\begin{figure}[t]
 \centering{
\includegraphics[width=0.3 \textwidth]{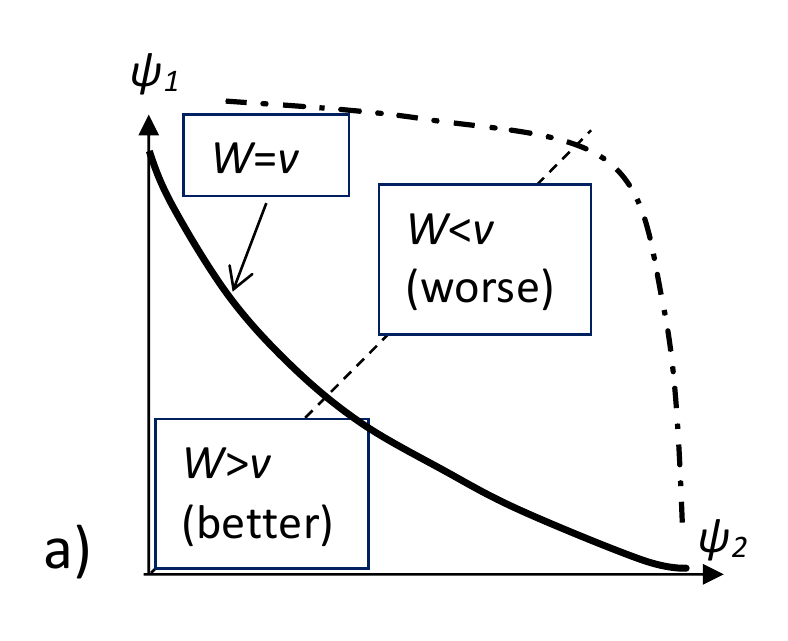}  \;\;\;
\includegraphics[width=0.3 \textwidth]{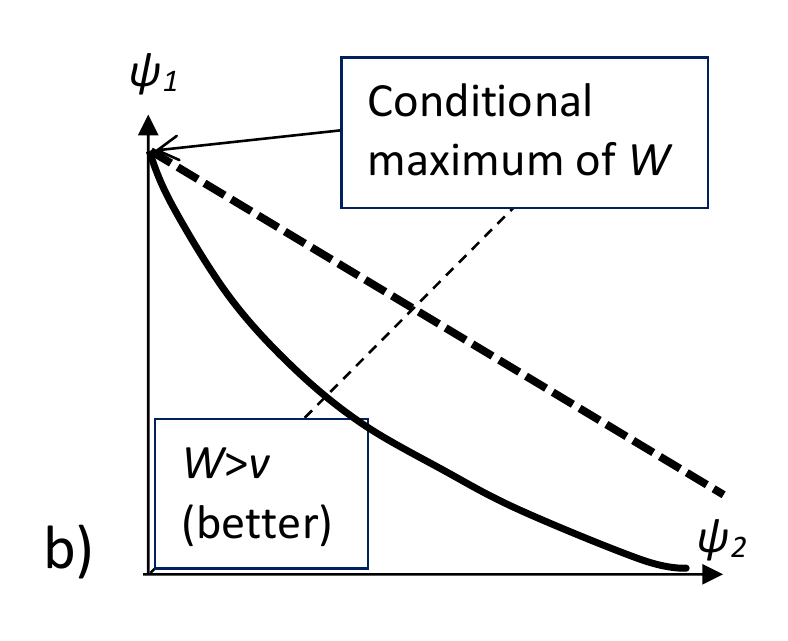}}
\caption{\label{Synergistic} (a) Synergistic system of harmful factors. Coordinates $\psi_1$,
$\psi_2$ are normalized values of non-compensated factors. The fitness function $W$ is quasiconvex that means convexity of the sublevel sets ($W\leq v$).  The dot
dash line shows the border of survival.  The bold solid line
is the level set of fitness, $W=v$. It separates the states with
better conditions (higher fitness) from the states with worse
conditions.  On the dashed line (the diagonal) the factors are equally important ($\psi_1=\psi_2$). (b)  Conditional maximization of $W$  with linear constrains. The dashed line  represents the plane of linear constrains $L$ with positive normal. The conditional maximum of $W$ on this plane in the positive orthant is achieved at the boundary of the  orthant  where some  factors   are completely compensated .}  
\end{figure}

An illustrative example of synergistic system of factors is presented in Fig.~\ref{Synergistic}. 
   
Formula (\ref{l_pFitness}) presents a synergistic system when $0<p<1$. The case $p=1$ is the bordercase: both sublevel and superlevel sets of $W$ are convex, and sensitivity to all factors is the same constant $1/\psi_0$ (when $W>0$). For $0<p<1$, the sensitivity $\frac{\partial W}{\partial \psi_i}$ is larger for smaller $\psi_i$ and increases to infinity when  $\psi_i \to 0$ (and when other $p_i$ are constant). Such a high sensitivity to small concentrations of some chemicals in combination with other chemicals has been widely discussed. 

For a synergistic system, adaptation tends to compensate some factors completely, and the dimensionality of the data cloud is expected to decrease with adaptation. Let us demonstrate this on the basic example -- the fitness   (\ref{l_pFitness}) with  $0<p<1$. A simple calculation shows that for a given vector of factor loads, $\boldsymbol{f}=(f_1,\ldots, f_q)$ and the available total resource $R$ the optimal distribution of the resource can be constructed by simple iterations. 
Let (after reordering) $f_1> \ldots > f_q$. 
\begin{enumerate}
\item If  $R< f_q$ then compensate   $f_q$ as much as it is possible: set $r_q=R$, $\psi_q=f_q-R$ and stop.
\item If $R\geq f_q$ then compensate $f_q$  completely, set $r_q=f_q$, $\psi_q=0$. 
\item Set $R:=R-f_q$  and iterate with $q-1$ factors $f_1> \ldots > f_{q-1}$.
\end{enumerate}

Optimization for general synergistic systems was analyzed in \cite{GorbanPokSmiTyu}. There is significant difference between adaptation to the  synergistic system of factors and   to the generalized Liebig system: in synergistic systems resource is distributed for complete neutralization of small factor pressures with  the highest sensitivity,  and  the dimensionality of data tends to decrease, while for a generalized Liebig system, adaptation tends to reduce the strongest differences between factor pressures and the dimensionality of data tends to increase in the course of adaptation.

Optimization principle is a basis of the analysis of trade-offs  between various functional traits under pressure of various environmental factors. The effect of shapes of the objective functions was clearly demonstrated by Tilman for ecological problems \cite{Tilman1980,Tilman1982}.  A general approach to evaluate the effect of different trade-off shapes on species coexistence based on the modeling of invasion experiments is presented in \cite{Ehrlich2017}.  Application of the general optimization principle (\ref{Optimality0}), (\ref{Optimality}) to physiological adaptation was proposed in
\cite{GorSmiCorAd1st}, and developed further in \cite{Sedov,GorbanSmiTyu2010,GorbanPokSmiTyu,GorbanTyukinaDeath2016}.

Qualitatively, the main difference is usually considered between convex and concave functions (or `trade-off curves' when analyzing pairwise relations). It is obvious that the concavity and convexity can be substituted by the quasiconcavity and quasiconvexity without the change of the results. These properties are invariant with respect to monotonically increasing  transformations of the scale of fitness. There are many other, combining versions of the trade-offs. For example, there can be groups of synergistic  factors with `Liebig's' relations between these groups.

\subsection{Alternative theories of low dimension and high correlations under stress \label{SubSec:AlternTheor}}

\subsubsection{Bifurcations, slow manifolds, critical retardations and critical fluctuations \label{subsec:Bifur}} 

\begin{figure}[t]
\centering
\includegraphics[width=0.7\textwidth]{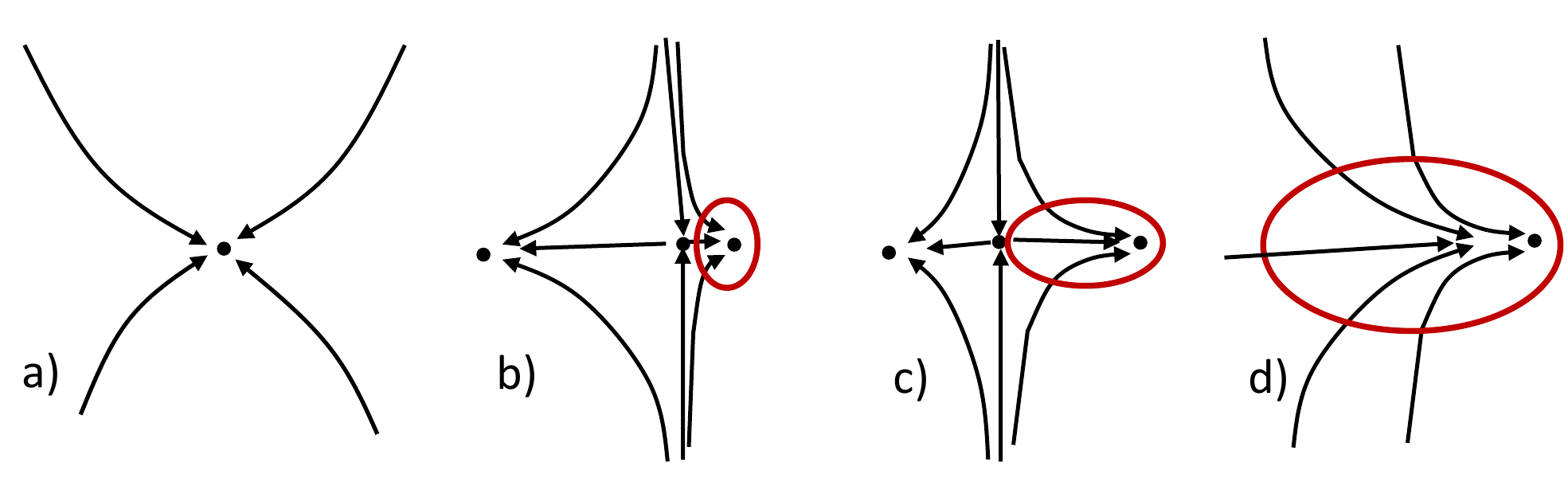}
\caption {Schematic representation of the bifurcation model: a) the globally stable healthy state; b) after the first bifurcation -- the disease state appeared (encircled by red); c) disease expansion -- attraction area of the healthy state decreases; d) after the second bifurcation -- the healthy state disappeared. }
\label{Fig:bifurMod}
\end{figure}

Some alternative theories suggest that observed increase in correlations is the result of bifurcation in the dynamical system of physiological regulation \cite{Chenetal2012}).
 This dynamical system is hidden behind the observed dynamics of physiological parameters. Before this bifurcation, a regime of normal functioning is stable in wide area, that is stable homeostasis. The hypothetical bifurcation leads to appearance of a new attractor, a locally stable regime of illness. As the simplest version, we can consider appearance new attractor as a saddle-node bifurcation \cite{Kuznetsov2004} (see Fig.~\ref{Fig:bifurMod}). With development of disease the basin of attraction of the normal homeostasis contracts and, finally, vanishes in the second saddle-node bifurcation. This evolution of phase portrait is illustrated in Fig.~\ref{Fig:bifurMod},  from  Fig.~\ref{Fig:bifurMod}a to Fig.~\ref{Fig:bifurMod}d.

 The increase of correlations near bifurcations means that there appear ``slow modes'' or, the dynamics (after some initial layer) approaches slow manifold (in the simplest cases this manifold is a curve), and  the systems near bifurcation are distributed along this low-dimensional manifold. The state space of hidden dynamics is not fully observed, but an increase in correlations that follows a decrease in effective dimensions near bifurcations can be observed in almost any set of relevant variables. 
 
\begin{figure}[t]
\centering
\includegraphics[width=0.9\textwidth]{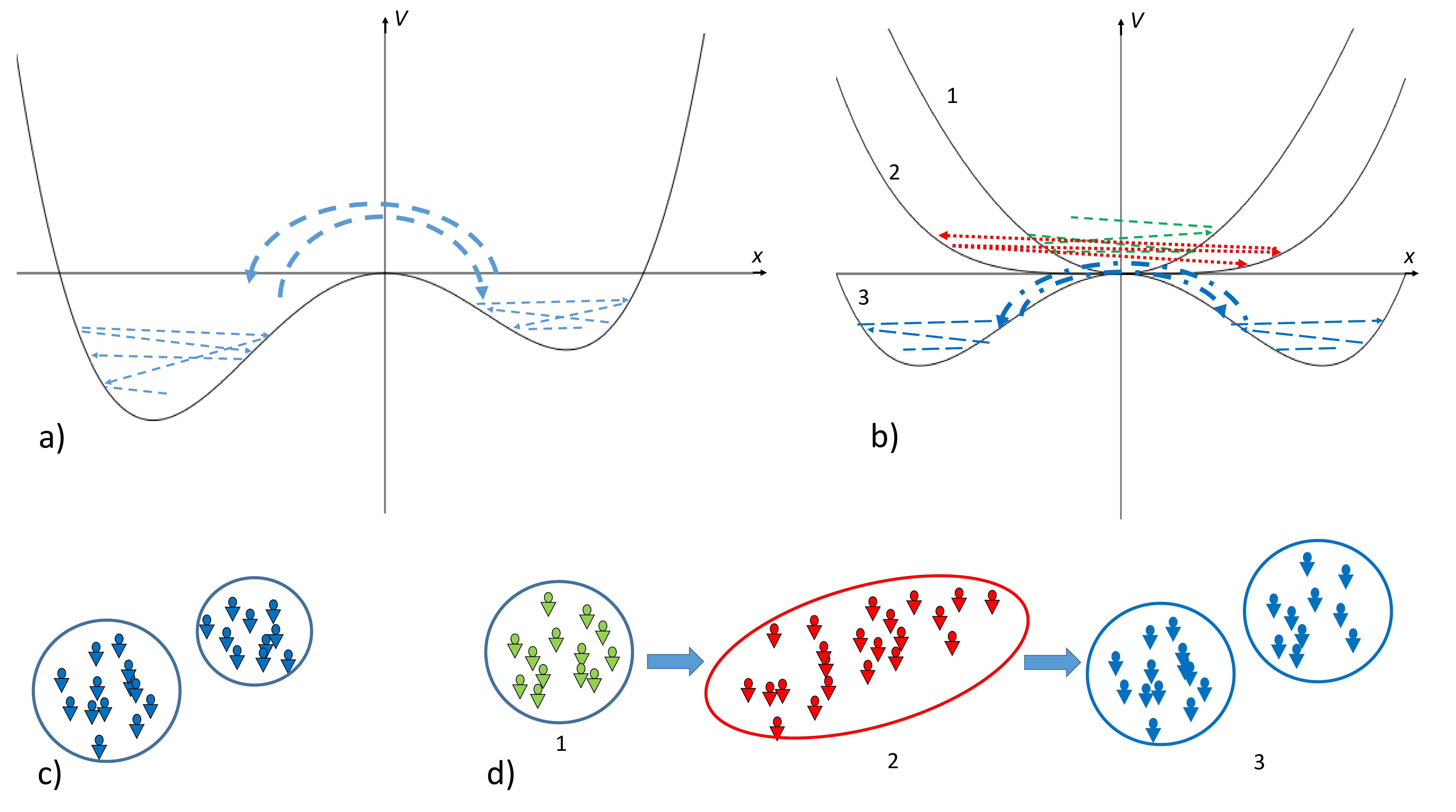}
\caption {Fluctuations in the bifurcation model. a) The potential wells for two stable and one unstable states. The thin dashed lines symbolize fluctuations in the potential wells near the stable states. The bold dashed lines illustrate jumps between potential wells due to  large fluctuations. b)~Disintegration of a stable equilibrium into three steady states: two stable and one unstable. Configurations of potential wells for three stages are presented: (1) one   stable state with non-degenerated quadratic term of the potential, (2) degeneration of the quadratic term, and (3) two potential wells for two symmetric stable steady states. Fluctuations are presented by dashed, dotted and  dash-dotted lines. c) Expected distribution of population in the potential wells of two steady states from the panel a); d) Evolution of the distribution of population for the evolution of the potential wells presented in panel b).\label{Fig:bifurModFluc}}
\end{figure}

 The approximately slow manifold in the illustration (Fig.~\ref{Fig:bifurMod}) is represented by the trajectories that lead from the unstable steady state (a saddle) to the stable nodes.
After some initial period, the motion approaches this slow manifold (a curve). One-dimensional  motion along this curve can always be represented as a gradient system in a potential well with `energy'  $V(x)$, where $x$ is the coordinate along the curve: $$\frac{dx}{dt}=-\frac{dV}{dx}.$$ This representation is illustrated in Fig.~\ref{Fig:bifurModFluc}a.

Various fluctuations in the environment are  imminent life conditions. There are two types of fluctuations in  Fig.~\ref{Fig:bifurModFluc}a: jiggling around the equilibria and large jumps between different basins of attraction. Such jiggling and jumps lead to a clustered stationary distribution (Fig.~\ref{Fig:bifurModFluc}c).
 
Increase of variance is a particular case of the famous effect of increase of fluctuations near bifurcations and phase transitions \cite{Dykman1980}. The relaxation to the attractor may have anomalously long critical delays \cite{GorbanDiss,GorbanDAN,GorbanSingularities} and, therefore, the random change of environment can add large fluctuations.   A brief digest of these works was published recently \cite{GorbanComment2020}.

The essence of the approach based on bifurcations, critical slowing down and critical fluctuations, can be formulated in the following way: {\em nothing special, just the dynamics}. The qualitative picture behind this theory is quite simple. Organism is a large-dimensional dynamical system that depends on many parameters. Parameters change in time due to many reasons (environment, illness, aging, etc.). Attributes are functions on the phase space. When the system approaches bifurcations, the slow modes appear,  the effective dimensionality decreases and the variance increases. It could be worth to mention that this point of view on the bifurcations and slow modes was developed first by L. Landau in his work about turbulence \cite{Landau1944}. The logic was very similar: the system is very multidimensional, and we do not know the details, but when the laminar flow loses stability and a new attracting (periodic) regime is born, then a slow mode appears, and we can study its dynamics separately, without consideration of the entire system. This slow mode can be defined as a measure of deviation from the laminar flow. 

The contradictions between this theory and the AE approach are not obvious: both consider low-dimensional dynamics under stress as a mechanism for increasing correlation. Near bifurcations, the AE can be considered as the coordinate along the slow manifold. Nevertheless, there exist some differences in details because the AE approach depends on the organization of the system of factors and gives different results for the Liebig and the anti-Liebig (synergistic) systems of factors.  The interpretation of the approach to death is also   unclear in the hidden bifurcation methods, whereas interpretation of death through exhaustion of the adaptation resource was introduced by Selye and supported by his experiments.  

A verifiable difference is in the details of the correlation increase. If after bifurcation two attractors appear at non-zero distance (see Fig.~\ref{Fig:bifurMod} and Fig.~\ref{Fig:bifurModFluc}a),  then the cloud of data is expected to have a clustered structure (Fig.~\ref{Fig:bifurModFluc}c). The apparent increase in correlations in this case just reflects the difference between attributes in clusters.  For systems with strong limitation such clusters are not expected (Fig.~\ref{Fig:LimFluct}a).

One important comment should be added for multidimensional systems: bifurcation has a significant effect only on a part of the variables. This is the relevant dominant group, and one of the first goals of the researcher is to identify these variables   \cite{Chenetal2012}.

Revealing of the relevant dominant group of attributes is important in a much broader context. When studying  any phenomena in a multidimensional system, we must identify the variables that are directly involved in this process. These variables can form a special group  of already known variables or must be constructed from them.  In physics of phase transitions, these are the classical `order parameters'. In the theory of bifurcations, the projection of a system on the relevant normal form selects the analog  of the order parameter (see, for example, \cite{Kuznetsov2004}). Selection of the relevant dominant groups is always needed in the adaptation research. Sometimes, it is performed explicitly, with relation to the specific theory about the nature of the dimensionality reduction (self-simplification or degeneration)   \cite{Chenetal2012} or, often, without specification of such a theory (for example, \cite{Censi2010,Censi2011,Censi2018,Giulianni2014}). Most often, the selection of relevant attributes is performed implicitly as a prerequisite for research. For example, the choice of lipid fractions for a comparative analysis of metabolism in healthy newborns in the temperate zone of Siberia and a Far North city was not accidental \cite{GorSmiCorAd1st} (see Fig.~\ref{Fig:lipid1987}).  At the beginning of this study, a biologically plausible hypothesis about the relevance of these variables  was accepted.
 
Sometimes the assumption about bifurcations (tipping points) can be supported by the data \cite{Chenetal2012}, but often, identifying hidden bifurcations is a non-trivial task, and they are postulated rather than demonstrated. In that sense, the hidden AE is substituted by another assumption,   hypothetical bifurcations.
However, explaining Selye's experiments \cite{SelyeAEN,SelyeAE1} in terms of bifurcation remains an open challenge. 
 
 It cannot be ruled out that the dominant path of the organism's dynamics, parametrized by the adaptive energy, gives  one more description of the hidden bifurcations, but such a theory still needs to be developed.

\subsubsection{Adaptation dynamics near a border of surviving \label{sec:adv}}

\begin{figure}[t]
\centering{
\includegraphics[width=0.7\textwidth]{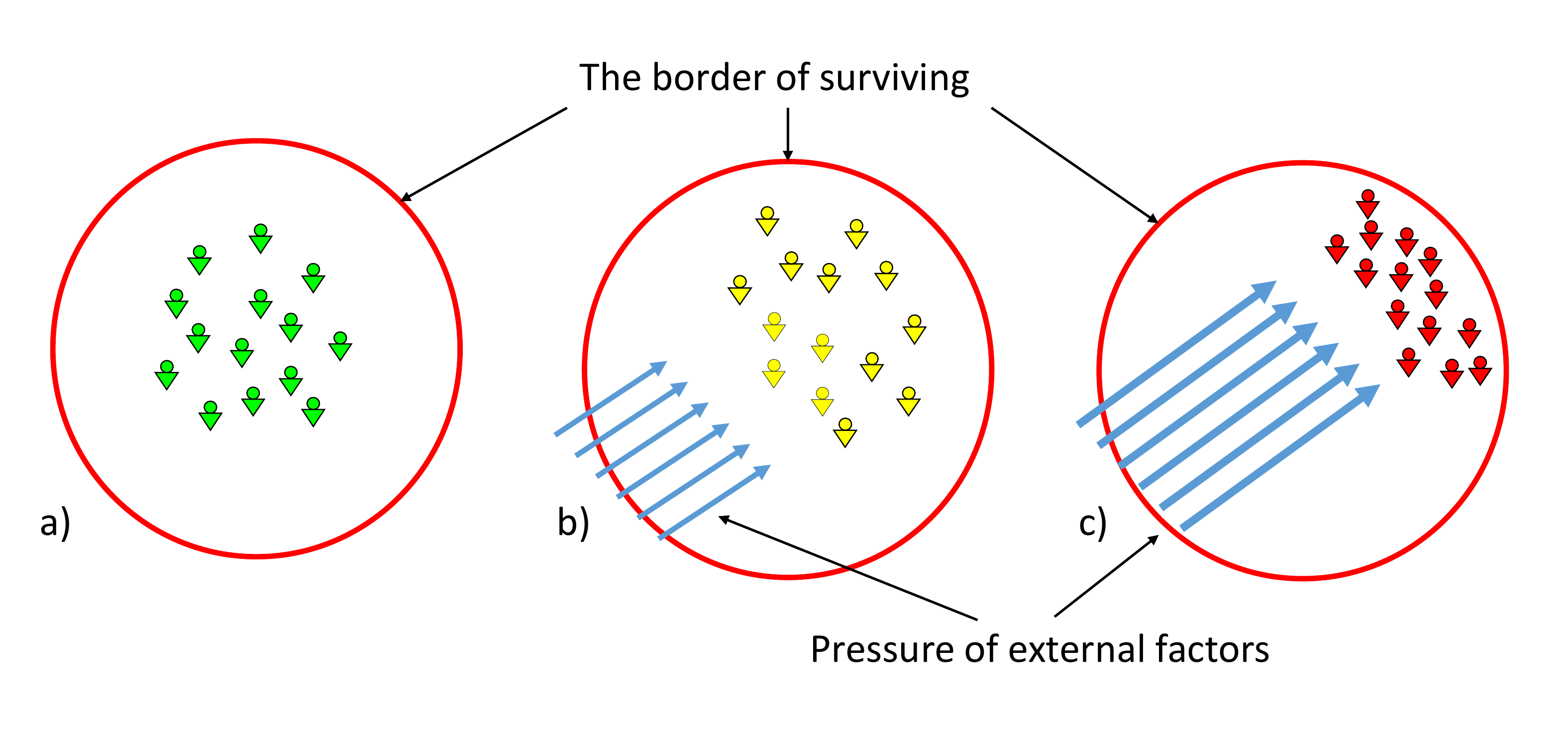}
}\caption{Under the pressure of external factors the population approached the border of existence and becomes effectively a cloud concentrated near a hyper-surface (one dimension less)
 \label{ModelFactorPressure}}
\end{figure}

We used the idea of adaptation resource to demonstrate how can the
models of adaptation explain the observable dynamics of correlation
and variance under environmental loads. Another theoretical approach
to explain these effects was proposed in
 \cite{RazzhevaikinZVMF} and developed further with applications in \cite{Razzhevaikin2008,Shpitonkov2017}. The idea is quite simple. Each organism is presented by a point in the $n$-dimensional space $\mathbb{R}^q$ of physiological attributes.  The `area of existence' is a bounded set  $U \subset \mathbb{R}^q$ with smooth boundary. The population is a cloud of points in $U$. At this scale, individual differences in properties and life conditions can be considered as random. The individual history can be represented by a random walk inside $U$. At the population level, this movement can be modeled as diffusion. In the absence of the pressure strong common factors diffusion distributes population in area of existence  (Fig.~\ref{ModelFactorPressure}a). 
 
The pressure of a common factor that acts similarly on all the organisms in the population can be modeled as an external `wind' (advection) (Fig.~\ref{ModelFactorPressure}b). It presses the population to the border of $U$ as it is represented in  (Fig.~\ref{ModelFactorPressure}b,c). if intensity of this external pressure increases then the population concentrates in close vicinity of the boundary   (Fig.~\ref{ModelFactorPressure}c). It looks as a disk near the border. The detailed analysis shows that for large intensity of the factor, $f$,   the thickness of this disk goes to zero as $1/f$, whereas its diameter decays slower, as $1/\sqrt{f}$ and it becomes essentially $n-1$-dimensional cloud. For formal detail we refer to \ref{Advection}.
 
 To check whether this mechanism is close  to reality we have to resolve the
``disk--or--bullet" alternative: How does the correlation increase:
(i) the largest eigenvalues of the correlation matrix increase and
the ellipsoid of concentration becomes stretched-out in small amount
of dimensions (presumably, in one, that is the bullet form) or (ii)
the smallest eigenvalue goes fast to zero, and the ellipsoid of
concentration becomes flattened out (the disk or the pancake form).
The existing data ensure us that this alternative should be resolved
in favor of bullet: under the pressure of environment factors, the
small amount of largest eigenvalues increases.It should be noted that the variance in this model decreases with increasing pressure of the factor. That also contradicts the majority of existing observation.

 \section{Selye--Goldstone kinetics: dynamics of individual adaptation \label{Sec:SelyeGoldKin}}
 
 \subsection{The type of model we need \label{SubSec:ModelRequir}}
 
Quasi-static modeling of adaptation as a chain of optimal allocation of resources to neutralize the pressure of harmful factors is simple, allows for a detailed analysis of models and provides a tool for analyzing adaptation experiments that are organized in the form of such a chain of adaptation  challenges and responses to them. Nevertheless, many problems remain open. For example: How long lasts the protection from a harmful factor provided by the allocated amount of AE? How much AE will be available for distribution at the next time step? The answers are impossible in a such reduced formalism because it does not include time. A system of dynamic models is needed. It is a shift from a view inspired by classical thermodynamics to the kinetics of production and expenditure of AE.
 
 The status of these models should be clearly represented. There are 
  two basic approaches to modeling, {\em bottom-up} and {\em top-down}, and the emerging {\em middle-out} approach. The bottom-up  approach starts from a detailed picture and many additional experimental  studies are usually required for identifying the details of the elementary events   \cite{Galle2009}.  The top-down approach to modeling  is closer to classical physiology  \cite{Hester2011}. It starts from very general integrative properties and then adds some details.  To combine the advantages of the bottom-up and the top-down approaches, the {\em     middle-out} approach was used by many authors and analyzed specially in several works  \cite{Brenner1998,Kohl2010}. With this approach, modeling  starts neither from the top level, nor from the bottom,    but from the level which is ready for formalization, where one can plausibly guess the main mechanism.  
    
We use and elaborate further in this section  Selye's empirical insights about the forms of AE, follow the  middle-out approach to modeling, and exploit the kinetic formalisms of mathematical chemistry and ecology.
 
\subsection{Kinetic factor--resource models \label{SubSec:KinFactResour}}
 
In most experiments and in many clinical situations, a dominant harmful factor can be identified. In particular, in Selye's experiments \cite{SelyeAEN,SelyeAE1}, rats were exposed to one noxious agent each time (but can be exposed to different agents sequentially). The dynamic model  presented in this Subsection  describes  adaptation to one factor with intensity $f$. 

\begin{figure}[t]
\centering{
\includegraphics[width=0.95\textwidth]{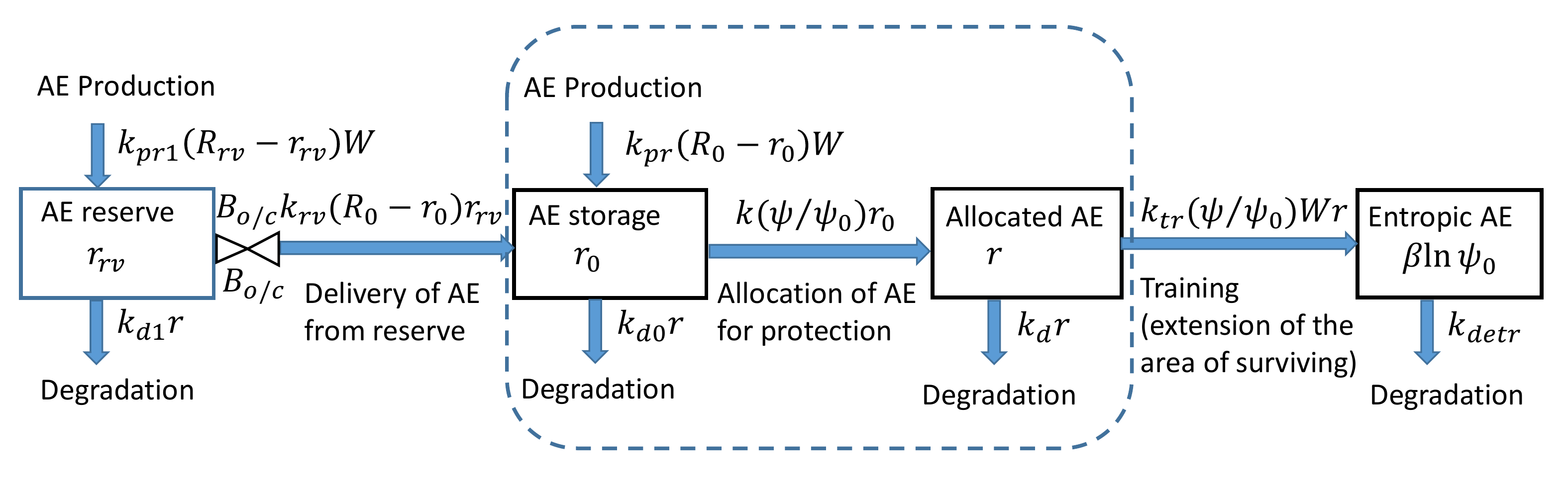}
}\caption{The scheme of kinetic model of AE production and delivery. Four quantities describe the distribution of AE ``main storage'' (resource) $r_0$, resource $r$, allocated to neutralization of the harmful  factor, ``reserve'' $r_{rv}$ and the ``entropic form of AE' accumulated in the  training results, $\beta \ln \psi_0$. AE flows are represented by arrows with the flow rate written above the arrows.  The state of the tap between the reserve and the main storage is described by the Boolean variable $B_{o/c}$. \label{Fig:EqScheme}}
\end{figure}

The scheme of the model (Fig.~\ref{Fig:EqScheme}) completely represents the system of equations, if we define all the quantities, mention that $W(\psi/\psi_0)=1-\psi/\psi_0$ (if $0\leq \psi/\psi_0 \leq 1$),  and formulate the rules of opening and closing the tap between the reserve AE storage ({\em deep AE } in Selye's terminology) and the main AE storage ({\em{superficial AE}}). Goldstone's comments about AE productions for realistic models of adaptation \cite{GP_AE1952,GP_AE1952C} are taken into account in this model (see Fig.~\ref{schemeGold}). The resource and reserve storages have limited capacity ($R_0$ and $R_{rv}$, correspondingly) and the production flux is proportional to the  product of fitness $W$ (the measure of wellbeing) by the  deficit of AE: $R_0-r_0$ for the main storage, and $R_{rv}-r_{rv}$ for the reserve. 

\begin{figure}[t]
\centering{
 \includegraphics[height=0.22 \textwidth]{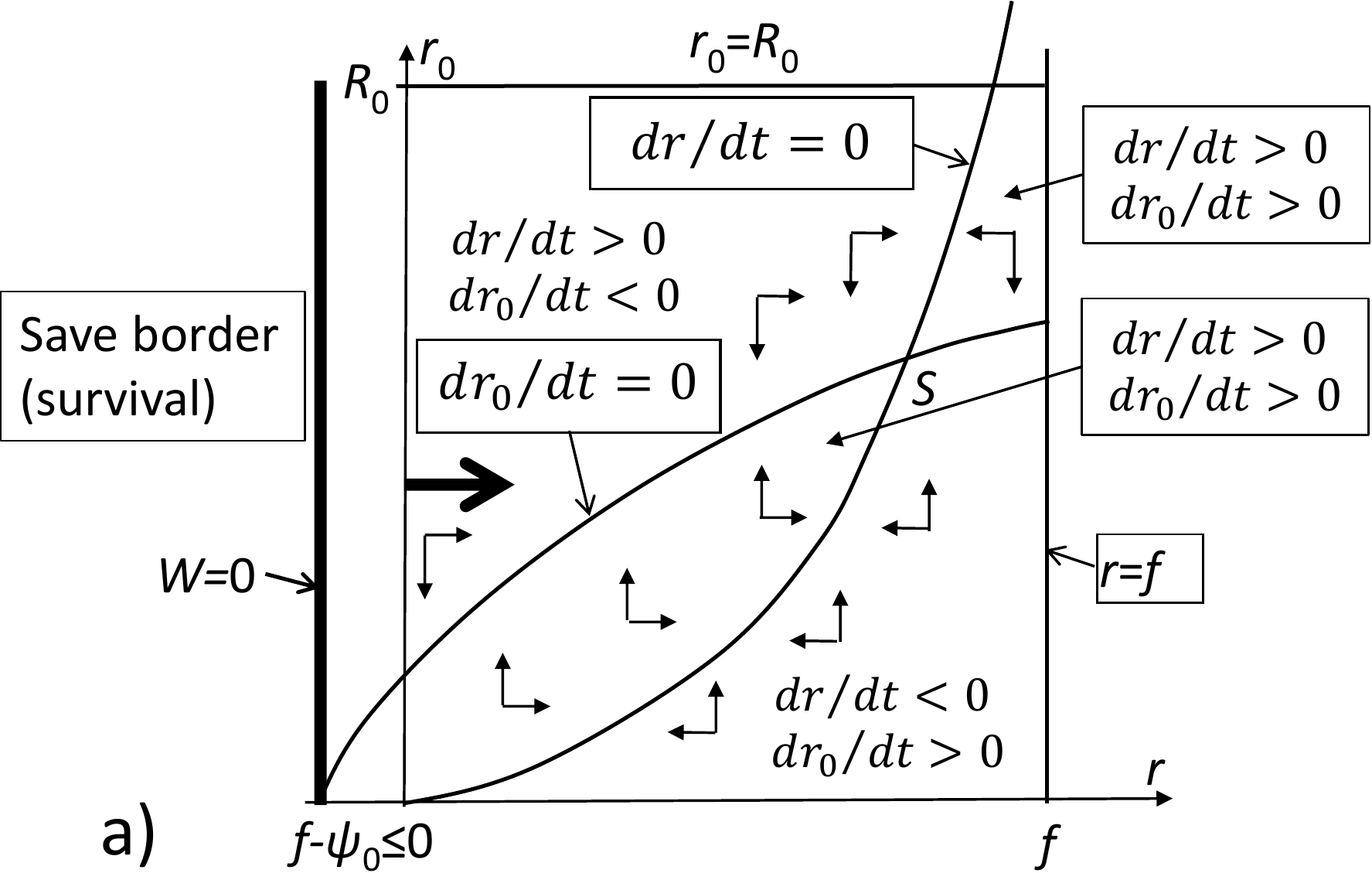}
 \includegraphics[height=0.22\textwidth]{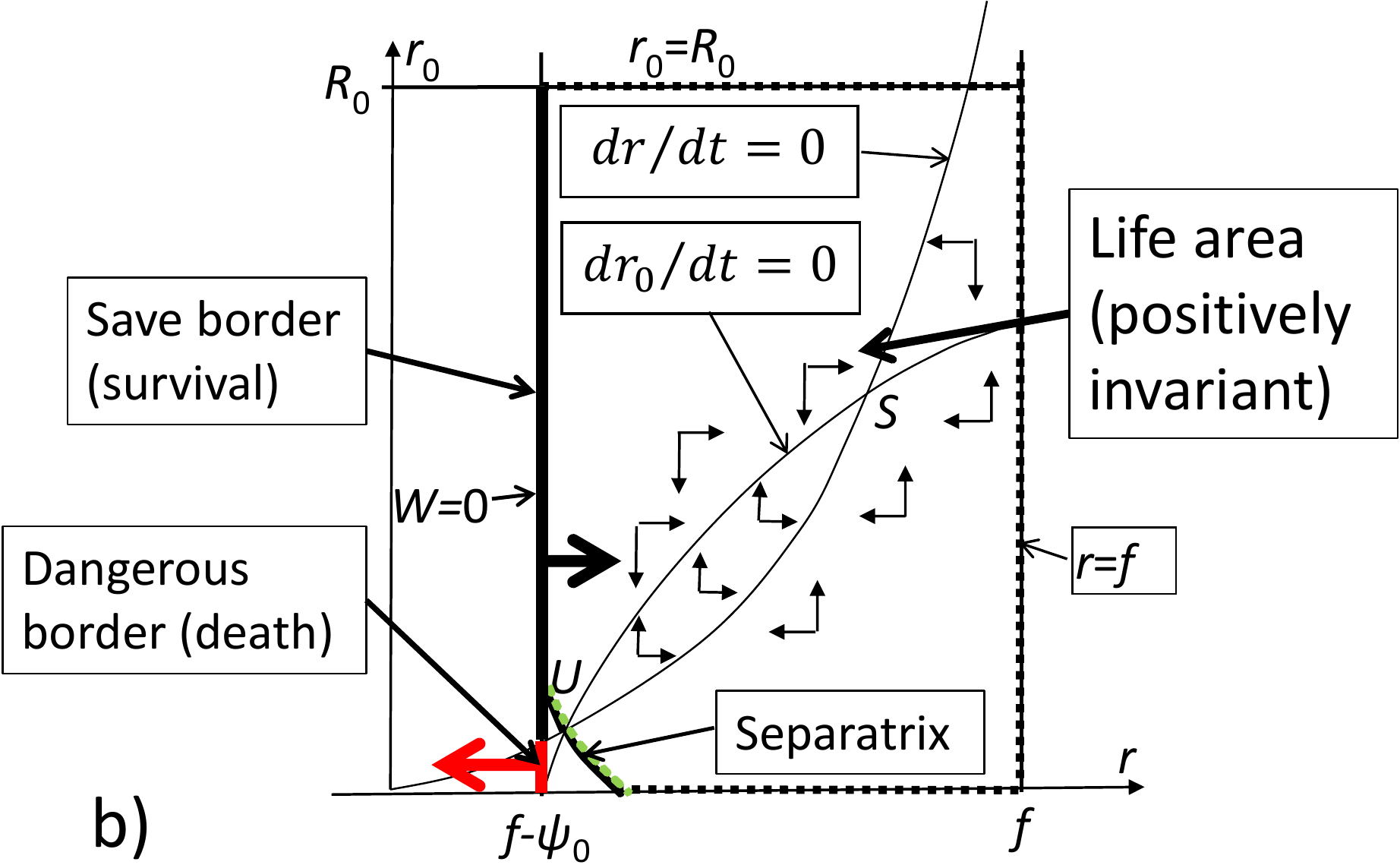}  
 \includegraphics[height=0.22\textwidth]{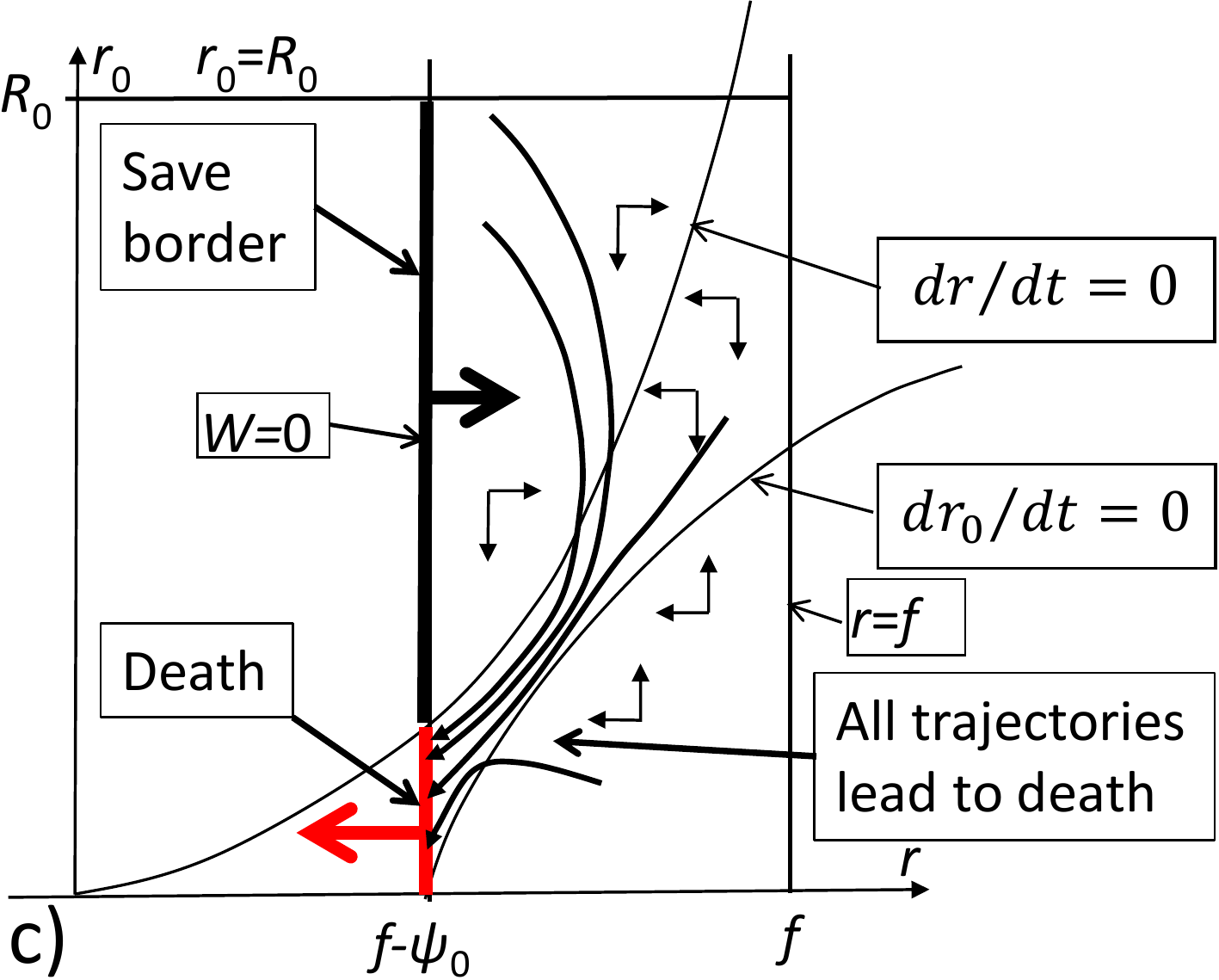}
}\caption{Safe and dangerous borders for dynamics of adaptation in the simplest model. The $r$-nullcline cuts the vertical line  $W=0$ (the border of death) into two parts: $\dot{W}<0$ (dangerous border, red) and $\dot{W}>0$ (safe border, black).  (a) If $f<\psi_0$ then the whole border is safe. The nullclines have unique intersection point $S$,  the globally stable node.  (b) If the $r$- and $r_0$-nullclines have two
intersections, the stable ($S$) and unstable ($U$) equilibria, then the separatrix of
the unstable equilibrium $U$ separates the attraction area of the dangerous border of death
from the area of attraction of the stable node (life area). (d) If there exists no intersection of the nullclines then all the trajectories are attracting to the dangerous border. \label{Border}}
\end{figure}

 Historically, the first set of ideas was formalized in a model with two variables: the stored AE $ r_0 $ and the amount of AE $ r $  allocated to neutralize the harmful factor.  In Fig.~\ref{Fig:EqScheme}, this part of the diagram is outlined with a dashed line. Recall that $\psi=f-r$ is the uncompensated pressure of the factor. $R_0$ is the maximal amount of AE in the storage. The production flux is proportional to the  product of fitness $W$ (the measure of wellbeing) by the  deficit of AE in the storage, $R_0-r_0$. The allocation flux is proportional to the normalized value of the uncompensated factor pressure, $\psi/\psi_0$, and the available AE. Kinetic constants are denoted by $k_{\ldots}$.  The phase portrait of this model is completely determined by the position of the two lines, where the time derivative of $ r_0 $ or $ r $ vanishes (nullclines)  (see Fig.~\ref{Border} where for certainty we assume that $k_{pr}>k$, which guarantees the concavity of the $r_0$ nullcline; the $r$ nullcline is always convex). 
 
 The vertical line $W=0$ (that is, $f-r=\psi_0$) is the border of death. If $f< \psi_0$ then the $r$-nullcline does not intersect the border of death, there exists a globally stable equilibrium and the factor pressure cannot lead to death  (Fig.~\ref{Border}a). If  $f> \psi_0$ then a dangerous (absorbing) fragment appears on the border of death (Fig.~\ref{Border}b,c, in red). 
 Up to a certain threshold of the factor pressure, the nullclines of $ r $ and $ r_0 $ intersect, and there are two stationary states: a (locally) stable node and a saddle point, the separatrix of which is the boundary between the basins of attraction of a stable node and of the dangerous (absorbing) part of the death border (Fig.~\ref{Border}b). Further increase of the factor pressure $f$ leads to the second critical effect: intersection between nullclines vanishes and the absorbing part of the border of death becomes globally attractive (Fig.~\ref{Border}c).

\begin{figure}[th]
\centering{
\includegraphics[width=0.3\textwidth]{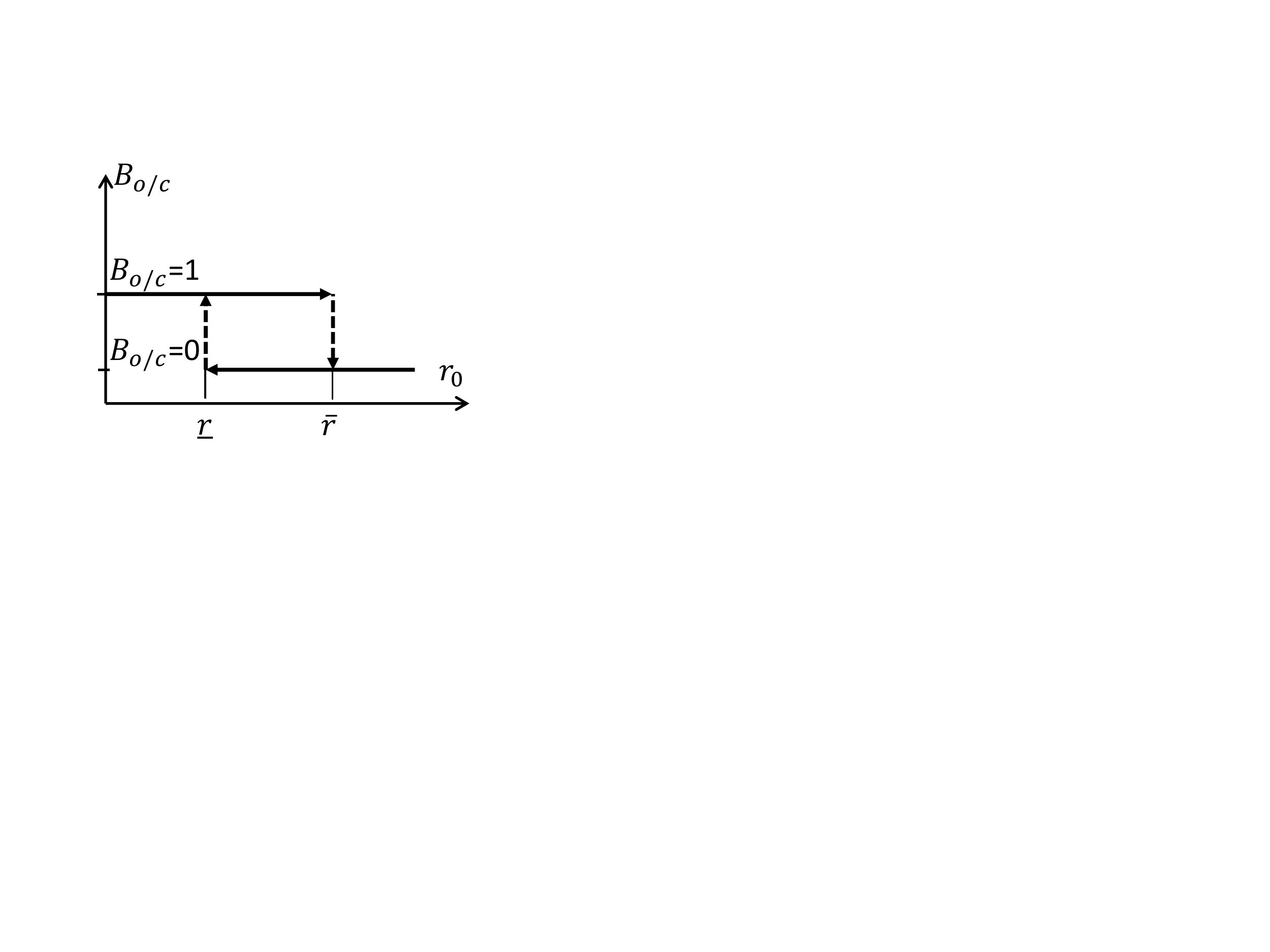}
}\caption{Resource -- reserve hysteresis. Hysteresis of reserve supply: if $B_{o/c}=0$ then
reserve is closed and if $B_{o/c}=0$  then reserve is open.  When   $r_0$ decreases and approaches
$\underline{r}$ then the supply or reserve opens (if it was closed). When
$r_0<\overline{r}$ increases and approaches $\overline{r}$ then the supply of reserve
closes (if it was open).  \label{Fig:Hysteresis}}
\end{figure}

Many physiologists and clinicians note the fact that when the adaptation load increases and approaches a certain threshold, a jump in adaptive abilities occurs. Selye and his successors fond that just the idea of AE supply cannot explain many such effects. Selye introduced the notion of ``deep AE'' (deep here means not another kind of AE, but rather its deeper storage) and used for the routinely accessible AE the term ``superficial AE''. The fourth   axiom of AE reads (accordingly to \cite{SchkadeOccAdAE2003}):
\begin{itemize}
\item AE is active at two     levels of awareness: a primary level at which creating the response occurs at a high
    awareness level, with high usage of finite supply of adaptation energy; and a
    secondary level at which the response creation is being processing at a sub-awareness
    level, with a lower energy expenditure.   
\end{itemize}
This  deep AE level is represented in the model diagram  (Fig.~\ref{Fig:EqScheme}) by the AE reserve (left). There are two storages of AE: resource (which is always available if it is not empty) and reserve (which becomes available when the resource becomes too low). The Boolean variable $B_{o/c}$ describes the state of the `tap' between the reserve and the main storage: if $B_{o/c}=0$ then the tap is closed and if $B_{o/c}=1$ then the tap is open. There are two switch lines on the phase plane $(r,r_0)$ (Fig.~\ref{Fig:Hysteresis}): $r_0=\underline{r}$ (the lower switch line that serves to opening the reserve storage) and $r_0=\overline{r}$ (the upper switch line that serves to closing the reserve storage). 

When  the available resource $r_0$ decreases and approaches $\underline{r}$ from above then the supply or reserve opens (if it was
closed). When the available resource $r_0<\overline{r}$ increases and approaches
$\overline{r}$ from below then the supply of reserve closes (if it was open). For
$r_0<\underline{r}$ the reserve is always open, $B_{o/c}=1$ and for $r_0>\overline{r}$
the reserve is always closed, $B_{o/c}=0$.  

\begin{figure}[t]
\centering{
a)\includegraphics[width=0.3\textwidth]{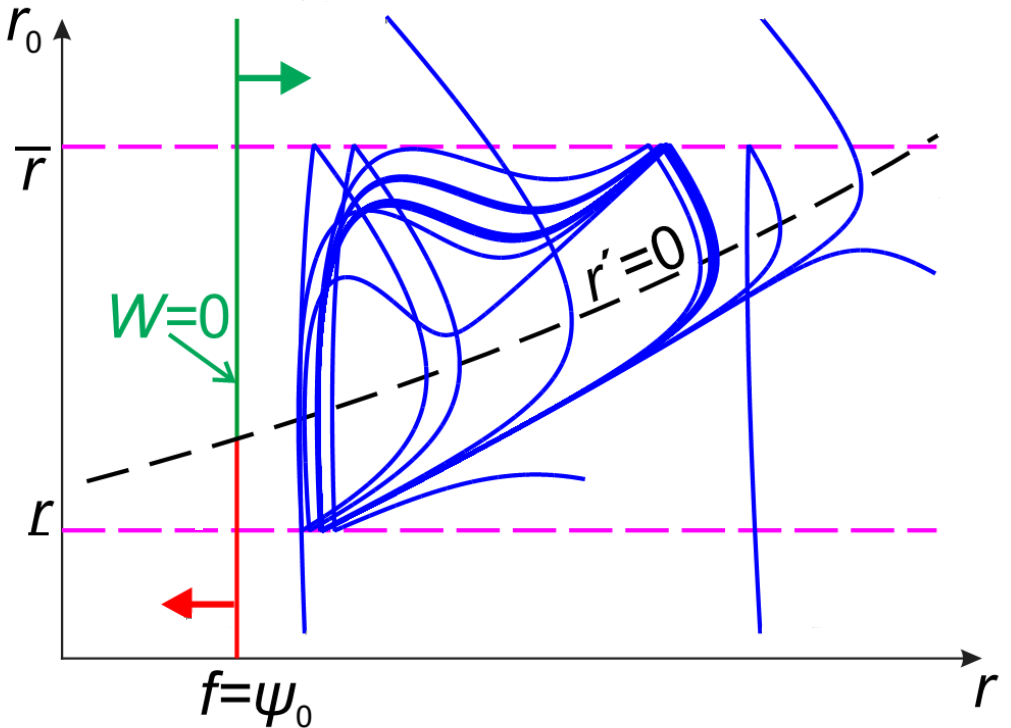}\;\;\;
b)\includegraphics[width=0.3\textwidth]{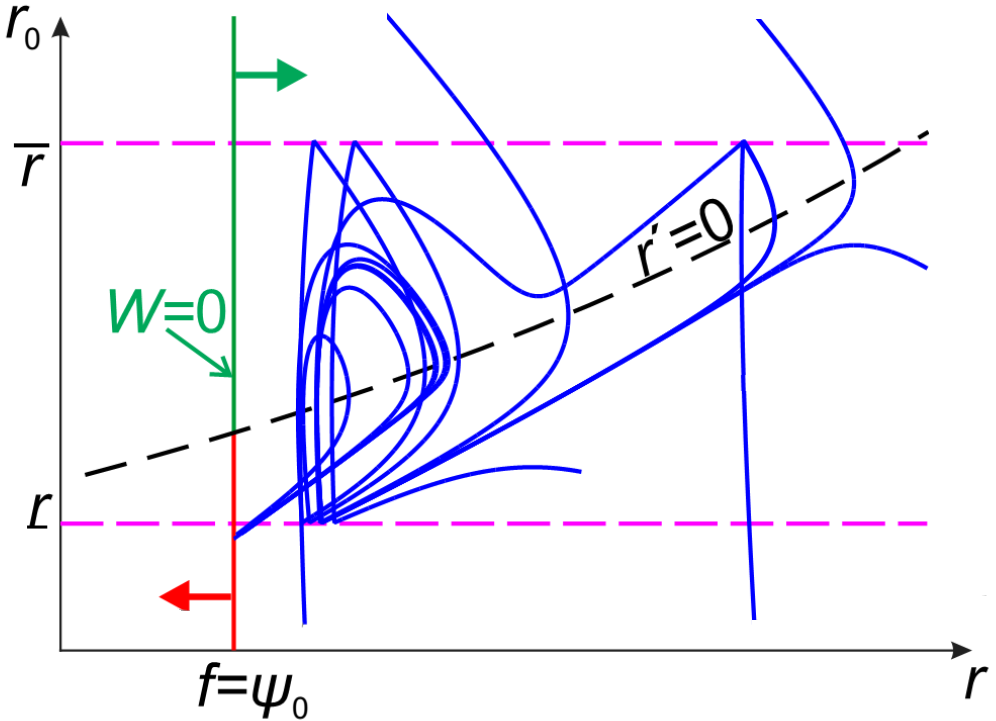} }
\caption{Oscillations near the border of death for 3D subsystem in projection onto the $(r,r_0)$ plane (the third coordinate, $r_{rv}$, is hidden). (a) Stable oscillations; (b) Oscillating death. Several trajectories for different initial states are plotted together for each case (a), (b). Obtained by numerical integration of the model. For detail and more examples see \cite{GorbanTyukinaDeath2016}). \label{cycle}} 
\end{figure}

A system of three blocks of the scheme shown in Fig. ~\ref{Fig:EqScheme}, from left to right: reserve, resource and allocated resource,  is a  3D  dynamic system with rich behavior. It seems that Selye and other physiologists underestimated the dynamic consequences of existence of additional reserve AE storage with on and off thresholds These thresholds lead to oscillations. 
For example, there are the stable closed orbits  (Fig.~\ref{cycle}a).   If we decrease the reserve recovering constant $k_{pr1}$ then the closed orbits loose stability and the oscillating death appears (Fig.~\ref{cycle}b). More examples are presented in \cite{GorbanTyukinaDeath2016}.

Fig.~\ref{cycle}b demonstrates an important effect: the trajectories spend a long time near the places where cycles appear for different values of constants (see Fig.~\ref{cycle}a) and go to the attractor (here it is death) after this delay. The delayed relaxation is a manifestation of the so--called `critical retardation': near a bifurcation with the appearance of new $\omega$-limit points, the trajectories spend a long time close to these points \cite{GorbanSingularities}.

 The phenomena of the oscillating remission and oscillating death  have been observed in clinic for a long time and now attract attention in mathematical medicine and
biology. For example, \cite{Zhang2014} demonstrated recently, on a more detailed
model of adaptation in the immune system, that cycles of relapse and remission, typical
for many autoimmune diseases, arise naturally from the dynamical behavior of the system.
The notion of `oscillating remission'  is used also in psychiatry \citep{Gudayol-2015}.

For extracting oscillating death and recovering from statistical data, synchronization is useful. Without phase synchronization, averaging  smears the periodic component.  Synchronization helps to extract the periodicity. There exist  some medical data sets with obvious synchronization. For example, in trauma, there exists a selected moment -- the trauma event. Of course, in severe injuries, mortality is high at the beginning, and this large peak with rapid decay shows no fluctuations. However, for low-severity injuries (for example, for NISS severity scores 1-8 or for NISS severity scores 9, where NISS = New Injury Severity Score), the daily mortality rate -- the estimated probability of a patient's death per day $t$, provided that he survived for a day $t-1$, is definitely not a monotonous function of time and includes a pair of periods of decaying fluctuations with a period of 12-16 days.
This was demonstrated by \cite{Mirkes2016} on data from TARN - the Trauma Audit \& Research Network that has the largest trauma database in Europe. Another example is the oscillating mortality after cancer  surgery (high severity cohort) \cite{MansurOnco}. Here the sync event is an operation.

The models based on Selye's idea of deep AE demonstrate  the oscillating
remission and oscillating  death. These effects  do not need additional exogenous reasons and are the core effects of adaptation process.  In further research \cite{Garkavi1979} additional  levels of adaptation processes were introduced at a lower intensity of stressors, and a ``periodic table'' of adaptation reactions was developed. According to them, the `chain of AE supply' in  Fig.~\ref{Fig:EqScheme} should be extended to the left to many storages and taps with thresholds.
  
Despite the interesting dynamical  effects, the AE resource and reserve model  (three left blocks in Fig.~\ref{Fig:EqScheme})    has an   important drawback.
They do not describe training. Training means, in particular, that the possible factor load that organism can resist may be increased after some experience.

 For example, in Selye's experiment, rats pretreated with a certain agent will resist such doses of this agent in the future. At the same time, their resistance to toxic doses of that one decreases \cite{SelyeAE1}. That is, the resistance to noxious agents can be trained and the training effect lasts for a rather long time. At the same time, this training decreases the ability to resist other agents.   In a recent review, the applicability of the concept of AE to the training process was questioned  \cite{Vasenina2020}:  ``It seems unlikely that adaptation energy can explain adaptation and acute responses to exercise. Instead, the concept of adaptation energy appears to share similarities with some theories of aging.'' We agree that the previously developed models do not describe training, but the rigthmost block of the model  Fig.~\ref{Fig:EqScheme} aims to solve this problem. AE is spending not only for neutralization of the harmful factor but to extending the zone of surviving measured by the logarithm of its volume, $\psi_0$. The logarithm of volume in the factor space where life is possible can be considered as entropy and training is an extension of this volume. The new terms are simple and it is worth to look on them separately. (For the complete system, see Fig.~\ref{Fig:EqScheme} and Appendix~\ref{Sec:AEkinur}.):
 
 \begin{equation}\label{Systraining} 
\begin{split}
&\frac{d r}{d t}=\ldots - k_{tr} r \frac{\psi}{\psi_0}W_0\left(\frac{\psi}{\psi_0}\right);\\
&\frac{d \ln \psi_0}{d t}=-k_{detr}+\alpha k_{tr} r \frac{\psi}{\psi_0}W_0\left(\frac{\psi}{\psi_0}\right),
\end{split}
\end{equation}
 where  $\alpha$ is the coefficient of training  efficiency ($0<\alpha <1$), $W_0(x)=1-x$ for $0\leq x \leq 1$.
 
 The main assumptions behind  this entropic approach are:
\begin{itemize}
\item Training is a medium-term adaptation process with transformation of the AE into entropic form, $\ln \psi_0$.
\item Following Selye, this transformation is irreversible and the transformed energy cannot be spent to other purposes.
\item Life length of this entropic form is determined by the detraining constant $k_{detr}$.
\end{itemize}
  
Mathematical modeling of athletic training and performance attracted much attention in physiology of sport and exercises. The models are considered as   effective conceptual and practical tools for evidence-based exercise (see the review \cite{ClarkeSport2013}).  Performance models  are used as the basis for training design and training planning. We expect some analogy between these models and the learning process models that can be extracted from the model  Fig.~\ref{Fig:EqScheme} and (\ref{Systraining}), in particular.
One of the key concepts in training models is `Critical Power' (CP), that is, the maximum speed of work that can theoretically be performed for an infinite time and corresponds to the maximum sustained aerobic power. During exercise at power above CP, there is a clear and progressive loss of metabolic homeostasis. This CP can be related to our $\psi_0$.
 It can be worth to mention that  for humans characteristic detraining time is evaluated as  $1/ k_{detr}\approx 8-10$ weeks in some athletic training exercises \cite{ClarkeSport2013}.
 
A natural question arises: if the resource allocation $ r $ is fixed, what is the  relationship between the training load $ \psi $ and the dynamics $ \psi_0 $, which measures the ability to survive under uncompensated pressure of the harmful factor?
The answer is given by the nullcline of the equation for $\psi_0$  (\ref{Systraining}):
 The positive effect of the training process (the growth of $\psi_0$) holds for 
\begin{equation}\label{trainiter}
\frac{1}{2}-\sqrt{\frac{1}{4}-\frac{k_{detr}}{\alpha k_{tr} r}}\leq \frac{\psi}{\psi_0} \leq \frac{1}{2}+\sqrt{\frac{1}{4}-\frac{k_{detr}}{\alpha k_{tr} r}}
\end{equation} 
if the available resource  $ r>4k_{detr}/\alpha k_{tr}$. If this inequality does not hold then  training is impossible and  $d\psi_0/dt \leq 0$ for all values of $\psi$. For a positive training effect (growth $\psi_0$), the uncompensated factor pressure $\psi$ must be in the interval (\ref{trainiter}) with the midpoint at $\psi_0/2$. The width of this interval depends on the allocated resource $r$. If this resource decreases, the interval width decreases and becomes zero at some final value of $r$. After that, the training may just slow down the decline of $\psi_0$. The optimal training load $\psi$ remains half of the maximum load
$\psi_0$. This example demonstrates the applicability of the model presented in Fig.~\ref{Fig:EqScheme}  to qualitative modeling of training. To obtain accurate quantitative results, it is necessary to work out in more detail the interpretation of the model in directly observed quantities and the estimation of constants.

 \section{Conclusion and outlook \label{Sec:Conclud}}

In many works of various groups of researchers, it has been shown that the dynamics of correlations and variances in the ensembles of systems that adapt to the pressure of harmful factors has universal properties:  under stress, both correlations and variance increase. This behavior lasts till some critical intensity of stressors. After that, the correlations may decrease.  This effect is supported by many observations in ecological physiology, medicine, economics and finance and may serve for early diagnosis of crises.
Analyzing the correlation graph between attributes provides a tool for studying adaptation, stress and critical transitions.

These effects appear for sets of relevant attributes. These attributes can be different for different factors. Selecting the appropriate attributes is an important part of the analysis. There are several approaches to this choice. It can be done on the basis of preliminary  hypotheses about the adaptation mechanisms or attributes with higher relative change of their values can be selected. The structure of the correlation graph can also be used with heuristics: internal connections in a set of relevant attributes under stress are (on average) stronger than external connections.

We presented several examples of the opposite behavior. The reasons for such deviations may be different, from the unsatisfactory choosing of the relevant attributes to the special interactions in the system of factors, like synergy. In any case, these examples warn us that effect validation should be performed for any type of system, harmful factors, and attribute selection before practical application. 
 
Several attempts have been made to develop a theory for this effect. We focused here on the classical physiological idea about adaptation resource (AE) expressed by Selye. Combined with the idea of limitation, this approach explains the observations and predicts other effects for different organization of systems of factors (for synergistic systems of  factors). It also allows to describe the effect of training in a natural way. Other approaches, such as the idea of bifurcations, critical retardations and critical fluctuations, also explain some effects but still fail  to explain phenomena that were described by Selye as spending and exhausting of adaptation resource.

The idea of AE and its optimal distribution to neutralize harmful factors has  analogy with thermodynamics. These general thermodynamic models are supplemented by kinetic models of the production and expenditure of AE in individual adaptation.  

 Selye's `AE' is an abstract adaptation resource, the universal currency for adaptation. Two types (at least) of the adaptation resource supply are needed for this modeling: from `checking account' of the available resource (the superficial energy) and from `saving account' of the reserve (the deep energy). Existence of these two types determine rich family of dynamical regimes including limiting cycles and oscillating death.

Analysis of training models leads to introduction of {\em entropic} form of adaptation
 energy. This energy  is irreversibly pumped into the entropic reservoir, extends the volume of comfort zone (this is the training effect) and cannot be reassigned. Models with AE and adaptation entropy capture main phenomenological effects and can be used in the top-down modeling of physiological adaptation.
This new quantity has a transparent geometric interpretation. Consider the area in the space of factor values, where the organism can survive without pumping AE. Formally, this `area of survival' can be defined by positivity of fitness $W$, as an area of $\boldsymbol{\psi}$ where $W(\boldsymbol{\psi}>0$. Let the volume of this area be $V$. Then the  AE pumped in the entropic reservoir is $\ln V$.

Our goal was to understand the classical experiments and to build the basic top-down models of physiological adaptation. 
Now, after 33 years of the first publication \cite{GorSmiCorAd1st} and efforts of many dozens of researchers, the main problems become clear.

{\em Let us  first mention applications.} Application practice will change the systems of models and create new questions. Three groups of applications should be noted:
\begin{itemize}
\item Development of early warning indicators. The potential of using the correlation graph and its characteristics is clearly demonstrated for early warning of stress and crisis for various systems, from individual health to the health system, social tensions and financial crashes (see Sec.~\ref{Sec:CorRiskCri}). 
\item Application of the models of individual adaptation (Sec.~\ref{Sec:SelyeGoldKin}) to training and sport physiology. These models should be incorporated into the existing model system, possibly with modifications, and used for training design and planning.
\item Detailed analysis of the oscillating recovery and oscillating death. These phenomena are recently predicted by the Selye--Goldstone kinetic AE models (Sec.~\ref{Sec:SelyeGoldKin}) \cite{GorbanTyukinaDeath2016}  and give a key to dynamic of individual adaptation and new branch of predictive personalized medicine. The kinetic models provide a tool for description, prediction, and work with these phenomena.
\item Application of the dynamic and thermodynamic models of adaptation to the analysis of aging and development of new methods of optimal aging control. 
Healthy and successful aging, is not connected  with excessive capacity to cope with stressors  but rather those with capability to afford intermediate  response  to  stress,  i.e.  neither  too  weak  nor  too  strong,  in  relation  to  the environment \cite{Franceschi2000a}. In terms of AE distribution, this moderation can be seen as optimal use of AE for the purpose of longevity.
\item Development of specific models of various systems, such as cell cultures and isolated organs, that allow detailed experimental testing. Using these models in planning experiments. A first example is a simple model of biological neural networks was developed  that explains the generation of bursts in the networks of living culture \cite{TyukIudin2019}. Each neuron is characterized by  a dynamic exogenous energy variable  and neuronal activation probability  depends on the energy and on the external activating signals from the network.  
\end{itemize}

{\em Secondly, the fundamental problem of interpretation and physical meaning of AE} has been around for a long time. Selye introduced this concept as `adaptability', and the analogy with thermodynamics highlighted his idea, but was not confirmed by physical experiments or solid theory  \cite{SelyeAEN,SelyeAE1}. This elusiveness of the AE raised many questions and doubts. In our research, we considered this quantity as a coordinate on the `dominant path' of adaptation processes and avoided direct physical interpretation \cite{GorbanTyukinaDeath2016}.

Surprisingly, the physical idea of AE (or rather, free adaptation energy) has recently been revived in the physical theory of adaptive systems \cite{Lan2012,Allahverdyan2016}. The relationship between Selye's AE and the expenditure of physical free energy on adaptation should be clarified.  Now we can hypothesize that this physical quantity can be considered as the lower bound of the adaptation energy, just as the physical efficiency coefficient is the upper bound of the efficiency coefficients of real devices.

{\em The third group of problems is related to the dynamics of correlations and variances `on another side of crisis'},  Fig.~\ref{Fig:CorrAdaptNonMon}. The   cases near fatal outcome are explored in much less detail than the initial stages of stresses, diseases and crises of adaptable systems. More attention is needed for this area because understanding of death is necessary for deep understanding of life.

{\em Detailed parametric analysis of phase portraits} for dynamics of individual adaptation with training   Fig.~\ref{Fig:EqScheme} and for the simpler resource--reserve dynamics     will answer many practically important question about general adaptation syndrome that remain open for decades \cite{GP_AE1952,GP_AE1952C}. On the other hand, this analysis can bring new ideas and benchmarks to dynamics of adaptive systems \cite{Tyukin2011}.

{\em Development of optimal control methods for adaptation dynamics}  will create a new discipline, ``adaptation engineering'', a special branch of applied physiology and personalized preventive health care.  
 
 {\em Analysis and classification of systems of factors} requires a lot of work. Special attention should be paid to such {\em universal physiological factors} as inflammation, which can be considered as a universal mechanism for accelerating aging (`inflammaging'  \cite{Franceschi2000}). The model ``universal resource-system of universal factors'' that describes the dynamics of the spending of AE for protection from several universal harmful factors  can be useful in various areas of physiology. A simple quasistatic (`thermodynamic')  theory of factor's interactions is presented in Section~\ref{SubSec:FactorResource}. For the future applications to monitoring and control of individual adaptation, it is necessary to develop dynamic models of the interaction of several factors and AE. The problem here is that there are too many possibilities for a formal description of this interaction and additional empirical clues are needed.
 
{\em Finally, a systematic search for counterexamples to the main effect is required.} The analysis of these contradictory cases will give us new ideas and will stimulate the development of theory and the discrimination between different theoretical ideas. It is now possible to indicate several potential reasons for the violation of the main effect 
(see Fig.~\ref{Fig:CorrAdapt}):
\begin{itemize}
\item Inappropriate choice of the relevant variables;
\item Approaching the `point of no return' on another side of crisis (see Fig.~\ref{Fig:CorrAdaptNonMon});
\item Synergy between several harmful factors (Sec.~\ref{SubSec:Synerg}).
\end{itemize}

The analysis of counterexamples will help in solving a really complex and difficult conceptual problem: choosing relevant attributes. There are many plausible algorithms and heuristics for identifying these variables, from simple common sense, intuition, and professional experience to some formalized algorithms with several tunable parameters.
In real multidimensional problems, the variety of solutions can be enormous. 

 This problem should not be considered as purely technical. It has an intrinsic connection with the deep problem of representing a system (an organism or other adaptable system) as a construction of relatively independent functional subsystems. Factor analysis and principal component analysis are used to identify  such subsystems. Independent component analysis (ICA) was specially invented for separation of a complex signal into signals from independent subsystems  \cite{Hyvarrinen2001} (For  the technical aspects of ICA applications such as  determining the optimal number of components  and improving reproducibility of the results see \cite{Sompairac2019}).  The theory of functional systems in biology was developed before the big data revolution came  \cite{Anokhin1964,Anokhin1968,Sudakov2004}. The configuration of subsystems is dynamic and changes over time and under the pressure of various factors. Now is the time to revive functional systems theory as a data-driven discipline closely related to machine learning  \cite{Redko2004}.
 
 The selection of relevant features can have a profound sense of revealing relevant subsystems.  Representation of the organism as a `universe' of flexibly joining and splitting functional subsystems is very attractive. It could be suitable to cite here the old wisdom:   ``The land under heaven, after a long period of division, tends to unite; after a long period of union, tends to divide.'' (The ``Romance of the Three Kingdoms'', a historical epic   fiction written in the fourteenth century.)

\section{Acknowledgments}
The work was supported by the University of Leicester and the Ministry of Science and Higher Education of the Russian Federation (Project No. 14.Y26.31.0022).  
 
\appendix

\section{Thermodynamic cost of homeostasis \label{SubSec:ThermodCost}}

 Selye experiments added a seminal idea to the homeostasis control loops:  adaptation has a cost. Surprisingly,  a simple thermodynamic estimate  of the cost of homeostasis in the terms of free energy is possible.
 
 Consider a non-equilibrium physicochemical system. Let its internal state be described by a vector of extensive variables $N$. If the system is closed and free energy is not added, then it goes to equilibrium. For living systems, this necessarily means death. For dynamics of the system without external fluxes we can write a general dynamic equation:
 \begin{equation}\label{simpleDyn}
 \frac{dN}{dt}=W(N).
 \end{equation}
This equation must satisfy the first and second laws of thermodynamics: the conservation laws must be satisfied, and the production of entropy must be positive. For isothermal conditions the second law means a decrease in free energy: the convex function $G(N)$ is defined and
 \begin{equation}\label{simple2nd}
 \frac{dG}{dt}\leq 0,
 \end{equation}
the time derivative  of $G$ by virtue of the system (\ref{simpleDyn}) is non-positive. Let us denote this decrease rate of $G$ in the closed system by $-\sigma(N) $: 
$$\sigma (N)=-(\nabla G\left|_{N}\right.,W(N))\geq 0.$$

Suppose that we aim to stabilize the non-equilibrium state $N^{\rm st}$ by external fluxes. For this purpose, $N^{\rm st}$ should be a steady state of the system  (\ref{simpleDyn}) with added external fluxes:
  \begin{equation}\label{simpleFlux}
 \frac{dN}{dt}=W(N)+v(N_{\rm in}- N),
 \end{equation}
  where $v$ is an intensity of flux, and $N_{rm in}$ is a vector of input composition.
If  $N^{\rm st}$ is a steady state of this system then the time derivative  of $G$ by virtue of the system (\ref{simpleFlux}) must vanish. This means that
\begin{equation}\label{sigma}
\sigma(N^{\rm st})=  v(\nabla G\left|_{N^{\rm st}}\right.,N_{\rm in}- N^{\rm st}).
\end{equation}
  According to Jensen's inequality for convex function $G$, 
 \begin{equation}\label{Jensen}
 ( \nabla G\left|_{N^{\rm st}}\right.,N_{\rm in}- N^{\rm st}) \leq
G(N_{\rm in})-G(N^{\rm st}).
\end{equation}
  Combining (\ref{sigma}) and (\ref{Jensen}) we obtain the inequality
 \begin{equation}\label{cost}\boxed{
 v(G(N_{\rm in})-G(N^{\rm st}))\geq \sigma.}
 \end{equation}
 The left part of (\ref{cost}) is the inflow of free energy into the system, and the right part is the dissipation rate for a closed system in the same state $N^{\rm st}$ (this state is, of course, the transition state of the closed system).

That is, the inflow of free energy required to stabilize the $ N^{\rm st} $ state in an open system must be greater than or equal to the rate of dissipation of free energy in a closed system at the same state  $N^{\rm st}$.

 It should be stressed that the inequality (\ref{cost}) is necessary but not sufficient condition for stabilization. Various versions of this inequality are  used in many works about thermodynamics of homeostasis.

\section{Advection-diffusion model of adaptation dynamics near a border of existence \label{Advection}}

We used the idea of adaptation resource to demonstrate how can the
models of adaptation explain the observable dynamics of correlation
and variance under environmental loads. Another theoretical approach
to explain these effects was proposed in
 \cite{RazzhevaikinZVMF} and developed further with applications in \cite{Razzhevaikin2008,Shpitonkov2017}. The idea is quite simple and
attractive. Let us represent the population of systems by points in
a bounded set  $U \subset \mathbb{R}^q$ with smooth boundary.
Coordinates correspond to the internal characteristics of the
systems. The whole population is described by a continuous density
function $u(x)$ in $U$.

The hypothetical behavior of the individual system can be described as a random walk in
$U$, and the density dynamics satisfies the correspondent Fokker--Planck equation
\begin{equation}\label{FP}
\partial_t u= - (\nabla, b u)+ a \triangle u \ ,
\end{equation}
where $a>0$ is a constant diffusion coefficient and $b$ is a
drift vector. This is a simple advection-diffusion equation with the constant advection speed $b$. After analysis of various optimality conditions, the
asymptotic behavior of solution of Eq.~(\ref{FP}) was studied for
$b=fb_0$, $b_0= {\rm const}$ ($\|b_0\|=1$), $ f \to \infty$, and for
no-flux boundary conditions $(bu-a \nabla u, \nu)_{\partial U}=0$
(where $\nu$ is the outer normal to $\partial U$). The parameter $f$
models the pressure of environmental factors. For large $f$, the
drift presses the distribution to the boundary of $U$, dimension
decreases, and correlations increase. 
This picture is schematically represented in Fig.~\ref{ModelFactorPressure}.

The asymptotics for large $f$
is determined by a tangent paraboloid to $\partial U$ at the point
$x_b$, where the linear function $(b,x)$ achieves the maximal value
on $\partial U$ (generically, such a point is unique). At this point
$\nu= b_0$. The asymptotic of the solution coincides in the main
term with exact analytical solution to the Fokker--Planck equation
(\ref{FP}) with constant coefficients in a paraboloid.

Let us select a rectangular coordinate system with origin at the
point $x_b$, such that in this system $b=(0,...,0,-A)$ and $U$ is
given by the inequality $$x_q \geq \sum_{i=1}^{q-1} \lambda_i x_i^2 +
o\left(\sum_{i=1}^{q-1} x_i^2\right).$$ In this coordinate system,
the solution to the problem is $$u(x)=v \exp \left(-{f
x_q}/{a}\right)$$ with constant $v$. For large $A$, the variance of
$x_q$, $\sigma_q^2\sim  {a^2}/{f^2}$, the marginal distribution
of $x_1,...x_{q-1}$ is almost Gaussian  with
$\sigma_i^2={a}/{f\lambda_i}$, and these variables are almost
independent. We can see that the concentration ellipsoid for
$x_1,...x_{q-1}$ does not change its form with increase of $A$: the
ratio ${\sigma_i^2}/{\sigma_j^2}={\lambda_j}/{\lambda_i}$ is
independent of $f$. On the other hand, contraction in direction
$x_q$ goes faster, and ${\sigma_q^2}/{\sigma_i^2} \to 0$ as
${a \lambda_i}/{f}$ ($i=1,...q-1$).

To check which mechanism is closer to reality we have to resolve the
``disk--or--bullet" alternative: How does the correlation increase:
(i) the largest eigenvalues of the correlation matrix increase and
the ellipsoid of concentration becomes stretched-out in small amount
of dimensions (presumably, in one, that is the bullet form) or (ii)
the smallest eigenvalue goes fast to zero, and the ellipsoid of
concentration becomes flattened out (the disk or the pancake form).
The existing data ensure us that this alternative should be resolved
in favor of bullet: under the pressure of environment factors, the
small amount of largest eigenvalues increases. It is necessary to
mention that the variance decreases in the Fokker--Planck model, that also contradicts the majority of existing observation.

\section{AE kinetic equations \label{Sec:AEkinur}}

Here, we present the system of equations that correspond to the scheme Fig.~\ref{Fig:EqScheme}. The main variables are three amounts of AE: $r$ --  AE allocated for factor neutralization; $r_0$ -- AE available at low level of alarm (`superficial'), $r_{rv}$ -- AE accumulated in reserve, and $\beta \ln \psi_0)$ -- entropic form of AE, proportional to logarithm of the area of surviving in the space of factor values.  The system is presented below:
 
 \begin{equation}\label{SysEntropy}\boxed{
\begin{split}
&\frac{d r}{d t}=-k_d r+kr_0\frac{\psi}{\psi_0}-k_{tr} r \frac{\psi}{\psi_0}W_0\left(\frac{\psi}{\psi_0}\right);\\
&\frac{d r_0}{d t}=-k_{d0}r_0 - kr_0\frac{\psi}{\psi_0}+k_{rv}B_{o/c}r_{rv}(R_0-r_0)+
k_{pr}(R_0-r_0)W_0\left(\frac{\psi}{\psi_0}\right);\\
&\frac{d r_{rv}}{d t}=-k_{d1}r_{rv}-k_{rv}B_{o/c}r_{rv}(R_0-r_0)+
k_{pr1}(R_{rv}-r_{rv})W_0\left(\frac{\psi}{\psi_0}\right); \\
&\frac{1}{\psi_0}\frac{d \psi_0}{d t}=-k_{detr}+\alpha k_{tr} r \frac{\psi}{\psi_0}W_0\left(\frac{\psi}{\psi_0}\right),
\end{split}}
\end{equation}
where $W_0(x)=1-x$ for $0\leq x \leq 1$ (this is the scale-invariant form of $W$),   $\alpha$ is the coefficient of efficiency of this training ($0<\alpha <1$), $k_{\ldots}$ are the rate constants.  The Boolean variable $B_{o/c}$ describes the state of the tap between the  reserve and the resource storages  (Fig.~\ref{Fig:Hysteresis}). 
 
System (\ref{SysEntropy}) formalizes the main general ideas of Selye, Goldstone and other physiologists and describes phenomena related to  dynamics of adaptation to a harmful factor. The detailed parametric analysis of this system and evaluation of its parameters for real life examples is now a priority task. Goldstone's `GP challenge' to the theory of general adaptation syndrome \cite{GP_AE1952,GP_AE1952C} should be met.  To meet them completely, we must also study the dynamic interaction of adaptation to several factors. However, to create such dynamic models, it is necessary to analyze, identify, and validate the dynamic model of adaptation to a single factor (\ref{SysEntropy}) in more detail.

\end{document}